\documentclass[11pt,a4paper]{article}
\pdfoutput=1
\usepackage{jheppub}

\usepackage{graphicx}
\usepackage{amsmath}
\usepackage{amssymb}
\usepackage{braket}
\usepackage{comment}
\usepackage{float}
\usepackage{color}

\usepackage[utf8]{inputenc}

\usepackage{subfigure}
\newcommand{\bea}{\begin{eqnarray}}
\newcommand{\eea}{\end{eqnarray}}
\newcommand{\be}{\begin{equation}}
\newcommand{\ee}{\end{equation}}
\newcommand{\ba}{\begin{align}}
\newcommand{\ea}{\end{align}}
\newcommand{\V}{\mathcal{V}}

\newcommand{\Kahler}{\ensuremath{\text{K}\ddot{\text{a}}\text{hler}\,}}

\newcommand\Tstrut{\rule{0pt}{2.6ex}}         
\newcommand\Bstrut{\rule[-0.9ex]{0pt}{0pt}}   

\preprint{LMU-ASC 35/18}

\title{Moduli Stars}

\author[1]{Sven Krippendorf}
\author[2]{, Francesco Muia}
\author[2,3]{and Fernando Quevedo}

\affiliation[1]{\small \it Arnold Sommerfeld Center for Theoretical Physics, LMU, Theresienstr. 37, 80333 M\"unchen, Germany}
\affiliation[2]{\small \it ICTP, Strada Costiera 11, Trieste 34014, Italy}
\affiliation[3]{\small \it DAMTP, Centre for Mathematical Sciences, Wilberforce Road, Cambridge, CB3 0WA, UK}

\emailAdd{sven.krippendorf@physik.uni-muenchen.de}
\emailAdd{fmuia@ictp.it}
\emailAdd{fq201@damtp.cam.ac.uk}

\abstract{
We explore the possibility that (Bose-Einstein) condensation of scalar fields from string compactifications can lead to long-lived compact objects. Depending on the type of scalar fields we find different realisations of star-like and solitonic objects. We illustrate in the framework of type~IIB string compactifications that closed string moduli can lead to heavy microscopic stars with masses of order $\mathcal{V}^\alpha M_{\rm Planck}$, $\alpha=1,3/2,5/3$ where $\mathcal{V}$ is the volume of the extra dimensions.  Macroscopic compact objects from ultra-light string axions are realised with masses of order $e^{\mathcal{V}^{2/3}}M_{\rm Planck}.$ Q-ball configurations can be obtained from open string moduli whereas the closed string sector gives rise to a new class of  solutions, named PQ-balls, that arise in the two-field axion-modulus system. The stability, the potential for the formation, and the observability of moduli stars through gravitational waves are discussed. In particular we point out that during the early matter phase given by moduli domination, density perturbations grow by a factor $\mathcal{V}^{\beta}$ with $\beta>2$ and non-linear effects cannot be neglected.
}


\arxivnumber{}

\begin{document} 
\maketitle

\section{Introduction}

The recent direct detection of gravitational waves (GWs)~\cite{Abbott:2016blz} has opened a new window of observation for physical phenomena in which gravity is the dominant interaction. Collisions of black holes~\cite{Abbott:2016nmj, Abbott:2017vtc, Abbott:2017gyy, Abbott:2017oio} and neutron stars~\cite{TheLIGOScientific:2017qsa} have been observed and a plethora of new events, even involving new physics, are expected to be detected in the next few years. Furthermore, the on-going search for new compact objects such as exo-planets will provide vast amount of new data over the upcoming decades (e.g.~from GAIA~\cite{gaia}). It is natural to explore alternative physical objects that may exist which are different from the standard stars and black holes and that could lead to particular imprints on the GW spectrum. Boson stars, Bose-Einstein condensates of gravitationally coupled scalar fields, are well motivated alternatives that have been discussed for several decades. A relatively vast literature already exists  on different realisations of boson stars (for reviews see~\cite{Jetzer:1991jr,Liddle:1993ha,Schunck:2003kk,1202.5809}). Similar arguments hold for fermion stars formed from fermions in beyond the standard model physics~\cite{Narain:2006kx}.

String theory, being a fundamental theory of gravity, has the potential to make physical predictions that can be tested only through gravitational couplings. The (complex) closed string moduli superfields of string theory provide a generic sector that couples only gravitationally. The corresponding real scalar particles tend to survive at low energies with masses of the order of the gravitino mass and below (e.g.~\cite{deCarlos:1993wie, GomezReino:2006dk}), phenomenologically required to be larger than $	\mathcal{O} (30 \, \text{TeV})$~\cite{Coughlan:1983ci, Banks:1993en, deCarlos:1993wie}\footnote{We comment in due course on how the relaxation of this constraint affects the spectrum of exotic compact objects.}. The axion components can have a much larger range of masses, in particular they can be lighter and can lead to a realisation of ultra-light axions. Furthermore fermionic partners of the moduli fields tend to have masses of order the gravitino mass and may in principle contribute to fermion stars. Open string modes may also provide candidates for boson and fermion stars.

A typical string compactification has hundreds or thousands of complex moduli fields that have different properties and each of them may lead to completely different physics (e.g.~dilaton, complex structure moduli, K\"ahler moduli which in turn can be blow-up modes, fibre moduli, etc.). In this work we explore the possibility that moduli can compose compact objects and in particular we focus on star-like solutions from string moduli which can be called \textit{moduli stars}. We briefly comment on the possibility for their fermionic partners, the `modulini' as well as the gravitino can give rise to a new class of fermion stars.

The effective field theory emerging for string moduli often leads to non-generic couplings from a standard QFT point of view and may point at regions in theory space which might be neglected otherwise. One scope of this article is to explore which exotic compact objects can be achieved from string effective field theory. In this sense, the potential discovery of other types of compact objects (e.g.~relying on EFTs not attainable in string theory) would challenge current scenarios of string compactifications, similar in spirit to the swampland discussion in the context of large field inflation~\cite{Vafa:2005ui}.

Stars composed of bosonic particles have been studied using a hypothetically long-lived complex or real scalar field~\cite{Kaup:1968zz, Ruffini:1969qy}. In this case the stability is not due to the Pauli exclusion principle as in fermion stars, but due to the Heisenberg uncertainty principle that constrains momenta to be bound by the inverse radius of the star. In each case the structure of the boson star (i.e.~radius and mass) varies substantially depending on the self couplings of the boson fields. For a free massive complex scalar field $\Phi$, the maximum mass of the boson star is $M_{\rm max}\simeq\left(10^{-10}\,{\rm eV}/m_{\rm boson}\right) \, M_{\odot}$ and these objects are called \textit{mini-boson stars}. This value of the mass can be enhanced if the scalar field has attractive interactions of the form $V_{\rm interaction} = - \frac{g}{4!} \left|\Phi\right|^4$ that dominates over the mass term, i.e.~if $\frac{\left|V_{\rm interaction}\right|}{m^2 \left|\Phi\right|^2} \gg 1$. In this case the maximum mass can be enhanced to $M_{\rm max} \simeq \left(\frac{10 \, \text{MeV}}{m_{\rm boson}}\right)^2 \, \sqrt{g} \, M_{\odot}$~\cite{Colpi:1986ye}.
 
Additional stability to compact objects can be achieved if an additional symmetry (e.g.~a global U$(1)$ symmetry) protects the bosonic condensate. The corresponding Noether charge is conserved and prevents the condensate to decay. Non-topological solitons such as Q-balls are the prime examples\,\cite{Coleman:1985ki}, that are already stable before turning gravity on. Furthermore the global U$(1)$ symmetry allows for a time dependence of $\Phi$, $\Phi(x,t)=\varphi(x) e^{i\omega t}$ while keeping a static spacetime metric (a constant time translation is compensated by a U$(1)$ transformation). For real scalars there is no conserved charge and stability is not automatic. 

In order to give rise to a long-lived compact objects, the corresponding particle has to be quasi-stable and could contribute to dark matter. String theory offers many dark matter candidates and quasi-stable particles, to name a few: matter fields from a hidden sector, moduli (including the dilaton and the many axions), the gravitino (see for instance~\cite{Halverson:2018xge}). But even if the particle decays relatively early in the history of the Universe it may still give rise to (relatively) long-lived compact objects that contribute to the energy density of the Universe for some time and may leave observational signatures such as GWs. We explore here this wide arena by giving explicit examples of axion stars, moduli stars and by discussing the realisation of Q-balls end extensions thereof in string theory.

The rest of this paper is organised as follows. In Section~\ref{sec:bosonstarsinfieldtheory} we review the basics of stars and Q-balls. Section~\ref{sec:StringCOs} is devoted to the string theory realisation of such objects and their phenomenology, including possible GW signatures. In Section~\ref{sec:formationall} we discuss the possible formation of compact objects, describing a possible new solution to the moduli field equations that could lead to the formation of compact objects (\ref{sec:SpinningAxion}), and discussing the possible formation of compact objects during an early matter era, which is generic in string models (\ref{sec:EarlyMatterEra}). We present our conclusions and outlook in Section~\ref{sec:Conclusions}.\\

Concerning the notation, we always use the $(-,+,+,+)$ convention for the metric signature. The Planck mass $m_{\rm p}$ is defined in terms of the Newton constant $G$
\begin{equation}
m_{\rm p} = \sqrt{\frac{\hbar c}{G}} \simeq 1.2 \times 10^{19} \, \text{GeV} \simeq 2 \times 10^{-5} \,g \,,
\end{equation}
and we will always take $\hbar = c = 1$. The reduced Planck mass $M_{\rm P}$ is defined through the relation
\begin{equation}
m_{\rm p}^2 = 8 \pi M_{\rm P}^2 \,.
\end{equation}
The Planck length is
\begin{equation}
\ell_{\rm p} = \sqrt{\frac{\hbar G}{c^3}} \simeq 1.6 \times 10^{-33} \,\text{cm} \,.
\end{equation}
The solar mass is
\begin{equation}
M_{\odot} \simeq 2 \times 10^{33} \, \text{g} \simeq 10^{57} \, \text{GeV} \,.
\end{equation}

We also report the value of one parsec
\begin{equation}
1 \, \text{pc} = 3 \times 10^{16} \, \text{m} \,,
\end{equation}
and some conversion rules
\begin{equation}
1 \, \text{GeV} \simeq 1.8 \times 10^{-24}  \, \text{g} \simeq 5 \times 10^{13} \, \text{cm}^{-1} \simeq 1.5 \times 10^{24} \, \text{Hz} \,.
\end{equation}
 
\section{Compact Objects in Field Theory}
\label{sec:bosonstarsinfieldtheory}

In this section we briefly review different types of compact objects which have been discussed in field theory and we classify them according to the mechanism that makes them stable against small perturbations. The first obvious example that we review is that of \textit{fermion stars} in which gravitational attraction is compensated by the fermion pressure coming from Pauli's principle, as in neutron stars. We then start the discussion of bosonic compact objects from non-topological solitons called \textit{Q-balls}~\cite{Coleman:1985ki, Lee:1991ax}, that exist for a complex scalar field with a global\footnote{See~\cite{Rosen, Lee:1988ag, Kusenko:1997vi} for gauged Q-balls.} U$(1)$ symmetry. The real countepart of Q-balls corresponds to pseudo-solitonic objects called \textit{oscillons}~\cite{Gleiser:1993pt} that we already studied in the string theory context in~\cite{Antusch:2017flz}. In both Q-balls and oscillons the repulsive gradient pressure is balanced by an attractive self-interaction. Moreover, the global U$(1)$ symmetry makes Q-balls composed of $\Phi$ particles absolutely stable, i.e.~they cannot decay into free $\Phi$ particles (although they can decay via couplings to other fields, see e.g.~\cite{Cohen:1986ct}). The absence of an analogous symmetry that stabilizes the localized configuration (see however~\cite{Mukaida:2016hwd}) makes oscillons long-lived but eventually they have to classically decay~\cite{Segur:1987mg, Fodor:2009kf}, radiating scalar waves to infinity\footnote{See~\cite{Hertzberg:2010yz} for an analysis of quantum effects.}. Q-balls and oscillons have a rich phenomenology in terms of GW production~\cite{Zhou:2013tsa, Antusch:2016con, Antusch:2017flz, Amin:2018xfe}, baryogenesis~\cite{Riotto:1999yt, Dine:2003ax, Buchmuller:2005eh, Lozanov:2014zfa}, dark matter~\cite{Kusenko:1997si, Burgess:2000yq}. While Q-balls and oscillons can exist in the regime in which gravity is negligible, for other classes of compact objects gravity is the force that stabilizes the repulsive gradient pressure, while self-interactions can be absent. This is the case of \textit{boson stars}~\cite{Kaup:1968zz, Ruffini:1969qy}, arising from a single complex scalar field, and of \textit{oscillatons}~\cite{Seidel:1991zh}, in the case of a real scalar field. To make a comparison with fermion stars the balance between gravity and gradients pressure in these objects can be thought as a consequence of Heisenberg's uncertainty principle. Analogously to Q-balls, boson stars can be absolutely stable in the presence of a global U$(1)$ symmetry, while oscillatons can be long-lived but they eventually have to decay like oscillons. One particularly interesting example of oscillatons is represented by axions stars~\cite{Kolb:1993zz, Kolb:1993hw}, that can arise in different contexts. Occasionally we will collectively refer to all the star-like solutions described here as ‘boson stars’, contrary to what is usually done in the literature, where the name boson stars refers specifically to the complex scalar field case.

Finally, axion-like particles provide an additional type of compact object, which is unrelated to the star-like or solitonic solutions mentioned so far. If the Peccei-Quinn (PQ) U$(1)$ symmetry is broken after inflation (i.e.~if the axion decay constant $f < H_{\rm inf}$, where $H_{\rm inf}$ is the Hubble parameter during inflation), the symmetry breaking mechanism generates large fluctuations in the axion field. In the case of the QCD axion ($m_{\rm QCD} \simeq 10^{-5} \, \rm eV$) it has been shown that these are so large that collapse before the start of matter domination (at $T \sim 0.75 \, \text{eV}$) giving rise to \textit{axion miniclusters}~\cite{Hogan:1988mp, Kolb:1993zz, Kolb:1993hw, Enander:2017ogx}. These objects would be much denser than the dark matter halos (even by a factor of $10^{10}$) but also quite rare, so that the probability of direct detection due to an encounter is very small. The typical size for QCD axion miniclusters is roughly $R_{\rm minicl.} \simeq 10^{13} \, \rm cm \sim 1 \, \rm UA$, while their typical mass would be $M_{\rm minicl.} \simeq 10^{-11} \, M_{\odot}$.

\subsection{Fermion Stars}
\label{sec:FermionStars}

Standard white dwarfs or neutron stars are understood in terms of a gas of fermions for which their degeneracy compensates for the gravitational attraction.
Following the argument of Landau for neutron stars~\cite{Narain:2006kx} for $N$ free fermions of mass $m_f$,  the total energy in a sphere of radius $R$ takes the form
\be
E(R)=-\frac{GMm_f}{R} + \left(\frac{9\pi}{4}\right)^{1/3} \frac{N^{1/3}}{R}\,,
\ee
where the first term describes the attractive gravitational potential and the second one is the (relativistic) kinetic energy of the fermion on the surface of the star. Here $M=Nm_f$ is the total mass of the star. We have used the relativistic limit in which the kinetic energy is roughly $k_f\gg m_f$ for a relativistic fermion of momentum $k_f.$ $k_f$ is determined by its relation to the number density in Fermi statistics: $N/(\frac{4}{3}\pi R^3)=k_f^3/(3\pi^2)$. If the second term of the equation above dominates the star expands until the fermion density is so small that the kinetic energy term becomes of  order $m_f$ and the gravitational interaction stabilises it. 

A rough estimate of the maximum mass and minimum radius of the star can be made by noticing that both energies are of the same order ($E(R)=0$) for a maximum  value of $N=N_{\rm max}$ 
giving a total mass:
\be
M_{\rm max}\sim \frac{M_{\rm P}^3}{m_f^2}~,
\ee
which gives the standard Chandrasekhar limit. The corresponding minimum radius can be estimated  by taking $k_f\sim m_f$:
\be
R_{\rm min}\sim \frac{M_{\rm P}}{m_f^2}.
\ee 
For a neutron with mass $m_{\rm N}\sim 1$ GeV these expressions give the standard results of $M_{\rm max}\sim M_\odot$ and $R_{\rm min}\sim 2$ Km.

\subsection{Boson Stars}
\label{sec:BosonStars}

Boson stars are solitonic-like solutions of the coupled Einstein-Klein-Gordon equations. The simplest case corresponds to a massive complex scalar $\Phi$ of mass m. The action is of the type:
\be
S=\int \sqrt{-g}\left(\frac{M_{\rm P}^2}{2} R-g^{\mu \nu}\partial_\mu \Phi \partial_\nu \Phi - V(|\Phi|)\right) \,,
\label{eq:actionbs}
\ee
with $V=m^2|\Phi|^2$.
A boson star would correspond to a spherically symmetric configuration with metric:

\be
ds^2= -A(r)^2 dt^2 + B(r)^2 dr^2 + r^2 \left(d\theta^2 + \sin^2\theta d\phi^2\right).
\label{eq:metricansatz}
\ee
A static spherically symmetric configuration for the scalar field would not give solitonic solutions due to Derrick's theorem. However, a stationary spherically symmetric scalar field of the form
\be
\Phi(r,t)=\Phi_{\rm R} (r)\, e^{i\omega t}
\ee
allows for a solution of the Einstein-Klein-Gordon equations from eq.~\eqref{eq:actionbs} with a static metric as above (time translations in $\Phi$ are compensated by a global U$(1)$ transformation $\Phi\to e^{i\alpha}\Phi$).

Contrary to fermion stars the gravitational attraction is compensated by the Heisenberg principle to prevent collapse. Naively this implies that  $\Delta x \Delta p\geq \hbar $ with $\Delta x= R$ and $\Delta p=mc$ for a boson of mass $m$ then the minimal radius is
\be
R_{\rm min}\sim \frac{1}{m}~.
\ee
From this we can obtain the maximal mass by setting the radius to the Schwarzschild radius $R=R_S=2GM$
\be
M_{\rm max}\sim \frac{M_{\rm P}^2}{m}~.
\ee
Comparing with fermion stars, fermionic stars are much heavier and larger than boson stars for fermions and bosons of the same mass. For instance, for a boson with a mass of a neutron $m\sim 1$ GeV the corresponding star radius is of order $R_{\rm min}\sim 10^{-15}$~cm and mass $M_{\rm max}\sim 10^{36}\,\rm{GeV}\sim 10^{-21}\, M_\odot$. To highlight that these objects are typically much lighter than $M_{\odot}$, they are usually called \textit{mini-boson stars}~\cite{Kaup:1968zz, Ruffini:1969qy}. However, if interactions are relevant this naive estimate can be modified~\cite{Colpi:1986ye}. For instance for a scalar field with quartic couplings
\be
V(|\Phi|)= \frac{1}{2} m^2 |\Phi|^2 - \frac{g}{4!} |\Phi|^4 \,,
\ee
the mass of the star becomes:
\be
\label{eq:MassEnhancement}
M\sim \tilde{g}^{1/2} \frac{M_{\rm P}^2}{m}\sim \frac{M_{\rm P}^3}{m^2} \,,
\ee
with $\tilde{g} = \frac{g M_{\rm P}^2}{m^2}$ the dimensionless quartic coupling. In this case the boson star mass takes the same form as the Chandrasekhar limit for fermion stars if $g\sim 1$, therefore allowing for macroscopic stars for scalar masses in the GeV range. 
 
\subsection{Oscillatons}
\label{sec:Oscillatons}
 
The pattern of bosonic compact objects may be substantially expanded by considering real scalar fields that we denote by $\varphi$~\cite{Seidel:1991zh}. As it is not possible to find a background field ansatz that makes the metric time-independent~\cite{UrenaLopez:2001tw, UrenaLopez:2002gx, UrenaLopez:2012zz}, the $t$-$t$ and $r$-$r$ components of the metric in eq.~\eqref{eq:metricansatz} become time-dependent. An equilibrium configuration of the star can be found by expanding the background field $\varphi(r,t)$ as well as the metric functions $A(r,t)$ and $B(r,t)$ in Fourier series. The corresponding solutions, called \textit{oscillatons}, have been found numerically and studied in different contexts~\cite{UrenaLopez:2001tw, UrenaLopez:2002gx, Alcubierre:2003sx, Guzman:2004wj}. The solutions depend crucially on the amplitude of the background field oscillations $\varphi_{\rm core} \equiv \text{max}\{\varphi(0,t)\}$. In this section we briefly describe the known results already contained in the literature and how they need to be modified to be extended to the case of string potentials.\\

We denote by $\Lambda$ the typical field range of the canonically normalized field in the potential under study. As an example, for an axion potential the scale $\Lambda$ would typically be $ \Lambda = 2 \pi f$, where $f$ is the axion decay constant. Along with the mass of the particle, the scale $\Lambda$ plays a crucial role as it determines the maximum energy that can be stored in a scalar field $\rho_{\rm max} \sim m^2 \Lambda^2$. As we will see more in detail in Sec.~\ref{sec:ModuliStars}, the scale $\Lambda$ sets the natural scale for the mass of a star composed by mass $m$ scalar particles
\begin{equation}
M/\tilde{M} \simeq \Lambda^2/m \,,
\end{equation}
where the dimensionless parameter $\tilde{M}$ has to be computed numerically and can span a few orders of magnitude. At the same time, the natural scale for the radius of the star is set by the scalar mass $m$
\begin{equation}
R/\tilde{R} \simeq 1/m \,,
\end{equation}
where again the dimensionless parameter $\tilde{R}$ has to be computed numerically and can span a few orders of magnitude. 
The density of the star is encoded in the compactness parameter
\begin{equation}
\label{eq:CompactnessDef}
C = \frac{M}{R} = \frac{\tilde{M}}{\tilde{R}} \Lambda^2 \,,
\end{equation}
where $\tilde{C} \equiv \tilde{M}/\tilde{R}$ is the dimensionless compactness\footnote{For instance the compactness of a black hole with Schwarzschild radius $R_{\rm S} = 2 G M$ is $C_{\rm BH} = 4 \pi$. Notice that in $m_{\rm p}$ units it would be $C_{\rm BH} = \frac{1}{2}$~\cite{Giudice:2016zpa}.}. It follows that the overall densest objects are typically those with $\Lambda = M_{\rm P}$: for a fixed value of $\tilde{C}$ the compactness is suppressed by a factor of $\left(\Lambda/M_{\rm P}\right)^2$. Given a fixed value of the scale $\Lambda$, stars with larger core amplitude are denser, as we will show explicitly in Sec.~\ref{sec:ModuliStars}. We then distinguish two regimes depending on the ratio between the $\varphi_{\rm core}$ and $\Lambda$~\cite{UrenaLopez:2002gx, Visinelli:2017ooc}:
\begin{enumerate}
\item{} {\it Dilute regime:} $\varphi_{\rm core} \ll \Lambda$.\\
In this regime self-interactions of the form $\left(\lambda_n/n!\right) \varphi^n$ (with $n \geq 3$) of the field are negligible. The system can be studied in the weak gravity approximation, in which case the metric components can be assumed to be static and can be expanded as $A^2(r) \sim 1+2\,\phi(r)$ and $B^{2}(r) \sim 1-2\,\phi(r)$ where $\phi(r)$ is the Newtonian potential~\cite{UrenaLopez:2002gx, Visinelli:2017ooc}. The equations of motion are then the Klein-Gordon equation for a massive real scalar field in the weak gravity regime coupled to the Poisson equation for $\phi(r)$. Equivalently, taking the field theory approach in the non-relativistic approximation the system reduces to a Schr\"odinger equation coupled to the Poisson equation~\cite{Guzman:2004wj, Guth:2014hsa}. In this limit the system of equations features a scale symmetry that makes the analysis particularly simple, see Sec.~\ref{sec:ModuliStars}.
 
\item{}{\it Dense regime:} $\varphi_{\rm core} \sim \Lambda$.\\
In this regime the self-interaction terms - if present - are important. The dense regime of the free massive case is fairly well understood: if $\Lambda = M_{\rm P}$ it features both a stable and an unstable branch~\cite{UrenaLopez:2002gx}. The oscillaton mass depends on the core amplitude and the maximum mass of a stable oscillaton in the free massive field case is
\begin{equation}
\label{eq:MaxOscillatonMass}
M_{\rm max} \simeq 0.607 \, \frac{m_p^2}{m} \simeq 15.26 \, \frac{M_{\rm P}^2}{m} \,,
\end{equation}
where $m$ is the mass of the real scalar field. This maximum value of the star mass corresponds to a core amplitude $\varphi_{\rm core}^{\rm max}/M_{\rm P} \sim 0.48$. Oscillatons with core amplitude smaller than $\varphi_{\rm core}^{\rm max}$ belong to the stable branch while those with larger core amplitude belong to the unstable branch. Oscillaton configurations perturbed around the unstable branch can either collapse to black holes or radiate energy and migrate back to the stable branch, depending on the perturbation. If self-interaction terms are present and $\Lambda = M_{\rm P}$ the numerics become extremely more involved and the study of a generic interacting potential is currently missing. Equilibrium configurations in the case of a repulsive quartic interaction has been studied in~\cite{ValdezAlvarado:2011dd, UrenaLopez:2012zz} for moderately large values of the dimensionsless coupling $\tilde{g} = \frac{g M_{\rm P}^2}{m^2}$ in the range $\tilde{g} \sim 1\text{-}4$. In this case the expected maximum oscillaton mass is enhanced but to numerically check the behaviour in eq.~\eqref{eq:MaxEnhancement} it would be necessary to probe the region of parameter space $\tilde{g} \gg 1$. Finally, dense solutions with $\Lambda \ll M_{\rm P}$ correspond to the regime in which gravity is negligible. In this case compact objects corresponding to oscillons can be formed in the presence of attractive self-interactions. As an example, oscillons formed in blow-up potentials studied in~\cite{Antusch:2017flz} belong to this case. In particular we stress that it is self-consistent to neglect gravity in that case.
\end{enumerate}

Oscillatons include the important case in which the real scalar is an axion-like particle giving rise to axion stars (see~\cite{Visinelli:2017ooc} and references therein for the state of the art). The Lagrangian is
\be
\mathcal{L}= - \frac{1}{2}\partial^\mu\theta \partial_\mu\theta -\mu^4\left(1-\cos \left(\frac{\theta}{f}\right)\right)\,,
\label{eq:bound}
\ee 
where $\mu$ is an energy scale generated by non-perturbative effects that break the original PQ shift-symmetry. If the leading interaction term is an attractive quartic term (e.g. $V_{\rm interaction} = - \left(g/4!\right) \varphi^4$) as for axion-like particles there is an additional regime for which $\frac{f}{8 \pi M_{\rm P}} \lesssim \frac{\varphi_0}{2 \pi f} \lesssim 1$, called the \textit{critical regime}~\cite{Visinelli:2017ooc}. In the critical regime the amplitude of the background field is still small but large enough such that the leading order self-interaction is stronger than gravity and balance the kinetic pressure from the uncertainty principle. Configurations in the critical regime are unstable against small perturbations: they either disperse or collapse to denser objects~\cite{Chavanis:2016dab, Helfer:2016ljl, Levkov:2016rkk}. The critical regime exists only if the quartic order self-interaction is attractive: in the repulsive case there is a single branch with $\varphi_{\rm core}/\Lambda < 1$ that is always stable~\cite{Schiappacasse:2017ham}. The dense regime of axion stars has first been studied in the Thomas-Fermi approximation that resulted to be not well justified~\cite{Braaten:2015eeu}. Recently, the it has been properly studied in full GR~\cite{Helfer:2016ljl}: it turns out that axion stars have a different evolution depending on their mass and on the axion decay constant: they can be (meta-)stable, collapse to black holes or disperse. One particularly interesting application of axion stars appears for an ultralight axion-like particle (ULA) with mass $m_{\rm ULA} \sim 1\text{-}10 \times 10^{-22} \, \rm eV$, which constitutes a good dark matter candidate called \textit{fuzzy dark matter}~\cite{Hu:2000ke} or \textit{ultralight dark matter} (ULDM). Interestingly, ULDM could address several issues arising in the cold dark matter case~\cite{Hui:2016ltb}, even though $m \lesssim 1\text{-}2 \times 10^{-21} \, \rm eV$ are in tension with observations of the Lyman-$\alpha$ forest~\cite{Viel:2013apy}. In particular, numerical simulations show that in the presence of ULDM solitonic cores of $\mathcal{O}\left(\text{kpc}\right)$ size are formed in dark matter halos~\cite{Schive:2014dra, Veltmaat:2018dfz, Levkov:2018kau}, potentially addressing the cusp-core problem of cold dark matter~\cite{Marsh:2015wka}. Such cores could also give rise to specific signatures~\cite{Hui:2016ltb}.\\

Notice that, as for axion stars, in the case of string potentials (typically given by the sum of exponentials) the scale $\Lambda$ implies that for a core amplitude $\varphi_{\rm core} \sim \Lambda$ all the interaction terms have to be included in the analysis. We have already studied the dense regime for blow-up potentials ($\Lambda \ll M_{\rm P}$) in~\cite{Antusch:2017flz}: in this paper we focus on the dilute regime for moduli potentials (both with $\Lambda \sim M_{\rm P}$ and with $\Lambda \ll M_{\rm P}$), and we will report the analysis of the dense regime for $\Lambda \sim M_{\rm P}$ moduli potentials (via a full GR simulation as in~\cite{Helfer:2016ljl}) in a forthcoming publication.

\subsection{Q-Balls}
\label{sec:QBalls}

Q-balls are particular cases of non-topological solitons which have been originally proposed in~\cite{Coleman:1985ki}. Let us consider a four-dimensional complex scalar field $\Phi$ with Lagrangian symmetric under a global U$(1)$:
 \be
 \mathcal{S}= \int d^4 x \left(\frac{1}{2} \partial^\mu \Phi \partial_\mu \Phi^* \, -\, V(|\Phi|)\right).
 \ee 
 The U$(1)$ Noether current and charge are:
 \be
 J_\mu = \frac{1}{2i}\left( \Phi^*\partial_\mu \Phi-\Phi\partial_\mu\Phi^*\right) \,, \qquad Q=\int d^3 x\, J^0=\frac{1}{2i}\int d^3 x\left(\Phi^*\dot{\Phi}-h.c.\right).
 \ee
 Assuming that $\Phi=0$ at the minimum of the scalar potential, it provides a $Q=0$ vacuum state. Configurations with charge $Q\neq 0$ can be obtained by minimising the total energy subject to a constant $Q$ constraint. That is we need to extremise the quantity:
 \be
 E_\omega=\int d^3 x \left(\frac{1}{2}|\dot{\Phi}|^2+\frac{1}{2}|\nabla\Phi|^2+ V(|\Phi|) \right)+\omega\left(Q-\frac{1}{2i}\int d^3 x\left(\Phi^*\dot{\Phi}-h.c.\right)\right),
 \ee
  where $\omega$ denotes a Lagrange multiplier. This expression can be rewritten as 
  \be
 E_\omega=\int d^3 x \left(\frac{1}{2}|\dot{\Phi}-i\omega \Phi |^2+\frac{1}{2}|\nabla\Phi|^2+ \hat{V}(|\Phi|) \right)+\omega Q \,,
  \ee
where 
\be
\hat{V}_\omega(|\Phi|)=V(|\Phi|)-\frac{1}{2}\omega^2 |\Phi|^2.
\ee
The kinetic term vanishes for :
\be
\Phi(x,t)=\Phi_{\rm R}(x) e^{i\omega t} \,,
\ee
which for real $\Phi_{\rm R}(x)$ provides a stationary configuration with time-independent but non-vanishing energy and charge. 

The task of extremising with respect to $\Phi_{\rm R}$ is the same as finding the tunneling solution for a 3-dimensional Euclidean action with potential $\hat{V}(\Phi_{\rm R})$. To simplify this task Coleman~\cite{Coleman:1985ki} assumed large $Q$ or the thin wall approximation such that the field $\Phi_{\rm R}$ has a value $\Phi_0$ (to be determined by minimising the energy) inside a region of volume $\text{Vol}$ and $\Phi_{\rm R}=0$ (the true vacuum) outside. In this approximation gradients are neglected and extremising $E_\omega $ with respect to $\omega$ gives $\omega_0 = Q/(\Phi_0^2 \, \text{Vol})$ and substituting into $E_\omega$ implies:
\be
E_{\omega_0}=V(\Phi_0) \, \text{Vol}+\frac{Q^2}{2 \Phi_0^2 \, \text{Vol}} \,.
\ee
Extremising now with respect to the volume $\text{Vol}$ leads to $\text{Vol}=Q/\sqrt{2 \Phi_0^2 V}$ and:
\be
E=Q\sqrt{\frac{2 V(\Phi_0)}{\Phi_0^2}} \,.
\ee
Therefore the value of $\Phi_0$ can be obtained by extremising the quantity: $V/\Phi_{\rm R}^2$. This coincides with the minimum of $\hat{V}$  (and therefore solves the equations of motion) for the value of $\omega=\omega_0=\sqrt{2 V/\Phi_0^2}$ as it can be easily verified. Notice that for this value of $\omega_0$ the value of $\hat{V}$ vanishes at the minimum $\Phi_{\rm R}=\Phi_0$ and so the new minimum is degenerate with the one at $\Phi_{\rm R}=0$ which remains a minimum of $\hat{V}$ as long as $\omega^2< \mu^2=V''(0)$. 

We then have that a charge $Q$ configuration with constant energy localised in a finite volume (the Q-ball) exists as long as there is a non zero minimum of the quantity $V/|\Phi|^2$. Since the energy per unit charge is less than the mass of a single charged particle ($\omega^2< m^2$), the Q-ball is stable against decay to a gas of individual particles.

Beyond the thin-wall approximation, a proper solution with non-vanishing gradient terms solving the field equation for $\Phi_{\rm R}$
\be
\Phi_{\rm R}'' +\frac{2}{r}\Phi_{\rm R}'+ \partial_\Phi V = 0
\ee
can be found numerically but inferred by standard tunneling solution techniques working with the analogy of a particle in the inverted Euclidean potential. Several examples including the thick wall case have also been found in the literature~\cite{Kusenko:1997ad}.

\section{Compact Objects from Strings}
\label{sec:StringCOs}

Let us start with the fermion stars. In string compactifications there are several classes of low energy fermions of mass $m$ that could be dark matter candidates and then can be the basis for exotic compact objects of maximum mass beyond which they can collapse to a black hole $M\sim M_{\rm P}^3/m^2$ and minimum radius $R\sim M_{\rm P}/m^2$. From the model independent closed string sector, the gravitino has a mass $m_{3/2}=M_{\rm P} W_0/\V$ which can be in the mass range from TeV to $10^{-2}\, M_{\rm P}$ and therefore a gravitino star of mass $M\sim \V^2 M_{\rm P}/W_0^2$ and radius $R\sim \V^2/(M_{\rm P}W_0^2)$.  Modulini, the fermionic partners of moduli fields, also have a mass $m\sim m_{3/2}$ leading to similar compact objects. In summary for TeV fermions coupled only gravitationally the corresponding stars would have maximum masses of order $M\sim 10^{48} \,\rm{GeV}\sim 10^{-9} \,M_\odot$ and radius $R\sim 10^{-3} \,\rm{cm}$.  In general for the Large Volume Scenario (LVS)~\cite{Balasubramanian:2005zx,Conlon:2005ki}, in the range $10^3\leq \V\leq 10^9$ for which the effective field theory is valid and the cosmological moduli problem is not present, we may have fermion stars with maximal mass and minimum radii in the range
\be
1 \,\text{g} \lesssim M\lesssim 10^{15} \, \text{g}\,, \qquad 10^{-27}\, \text{cm} \lesssim R \lesssim 10^{-15} \,\text{cm}~.
\ee
We then may have objects of the size of an atomic nucleus and as heavy as Mount Everest. There may be a more diverse variety of candidates from open strings. First potential dark matter candidates from the visible sector corresponding to axinos and neutralinos they tend to have masses either of order $m_{3/2}$ or, if they are sequestered in LVS, they can be as light as $m\sim M_{\rm P}/\V^2$~\cite{Blumenhagen:2009gk} and therefore their corresponding stars of mass and radius $M\sim M_{\rm P}\,\V^4$,  $R\sim \V^4\, \ell_{\rm p}$. This could lead to compact objects as massive as $M\sim 10^{-2} \,M_\odot  $ and radii $R\sim 10 \, \rm m$ for $\V\sim 10^9$ in string units. \\

We now turn to the realisation of exotic bosonic compact objects in string theory. String theory features many gravitationally coupled scalar fields called moduli. A generic feature of string compactifications is also the presence of several axionic fields that appear both as phases of open string moduli and as imaginary parts of closed string moduli. In particular, concerning the closed string sector, we are particularly interested in the two-field system composed by a modulus $\tau$ and the corresponding axion $\theta$, whose physics is captured by the following action
\begin{equation}
\label{eq:GenericAction}
\mathcal{S} = \int d^4 x\, \mathcal{L} = \int d^4x\, \left[- f(\tau) \left[\partial_\mu \tau \partial^\mu \tau + \partial_\mu \theta \partial^\mu \theta\right] - V(\tau, \theta)\right] \,,
\end{equation}
where the two fields can be identified as the real and imaginary parts of a complex modulus $T = \tau + i \theta$ and $f(\tau) = K_{T \overline{T}}$ is the second derivative of the \Kahler potential $K \equiv K(T + \overline{T})$. Axions can obtain a potential and mass from non-perturbative effects that break the PQ shift-symmetry or can be absorbed by gauge fields in a St\"uckelberg mechanism. In the case of closed string moduli we can organise the scenarios as follows depending on how the PQ shift-symmetry is broken:

\begin{enumerate}
\item{}
The first case is for the complex moduli to appear directly in the superpotential $W(\Phi)$ where $W$ can come from tree-level effects such as fluxes in Type IIB strings for complex structure moduli or from non-perturbative effects like the blow-up \Kahler moduli (or the overall volume in KKLT).  In this case both scalar and pseudoscalar (axionic) components of the superfield receive a potential and masses of the same order. The effective potential has no conserved current but the fields oscillating around their minima can give rise to boson stars.

\item{}
The second case corresponds to the scenario in which the superpotential does not depend on the modulus field. Therefore the scalar component receives a potential from perturbative effects and the pseudoscalar $\theta$ remains flat. In this case, there is a remaining global shift-symmetry corresponding to the standard PQ shift-symmetry for the axion component $\theta\to \theta + c,$ which is broken non-perturbatively, similar to the situation for the QCD axion. 
This scenario appears naturally for all \Kahler moduli for which a non-perturbative superpotential is hierarchically smaller than perturbative contributions to the potential. Examples are the overall volume and fibre moduli in the LVS~\cite{Cicoli:2008va}, for which the non-perturbative superpotential either does not exist or is exponentially suppressed compared to perturbative contributions arising from the K\"ahler potential. At the level of the perturbative contributions the axions are essentially flat directions.

\item{}
While axions remain flat if the corresponding real field is stabilized perturbatively, they can receive a small mass by non-perturbative effects. The third case corresponds to the study of the lightest axion after integrating all heavier moduli and axions out. This is a simple axion system such as in eq.~\eqref{eq:bound} with the stringy input provided by the expressions of the coefficients in the scalar potential in terms of the integrated string moduli. This is the only case suitable to discuss ultra-light axions in string compactifications since otherwise the mass of the real component of the superfield would be of the same order (as in the first case) and would be ruled out by fifth force constraints.
\end{enumerate}

In general, many compact object configurations can be obtained classically from moduli. However, when including the decay of moduli fields, the lifetime of such objects is limited. As a back of the envelope calculation indicates, such a star can be stable until today if the modulus mass is
\be
m \lesssim 10^{-2} \, \text{GeV} \,,
\label{eq:bound2}
\ee
where we assumed a gravitational decay rate $\Gamma\sim m^3/M_{\rm P}^2$. As axions are phenomenologically allowed to satisfy the bound in eq.~\eqref{eq:bound2}, compact axionic objects can in principle be still present in our Universe. Nevertheless, star-like objects could have been relevant in the early Universe. The observational consequences are highly model dependent, at first hand, and the estimate of, for instance, a stochastic GW background will be heavily dependent on the assumptions of the Early Unvierse history. We discuss this further later in Section~\ref{sec:formationall}. Here we discuss several examples which realise the field theory configurations from the previous section. We start with the discussion of axion stars (Sec. \ref{sec:StringyAxionStars}) and continue with moduli stars appearing from real scalar fields (Sec. \ref{sec:ModuliStars}). These are two examples of oscillaton solutions realized in string theory models and can be obtained from the action in eq.~\eqref{eq:GenericAction} in the regimes in which one of the field can be neglected. In particular, axion stars can be easily obtained in the third case described above, after the heavy moduli have been integrated out. Moduli stars on the other hand can be obtained by choosing a specific initial condition that fixes the axion into the origin\footnote{Let us however mention that even an initial displacement in the axionic direction would not change dramatically the results in the subsequent sections, as the equations decouple almost completely in the small amplitude (dilute) regime considered in this paper.}. We then present a realisation of Q-balls using open string fields (Sec.~\ref{sec:OpenStringsQBalls}) and finally an attempt to extend the concept of Q-balls to the two-field system of the second case described above (Sec.~\ref{sec:PQBalls}).

\subsection{Axionic Compact Objects}
\label{sec:StringyAxionStars}

Axion-like particles are a widely studied topic and well-motivated both from the bottom-up perspective and from the top-down approach. In fact, they are a key ingredient of the PQ mechanism to solve the strong CP problem of QCD~\cite{Peccei:1977hh, Wilczek:1977pj, Weinberg:1977ma} and their existence is a generic prediction of string theory~\cite{Conlon:2006tq,Svrcek:2006yi,Arvanitaki:2009fg}. Despite most of the literature emphasise the QCD axion case, many other options for the axion couplings and mass have been considered. The QCD axion could be obtained in string theory compactification mainly as the phase of open string moduli~\cite{Cicoli:2012sz} as getting the axion mass in the right ballpark from closed string axions is non-trivial. In this case miniclusters can be formed as in the field theory scenario if the PQ U$(1)$ symmetry is broken after inflation.

We focus on closed moduli axions in the following, restricting to the third case listed at the beginning of Section~\ref{sec:StringCOs}. The LVS provides a concrete and consistent example of a ULDM. The volume axion of the LVS is in fact naturally very light\footnote{The volume axion has a rih phenomenology, as it can act as dark radiation~\cite{Cicoli:2012aq, Higaki:2012ar, Cicoli:2015bpq}.}, being its mass suppressed by a factor $e^{-\V^{2/3}}$. Hence, it is possible to consistently integrate out all the heavy fields and to be just left with the light volume axion.\footnote{Notice that even though it is relatively easy to get axion masses in the ULA range from non-perturbative effects in string theory, as discussed for instance in~\cite{Hui:2016ltb}, generically, the closed string moduli $\tau$ will receive a mass of the same order as the corresponding axion (since the superpotential is holomorphic) which would violate fifth force bounds. Therefore a different mechanism is required to give a larger mass to $\tau$. Since, unlike $\tau$, the axion mass is protected by the corresponding PQ shift-symmetry which is valid to all orders in perturbation theory, perturbative effects can be the dominant source for the $\tau$ mass whereas non-perturbative effects give mass to the axion. This is precisely what happens in the LVS for $\tau$ the overall volume or a fibre modulus (but not for blow-up modes), allowing the possibility to integrate $\tau$ out and consider only the effective field theory for the axion field.}

To be concrete, let us consider the simplest setup including just two \Kahler moduli $T_b = \tau_b + i \theta_b$ and $T_s = \tau_s + i \theta_s$. The EFT model can be described in terms of a \Kahler potential $K$ and a superpotential $W$:
\begin{equation}
K = - 2 \log\left(\V + \frac{\hat{\xi}}{2}\right), \, \qquad W = W_0 + A_s e^{- a_s T_s} + A_b e^{- a_b T_b}\,,
\end{equation}
where $\hat{\xi} = \xi \langle s\rangle^{3/2}$ ($s$ is the dilaton field) and $A_b$, $A_s$ are $\mathcal{O}(1)$ coefficients that depend on the details of the compactification.

The potential arising from such EFT is well-known~\cite{Balasubramanian:2005zx,Conlon:2005ki}:
\begin{equation}
V = \frac{g_s}{8 \pi} \left[\frac{8}{3} \left(a_s A_s\right)^2 \frac{\sqrt{\tau_s} e^{-2 a_s \tau_s}}{\V} - 4 a_s A_s W_0 \frac{\tau_s e^{-a_s \tau_s}}{\mathcal{V}^2} + \frac{3 \xi W_0^2}{4 g_s^{3/2} \mathcal{V}^3}\right] + \delta V_{\text{dS}} \,,
\end{equation}
where $\delta V_{\text{dS}}$ is an additional contribution needed to achieve a de Sitter vacuum and we have implicitly set $\theta_s = \pi$. The terms containing $\theta_b$ are usually omitted since they are very suppressed. The leading contribution that includes $\theta_b$ takes the form
\begin{equation}
V(\theta_b) \supset \frac{g_s}{2 \pi} a_b A_b \frac{e^{-a_b \tau_b}}{\tau_b^2} \cos\left(a_b \tau_b\, \theta_b\right) \,.
\end{equation} 
Therefore we have a realisation of the simple single-field axion Lagrangian in eq.~\eqref{eq:bound} and the mass of the axion is then\footnote{We assume that the term $\delta V_{\rm dS}$ in the potential uplifts the minimum of the full potential to the current value of the cosmological constant.}
\begin{equation}
\label{eq:VolumeAxionMass}
m_{\theta_b} = \sqrt{\frac{g_s A_b a_b^3}{2 \pi}} \, {e^{- \frac{a_b \tau_b}{2}}} \,.
\end{equation}
The effective field theory is valid for volumes of order  $\mathcal{V} \gtrsim 10^3$ ($\tau_b\gtrsim 10^2$) which implies that approximately $m_{\theta_b}\lesssim 10^{-22}$ eV (by taking e.g.~$A_b = 1$, $g_s = 0.1$ and $10^{-1} \lesssim a_b \lesssim 1$) and therefore $\theta_b$ is a good candidate to be ULDM, although lighter and less constrained masses are also possible. In the case $\tau_b \simeq 10^3,$ the volume of the compact dimensions is $\V \simeq 3 \times 10^4$. This value of the volume implies a high scale of supersymmetry breaking, with a gravitino mass of order $m_{3/2} \simeq 3 \times 10^{13} \, \rm GeV$. 

It is interesting to ask whether it is possible to get the analogue of axion miniclusters with this ULA. As we mentioned in Section~\ref{sec:bosonstarsinfieldtheory} the formation of miniclusters needs large fluctuations as initial conditions, that grow and collapse during radiation domination (or immediately after the start of matter domination). The first obstruction to this is the fact that there is actually no U$(1)$ symmetry linearly realized in the four-dimensional effective field theory that describes the two-field system composed by the modulus and the corresponding axion. In fact, the shift-symmetry of the volume axion is inherited from the higher dimensional gauge symmetry of the $C_4$ form, rather than coming from a U$(1)$ symmetry. Hence the large initial fluctuations needed for the formation of miniclusters cannot be obtained from PQ U$(1)$ symmetry breaking after inflation as in the QCD axion case. The large initial fluctuations could be generated by  a first order phase transition, as suggested in~\cite{Hardy:2016mns}. However this mechanism does not work for ULAs, since the energy scale $\mu$ of non-perturbative effects that give mass to the axion would be required to be $\mu < \text{MeV}$, which is highly constrained from bounds on the number of relativistic degrees of freedom during BBN~\cite{Hardy:2016mns, Feng:2008mu}. 

\subsection{Moduli Stars}
\label{sec:ModuliStars}

In this section we will show that the same solutions already obtained in~\cite{Ruffini:1969qy, UrenaLopez:2001tw, UrenaLopez:2002gx, Alcubierre:2003sx, Guzman:2004wj, Visinelli:2017ooc} imply that string moduli potentials support star-like solutions in the dilute regime. We will explore the properties and possible phenomenological features of these objects. The actual formation of such objects is partially discussed in Section~\ref{sec:formationall}. As briefly discussed in Sec. \ref{sec:Oscillatons}, the task of finding equilibrium solutions in the dense regime is extremely involved from the numerical point of view, in the case of generic potentials. We leave the numerical analysis of the dense regime including gravity for the future.\\

In the single field case we can canonically normalize the field, so that the action is simply given by
\begin{equation}
\label{eq:ScalarAction}
S = \int d^4x\, \sqrt{-g} \left[- \frac{g^{\mu \nu}}{2} \partial_\mu \varphi \partial_\nu \varphi - V(\varphi)\right]\,.
\end{equation}
We consider a toy model potential that mimics the moduli potential expanded around the minimum in $\varphi = 0$. For the analysis of the dilute regime an expansion up to fourth order is sufficient:
\begin{equation}
\label{eq:ToyPotential}
V(\varphi) = \frac{m^2}{2} \varphi^2 + \frac{\lambda}{3!} \varphi^3 + \frac{g}{4!} \varphi^4 \,.
\end{equation}
The stringy examples studied below have distinctive properties, first we always observe $\lambda < 0$ and $g > 0$\footnote{Note the dimensions of the couplings $[m] = [\lambda] = 1$ and $[g]=0$.}. This makes these models different from the axionic cases for which $\lambda=0,\,  g<0$. Second, the expansion in $\varphi$ is such that the scale of all couplings is of similar order and therefore the couplings are not strong enough to change substantially the expression for the mass $M\sim M_{\rm P}^2/m$ typical for mini-boson stars to $M\sim M_{\rm P}^3/m^2$. The main reason for this is that there is only one mass scale in the expansion of a potential in string compactifications and once this scale is factorised the dimensionless coefficients are naturally of $\mathcal{O}(1)$. This argument is similar to the argument against realising Starobinsky inflation from string moduli (see for instance the appendix of~\cite{Burgess:2016owb}). \\

In order to study moduli stars we first assume a single harmonic, spherically symmetric ansatz for the background field of the form
\begin{equation}
\label{eq:StarAnsatz}
\varphi(r,t) = \varphi_0(r) \cos\left(\omega t\right) \,,
\end{equation}
where $\omega = m\left(1+\epsilon\right)$ ($\epsilon < 0$) and $|\epsilon| \ll 1$. We neglect the expansion of the Universe (i.e.~we assume that $m \gg H$) and we include weak gravity effects, encoded in the Newtonian potential $\phi \ll 1$ ($\mathcal{O}\left(\phi\right) \sim \mathcal{O}\left(\epsilon\right)$) appearing in the metric
\begin{equation}
\label{eq:LineElement}
ds^2 = - (1+2 \phi) dt^2 + (1-2\phi) \, dr^2 + r^2 d\Omega^2 \,,
\end{equation}
where $d\Omega^2$ is the differential solid angle and $\phi$ satisfies the Poisson equation. It is useful to rewrite all the equations in terms of dimensionless variables: we rescale the coordinates $(t,x^i)$, the field $\varphi$ and the energy density $\rho$ as follows
\begin{equation}
\label{eq:Rescalings}
\tilde{t} = m t \,, \qquad \tilde{x}^i = m x^i \,, \qquad \tilde{\varphi} = \frac{\varphi}{\Lambda} \,, \qquad \tilde\rho = \frac{\rho}{m^2 \Lambda^2} \,, \qquad \tilde\omega = 1 + \epsilon \,,
\end{equation}
where the scale $\Lambda$ is defined as in Sec.~\ref{sec:Oscillatons}. In the limit $g \rightarrow 0$, neglecting the gradient energy\footnote{We approximate here the total energy as $\tilde\rho = \dot{\tilde{\varphi}}^2/2 + \tilde{\varphi}^2/2 \approx \tilde\varphi_0^2/2$ which along with the Poisson equation implies that $\phi$ can be taken to be static in the dilute appoximation.} and taking $\Lambda = M_{\rm P}$ for the moment\footnote{In the limit of vanishing interactions, the $(\Lambda/M_{\rm P})^2$ term that would appear in the Poisson equation if $\Lambda \neq M_{\rm P}$ could be reabsorbed through the rescaling of the field in eq.~\eqref{eq:Rescalings}.}, the physical system is described by the following equations
\begin{align}
\label{eq:SPField}
\tilde{\varphi}_0''(\tilde{r}) + \frac{2}{\tilde{r}} \tilde{\varphi}'_0(\tilde{r}) & = 2 \left(\phi(\tilde{r}) - \epsilon\right) \tilde{\varphi}_0(\tilde{r}) \,, \\
\phi''(\tilde{r}) + \frac{2}{\tilde{r}} \phi'(\tilde{r}) & = \frac{\tilde{\varphi}_0^2(\tilde{r})}{4}\,,
\label{eq:SPPoisson}
\end{align}
where all the derivatives are taken with respect to the rescaled variables. In the limit of vanishing interactions the solutions of this system obey a scaling relation~\cite{Ruffini:1969qy}
\begin{equation}
\label{eq:Scaling}
\left(\tilde{r}, \tilde{\varphi}, \phi, \epsilon\right) \quad \longrightarrow \quad \left(\tilde{r}/\zeta, \zeta^2 \tilde{\varphi}, \zeta^2 \phi, \zeta^2 \epsilon\right) \,.
\end{equation}
This can be used to find all solutions in the dilute regime. In particular, small amplitude solutions can be obtained from generic solutions by rescaling with $\zeta \ll 1$. The boundary conditions follow from requiring asymptotic flatness and a regular solution at $\tilde{r} = 0$
\begin{equation}
\label{eq:BCvarphi0}
\tilde\varphi(0) = \tilde\varphi_{\rm core} \,, \qquad \tilde\varphi'(0) = 0 \,, \qquad \tilde\varphi(\infty) = 0 \,,
\end{equation}
\begin{equation}
\label{eq:BCphip0}
\phi'(0) = 0 \,, \qquad \phi(\infty) = 0 \,.
\end{equation}
In practice, in the dilute regime one can use the scaling in eq.~\eqref{eq:Scaling} to fix $\tilde\varphi_{\rm core} = 1$ and then vary $\phi(0)$ and $\epsilon$ until the correct boundary conditions at $\tilde{r} \gg 1$ is found via a shooting method.\\

\noindent The solution to the system in eq.s~\eqref{eq:SPField},~\eqref{eq:SPPoisson} can be written in integral form as~\cite{UrenaLopez:2002gx, Guzman:2004wj}
\begin{align}
\label{eq:IntegralSystem1}
\tilde{\varphi}_0(\tilde{r}) & = 1 + 2 \int_0^{\tilde{r}} dr' \, r' \left(1 - \frac{r'}{\tilde{r}}\right) \left(\phi(r') - \epsilon\right) \tilde{\varphi}_0(r') \,, \\
\phi(\tilde{r}) & = \phi(0) + \int_0^{\tilde{r}} dr' \, r' \, \frac{\tilde{\varphi}^2_0(r')}{4} - \frac{\tilde{M}(\tilde{r})}{8 \pi \,\tilde{r}} \,,
\label{eq:IntegralSystem2}
\end{align}
where we defined $\tilde{M}$ of the star through the relations\footnote{We write the generic expression for the mass with $\Lambda \neq M_{\rm P}$ for future reference.}
\begin{equation}
\label{eq:MassDef}
M(r) = \left(\frac{\Lambda^2}{m}\right) \tilde{M}(\tilde{r}) \,, \qquad \tilde{M}(\tilde{r}) = 4 \pi \int_0^{\tilde{r}} d\tilde{r}'\, \tilde{r}'^2 \tilde{\rho}(\tilde{r}') \, .
\end{equation}
Notice that in the dilute regime, asymptotic flatness implies that at $\tilde{r} \gg 1$ the Newtonian potential scales as $\phi(\tilde{r}) \sim -\tilde{M}/\tilde{r}$ and this condition fixes the value of $\phi(0)$ in eq. \eqref{eq:IntegralSystem2}. We will parametrize the solutions using both the dimensionless total mass $\tilde{M}$ defined as in eq.~\eqref{eq:MassDef} (with $\tilde{r} \rightarrow \infty$) and the radius of the star $\tilde{R}_{90}$, defined as the radius that contains $90\%$ of the total mass of the star. It is straightforward to check that the rescaling in eq.~\eqref{eq:Rescalings} acts on $\tilde{M}$ and $\tilde{R}_{90}$ as follows
\begin{equation}
\label{eq:ScalingMassRadius}
\left(\tilde{M}, \tilde{R}_{90}\right) \quad \longrightarrow \quad \left(\zeta \tilde{M}, \zeta^{-1} \tilde{R}_{90}\right) \,.
\end{equation}

In order to marginally take into account the first interaction terms in the potential in eq.~\eqref{eq:ToyPotential} we rescale it and the total energy density
\begin{equation}
\label{eq:RescaledPotRho}
\tilde{V} = \frac{\tilde\varphi^2}{2} + \tilde{\lambda} \frac{\tilde{\varphi}^3}{3!} + \tilde{g} \frac{\tilde\varphi^4}{4!} \,, \qquad \tilde{\rho} = \frac{\left(\dot{\tilde\varphi}\right)^2}{2} + \frac{\left(\tilde\varphi'\right)^2}{2} + \tilde{V} \,,
\end{equation}
where we have redefined the dimensionless couplings
\begin{equation}
\tilde{\lambda} = \frac{\lambda \Lambda}{m^2} \,, \qquad \tilde{g} = \frac{g \Lambda^2}{m^2} \,.
\end{equation}
The equation of motion (dropping subleading terms) and the Poisson equation are
\begin{align}
\label{eq:EOMStarTBS}
\tilde\varphi'' + \frac{2}{\tilde{r}} \tilde\varphi' & = \left(2 \phi + 1 - \tilde\omega^2\right) \tilde\varphi + \frac{\tilde{\lambda}}{2} \tilde\varphi^2 + \frac{\tilde{g}}{6} \tilde\varphi^3 \,, \\
\phi'' + \frac{2}{\tilde{r}} \phi' & = \left(\frac{\Lambda}{M_{\rm P}}\right)^2 \frac{\tilde\rho}{2} \,.
\end{align}
Following~\cite{Visinelli:2017ooc}, after using the ansatz in eq. \eqref{eq:StarAnsatz} it is easier to solve the system by taking an average of the previous equations integrating over a period $2 \pi/\tilde\omega$. Interestingly, the contribution coming from the cubic term of the potential in eq.~\eqref{eq:RescaledPotRho} is averaged out: clearly this is a good approximation as long as the amplitude of the field amplitude is small $\tilde\varphi_0(\tilde{r}) \ll 1$. Using $\tilde\omega^2 \simeq 1 + 2 \epsilon$ we get
\begin{align}
\label{eq:FinalSystem1}
\tilde\varphi_0'' + \frac{2}{\tilde{r}} \tilde\varphi'_0 & = 2 \left(\phi - \epsilon\right) \tilde\varphi_0 + \frac{\tilde{g}}{8} \tilde\varphi_0^3 = 0 \,, \\
\phi'' + \frac{2}{\tilde{r}} \phi' & = \left(\frac{\Lambda}{M_{\rm P}}\right)^2 \left[ \frac{\left(\tilde\varphi_0'\right)^2}{8} + \frac{1 + 2\epsilon}{4} \tilde\varphi_0^2 + \frac{3 \tilde{g}}{16} \frac{\tilde\varphi_0^4}{4!}\right] \,.
\label{eq:FinalSystem2}
\end{align}

It is possible to understand the origin of the existence of these stable solutions by looking at the energy functional of this system, as suggested in~\cite{Visinelli:2017ooc, Schiappacasse:2017ham}. After averaging, assuming the star has radius $R$ and using the rescalings in eq.~\eqref{eq:Rescalings}, it takes the form
\begin{align}
E & = -\left(\frac{M}{M_{\rm P}}\right)^2 \frac{1}{R} + \int d^3 x\left[\frac{\left(\nabla \tilde\varphi_0\right)^2}{2} + \frac{3 \tilde{g}}{8} \frac{\tilde\varphi^4}{4!}\right]  \nonumber \\
& = \frac{\Lambda^2}{m} \left[-\left(\frac{\Lambda}{M_{\rm P}}\right)^2 \frac{\tilde{M}^2}{\tilde{R}} + \alpha \frac{\tilde{M}}{4 \tilde{R}^2} + \frac{1}{4!} \frac{3 \beta \tilde{g}}{8} \frac{\tilde{M}}{\tilde{R}^3}\right] \,,
\end{align}
where $\alpha$ and $\beta$ are coefficients to be determined by matching the energy functional with the numerical solutions and we hae used that $M \simeq \int d^3x\, m^2 \varphi_0^2 = \frac{\Lambda^2}{m} \varphi_0^2 \tilde{R}^3 = \frac{\Lambda^2}{m} \tilde{M}$.
It is possible to extremize the energy functional with respect to the radius $\tilde{R}$. The solution of $\partial E/\partial \tilde{R}$ is
\begin{equation}
\label{eq:RStable}
\tilde{R}_{\rm stable} = \frac{1}{4 (\Lambda/M_{\rm P})^2} \left[\frac{\alpha}{\tilde{M}} + \sqrt{\frac{\alpha^2}{\tilde{M}^2} + \frac{3 \beta \tilde{g}}{4} \left(\frac{\Lambda}{M_{\rm P}}\right)^2} \right]\,,
\end{equation}
and it can be easily checked that it is always a minimum of $E$, hence a stable solution\footnote{Stability in this section should be understood as stability against radial perturbations of the star. There is no physical law preventing the moduli composing the star from decaying gravitationally.}. The expression for $\tilde{R}_{\rm stable}$ is inversely proportional to the mass $\tilde{R} \propto 1/\tilde{M}$ for small values of the mass. In this regime the stability comes from the balance between the gradient energy (repulsive) and gravity (attractive). In the limit of large mass $\tilde{M}$ the radius tends to a constant, which depends on the coupling constant $\tilde{g}$
\begin{equation}
\label{eq:AsymptoticRadius}
R_{\rm stable} \rightarrow \frac{\sqrt{3 \beta \tilde{g}}}{8 (\Lambda/M_{\rm P})} \,.
\end{equation}
This limit is achieved in the regime in which the interaction terms are important, namely for a core amplitude $\tilde\varphi_{\rm core} \lesssim 1$. Since we will consider stringy potentials expanded around the minimum truncated at quartic order, we will never be allowed to explore this regime and trust the truncation at the same time: when the core amplitude is of order $\tilde\varphi_{\rm core} \lesssim 1$ all the higher order interactions should be included.

\subsubsection{Overall volume modulus in KKLT and the LVS}

We first consider the potential for the canonically normalized volume modulus $\tilde\varphi = \frac{\varphi}{M_{\rm P}} = \sqrt{\frac{2}{3}} \ln \V$ in the LVS. Following~\cite{Cicoli:2016olq}, we can write the uplifted potential for the volume modulus $\varphi$ (after having integrated out the blow-up moduli) in terms of two parameters\footnote{In principle there could be also the parameter $\gamma$ that denotes the power of the volume of the uplifting contribution to the scalar potential $V_{\rm dS} \propto \V^{-\gamma}$ (it has to be in the range $1 \lesssim \gamma < 3$). Given that the results do not depend on its value, in the following we set it to $\gamma = 2$.}, the overall normalization $V_0$ and the position of the minimum $\langle\varphi\rangle:$ 
\begin{equation}
V_{\rm LVS}(\varphi) = 3 \left(\frac{3}{2}\right)^{1/4} V_0\, e^{-3 \sqrt{\frac{3}{2}} \tilde{\varphi}} \left[\langle\tilde{\varphi}\rangle^{1/2} e^{\sqrt{\frac{3}{2}} \left(\tilde{\varphi} - \langle\tilde{\varphi}\rangle\right)} - \sqrt{\frac{2}{3}} \left(\tilde{\varphi}^{3/2} - \langle\tilde{\varphi}\rangle^{3/2}\right) - \langle\tilde{\varphi}\rangle^{1/2}\right]\,,
\end{equation}
where the normalization is $V_0 = \frac{3 P |W_0|^2}{4}$ and $P$ is a $\mathcal{O}(1)$ coefficient that depends on the details of the compactification space, see~\cite{Cicoli:2016olq} for details.\\

\begin{figure}
\subfigure{\includegraphics[width=7.2cm]{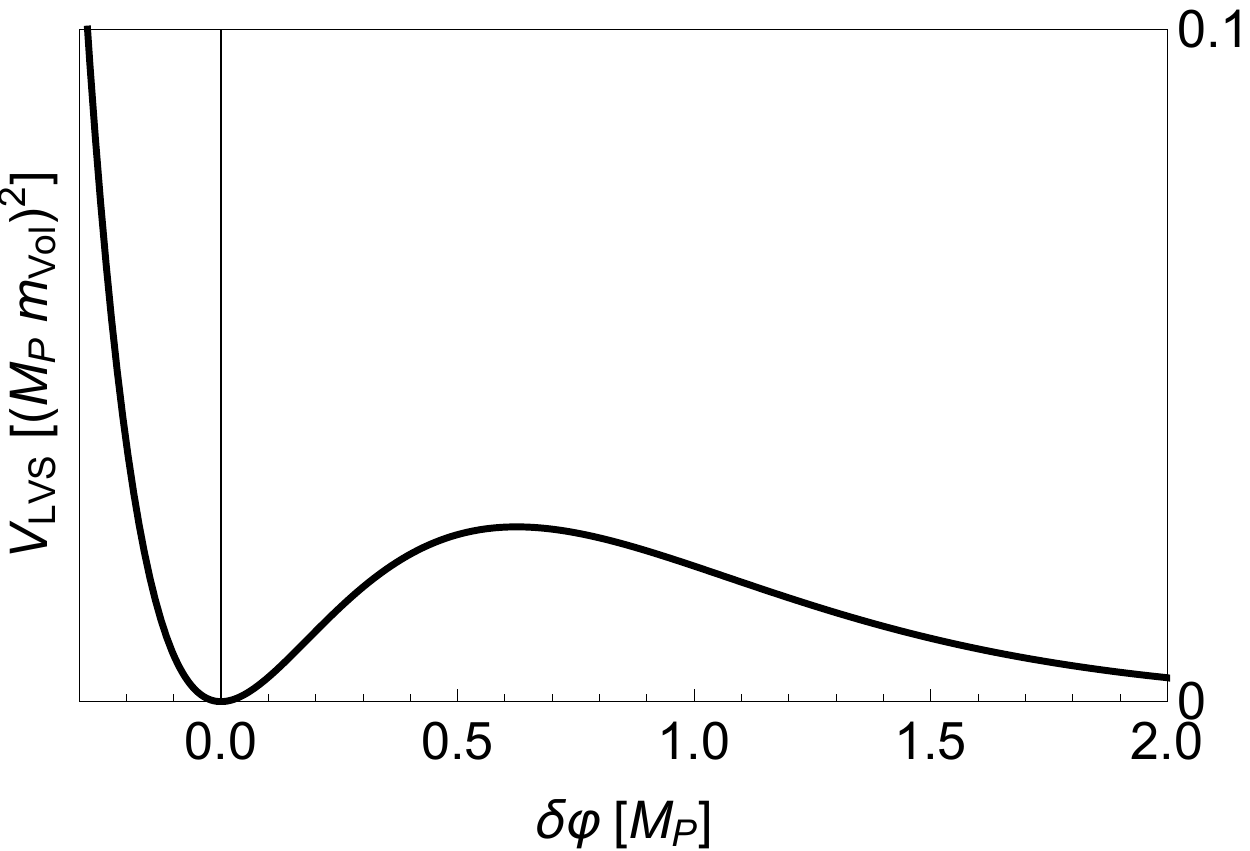}}
\hfill
\subfigure{\includegraphics[width=7.5cm]{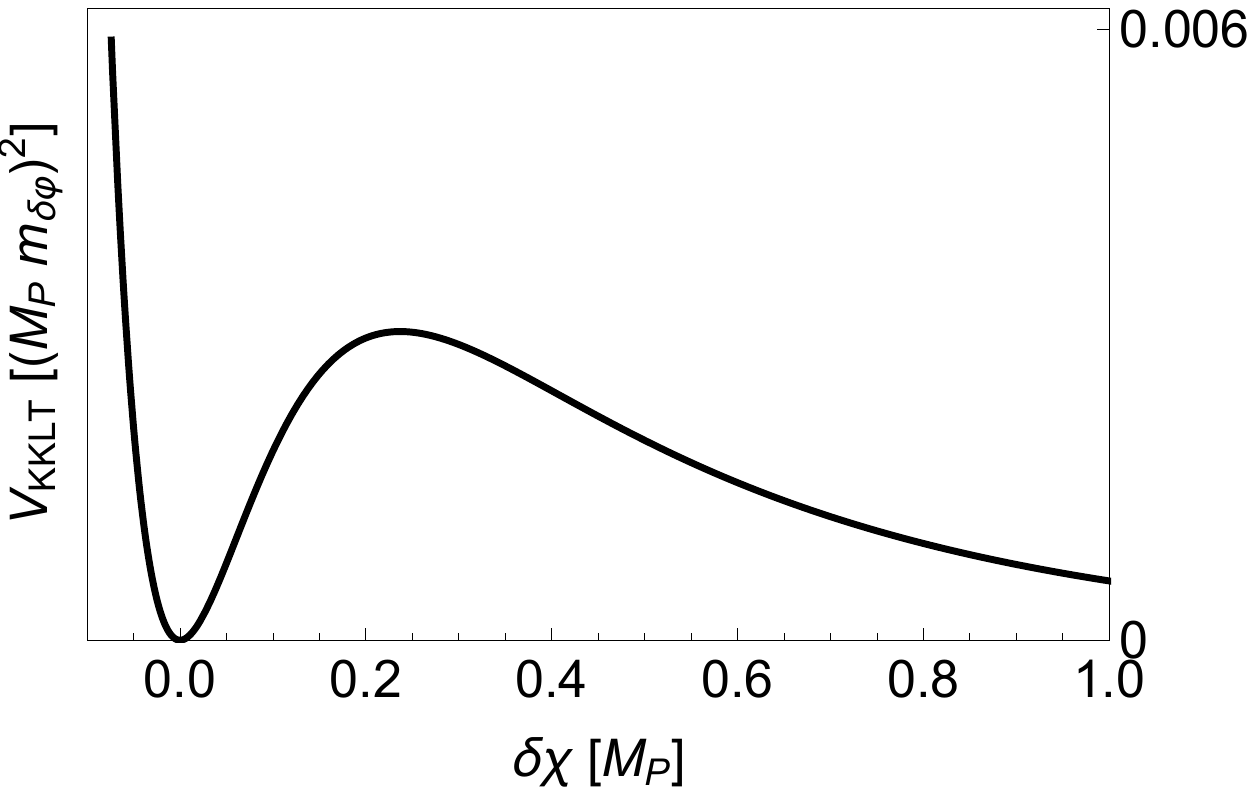}}
\caption{\textit{Left panel}: LVS potential in terms of the canonically normalized field $\delta \tilde{\varphi}$ in Planck units. \textit{Right panel}: KKLT potential in terms of the canonically normalized field $\delta \tilde{\chi}$ in Planck units.}
\label{fig:Potentials}
\end{figure}

We use as reference value $\langle\tilde{\varphi}\rangle = 14$, which corresponds to a volume of $\V \simeq 2.8 \times 10^7$, and we expand around the minimum of the potential $\tilde{\varphi} = \langle\tilde{\varphi}\rangle + \delta\tilde{\varphi}$ but the results are basically independent of the exact value of the volume. The potential is plotted in the left panel of Figure~\ref{fig:Potentials}. The expansion up to quartic order reads\footnote{Notice that even if the couplings (both in the LVS and in the KKLT case) look larger than $1$, the expansion is always under control as long as $\delta \tilde{\varphi} \lesssim 10^{-2}$ (or $\delta \tilde\chi \lesssim 10^{-2}$ in the KKLT case). This expansion is then not fully under control for the first numerical solution in Tab. \ref{tab:ParametersShooting} for which $\delta\tilde\varphi_{\rm core} = 10^{-1}$.}
\begin{equation}
\label{eq:LVSPotential}
\tilde{V}_{\rm LVS}(\delta\tilde\varphi) \simeq \frac{\delta\tilde\varphi^2}{2} - \tilde{\lambda}_{\rm LVS} \frac{\delta\tilde\varphi^3}{3!} + \tilde{g}_{\rm LVS}\frac{\delta\tilde\varphi^4}{4!} + \dots \,,
\end{equation}
where
\begin{equation}
\label{eq:LVSCoefficients}
\tilde{\lambda}_{\rm LVS} \simeq 9.75 \, \qquad \text{and} \, \qquad \tilde{g}_{\rm LVS} \simeq 63.8\,.
\end{equation}

\begin{figure}
\subfigure{\includegraphics[width=7.2cm]{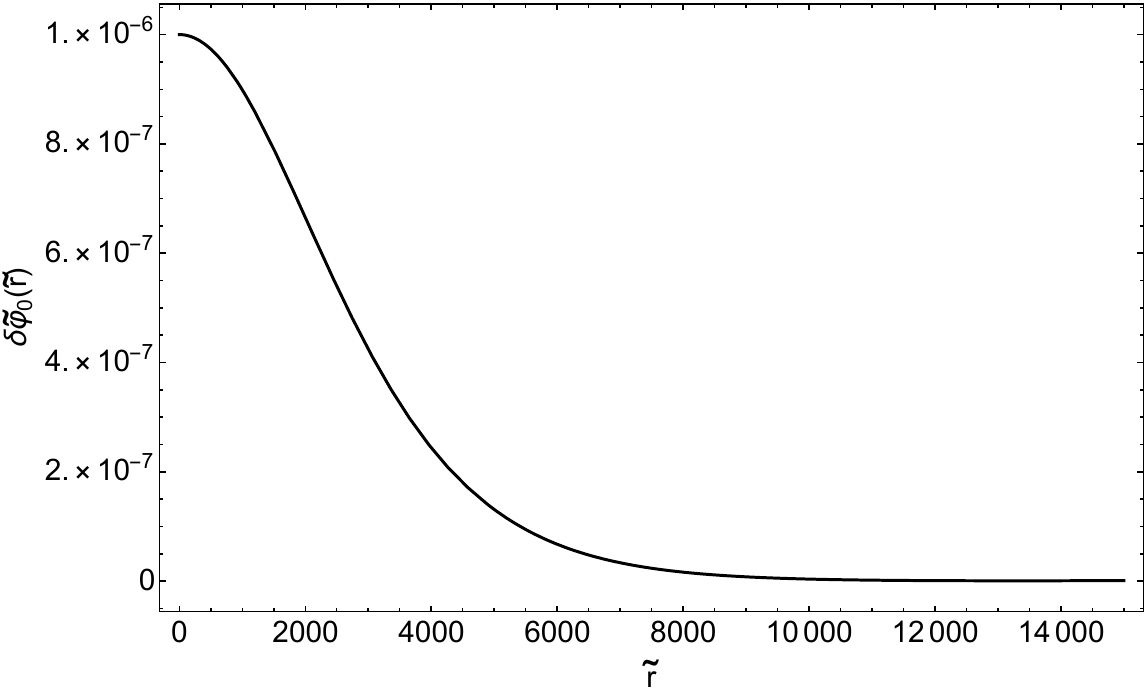}}
\hfill
\subfigure{\includegraphics[width=7.2cm]{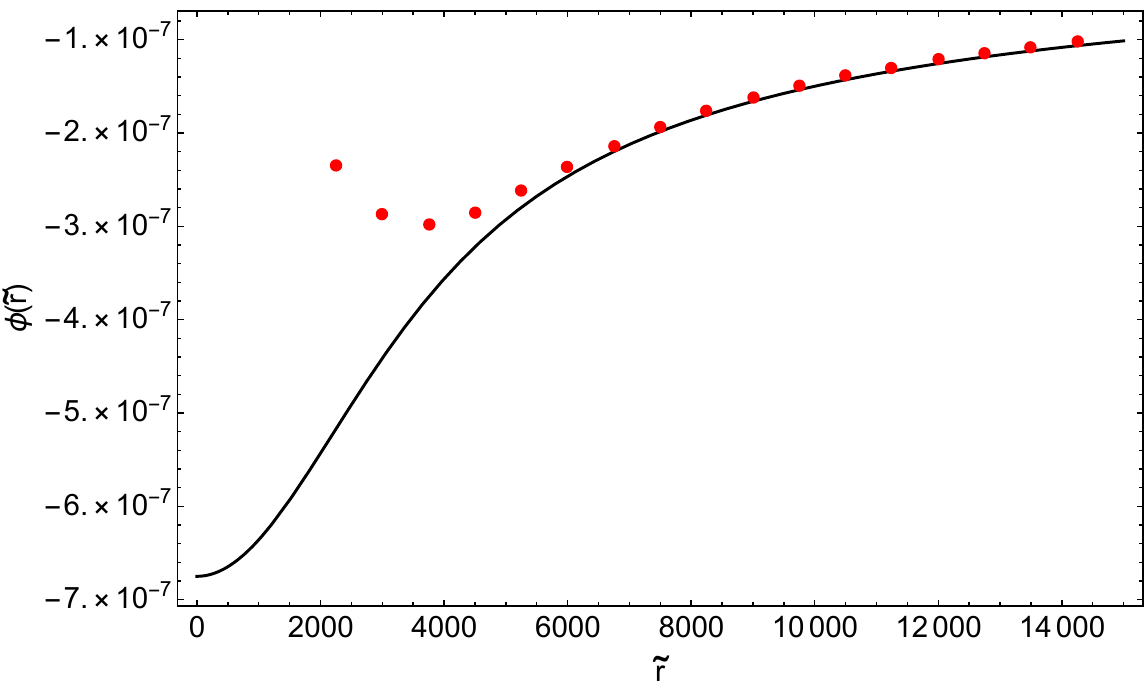}}
\caption{\textit{Left panel}: the solid line is the profile of the star composed by the LVS modulus obtained numerically, with background amplitude $\delta\tilde\varphi_{\rm core} = 10^{-6}$. \textit{Right panel}: the solid line is the Newtonian potential profile obtained numerically. Red dots correspond to the last term of eq. \eqref{eq:IntegralSystem2} $\phi(\tilde{r}) \simeq - \tilde{M}(\tilde{r})/\tilde{r}$, showing the expected behaviour of $\phi$ at large $\tilde{r}$.\label{fig:StarProfile}}
\end{figure}

\begin{figure}[h!]
\begin{center}
\includegraphics[width=\textwidth]{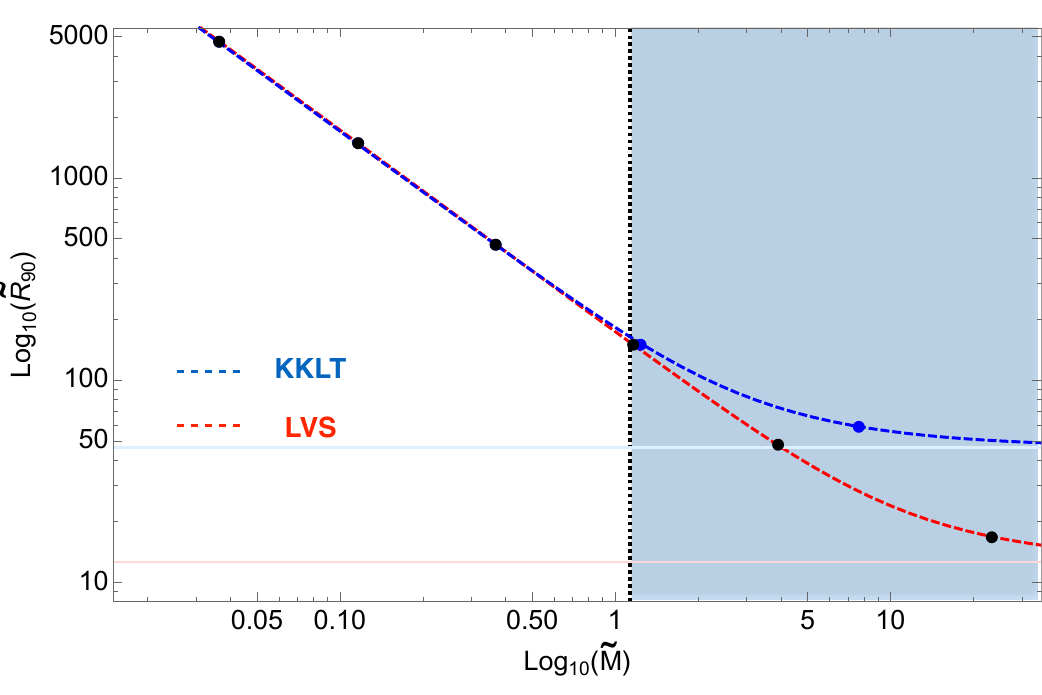}
\caption{We plot radius of the stars $\tilde{R}_{90}$ (in LVS and KKLT) as a function of the star mass $\tilde{M}$. Black dots correspond to the numerical solutions listed in Tab.~\ref{tab:ParametersShooting} obtained varying the core amplitude $\delta\tilde\varphi_{\rm core}$ from $10^{-6}$ (extreme left black dot) to $10^{-1}$ (extreme right black dot). The blue dots are the numerical solutions for the KKLT potential (for core amplitudes in the range $10^{-6}$ to $10^{-2}$) that coincide with the LVS results in the dilute regime $\delta\chi_{\rm core} \lesssim 10^{-3}$, as expected. The red and blue dashed lines are found matching the numerical data with the function defined in eq.~\eqref{eq:RStable}. The light red and blue solid lines correspond to the asymptotic values obtained from eq.~\eqref{eq:AsymptoticRadius}. The blue region corresponds to a background amplitude $\delta\tilde\varphi_{\rm core} \gtrsim 10^{-3}$ (or $\delta\tilde\chi_{\rm core} \gtrsim 10^{-3}$), where the approximations used are not fully reliable, as explained in the main text. \label{fig:MassRadiusPlot}}
\end{center}
\end{figure}

Concerning KKLT, we consider the potential~\cite{Kachru:2003aw}
\begin{equation}
V_{\rm KKLT}(\tau) = \frac{a A e^{-a \tau}}{2 \tau^2} \left[\frac{a A}{3} \tau e^{-a \tau} + W_0 + A e^{-a \tau}\right] + \frac{D}{\tau^3} \,,
\end{equation}
where $W_0 = - 10^{-4}$, $a=0.1$, $A=1$, $D = 3 \times 10^{-9}$ but again the results are independent of the exact value of the parameters, provided that the potential has a dS minimum. The de Sitter minimum of $V_{\rm KKLT}$ in terms of the canonically normalized field $\tilde\chi = \frac{\chi}{M_{\rm P}} = \sqrt{\frac{2}{3}} \log\,\tau$ is located at $\langle\chi\rangle \simeq 5.8$. This potential is plotted in the right panel of Figure~\ref{fig:Potentials} in terms of the field expanded around the minimum $\tilde{\chi} = \langle\tilde{\chi}\rangle + \delta\tilde{\chi}$. The expansion of the potential up to quartic order is
\begin{equation}
\tilde{V}_{\rm KKLT}(\delta\tilde\chi) \simeq \frac{\delta\tilde\chi^2}{2} - \tilde{\lambda}_{\rm KKLT} \frac{\delta\tilde\chi^3}{3!} + \tilde{g}_{\rm KKLT}\frac{\delta\tilde\chi^4}{4!} + \dots \,,
\end{equation}
where
\begin{equation}
\label{eq:KKLTCoefficients}
\tilde{\lambda}_{\rm KKLT} \simeq 30.6 \, \qquad \text{and} \, \qquad \tilde{g}_{\rm KKLT} \simeq 652.9\,.
\end{equation}
To clarify the notation, the ansatz in eq. \eqref{eq:StarAnsatz} takes the form
\begin{equation}
\delta\tilde\varphi(r) = \delta \tilde\varphi_0(r) \cos \left(\tilde\omega \tilde{t}\right) \,, \qquad \delta\tilde\chi(r) = \delta \tilde\chi_0(r) \cos \left(\tilde\omega \tilde{t}\right) \,,
\end{equation}
for the LVS and the KKLT volume moduli respectively. We numerically solve eq.s~\eqref{eq:FinalSystem1} and~\eqref{eq:FinalSystem2} as described above, varying the initial core amplitude of the field in the ranges $(10^{-6},10^{-1})$ in the LVS case\footnote{As already mentioned, the truncation in eq. \eqref{eq:LVSPotential} is not a good approximation for the solution with $\delta\tilde\varphi_{\rm core} = 10^{-1}$. We however include it to show that, assuming the potential is exactly the one in eq. \eqref{eq:LVSPotential} we get the flattening expected in the case of repulsive interactions~\cite{Schiappacasse:2017ham}.} and $(10^{-6},10^{-2})$ for the KKLT potential\footnote{As the core amplitude gets larger and larger the numerics become more and more difficult especially in the KKLT case for which the interaction coupling $g_{\rm KKLT}$ is large.}. We find that both potentials support star-like solutions and that they coincide in the dilute regime where basically only the mass term in the potential is relevant. We report the values of the parameters for the LVS case in Tab.~\ref{tab:ParametersShooting}. Notice that the scaling in eq.~\eqref{eq:Scaling} and eq.~\eqref{eq:ScalingMassRadius} is manifest in the dilute regime where $\delta\tilde{\varphi}_{\rm core} \lesssim 10^{-3}$. We also report as an example the field profile $\delta\tilde\varphi_0(r)$ and the Newtonian potential in the LVS case with $\delta\tilde\varphi_{\rm core} = 10^{-6}$ in Figure~\ref{fig:StarProfile}. In the Newtonian potential we also plot (red dots) the last term in eq. \eqref{eq:IntegralSystem2} to show the asymptotic behaviour $\phi(\tilde{r}) \sim - \tilde{M}(\tilde{r})/\tilde{r}$ at large $\tilde{r}$.
The values of mass and radius are reported in Fig.~\ref{fig:MassRadiusPlot}. The dashed blue line corresponds to the fit of numerical data using the function in eq.~\eqref{eq:RStable}, varying the parameters $\alpha$ and $\beta$. Even though we use $\delta\tilde\varphi_{\rm core} = 10^{-1}$ as maximum value for the core amplitude, the results should be trusted up to a core amplitude of $\mathcal{O}\left(10^{-3}\right)$. This follows from previous studies of the dilute regime for the free field case (or Newtonian oscillatons)~\cite{UrenaLopez:2002gx} and from the observations that for larger core amplitudes the deviation from the single harmonic approximation in eq.~\eqref{eq:StarAnsatz} piles up quickly, invalidating the solution. In this case the single harmonic approximation in eq.~\eqref{eq:StarAnsatz} should be replaced by a Fourier expansion~\cite{UrenaLopez:2001tw, UrenaLopez:2002gx} and higher order interaction terms should be included in the potential. This procedure has the drawback that stable solutions can be found only for specific interaction potentials (quartic) and for small values of the couplings~\cite{UrenaLopez:2012zz}. As we will show in a forthcoming publication\footnote{And as it has already been shown for the axion potential in~\cite{Helfer:2016ljl}.}, it is more fruitful to directly study the evolution of the system using a full GR simulation code~\cite{Clough:2015sqa}\footnote{This has also drawbacks: first, as the simulation starts from an arbitrary initial condition, it is not certain that equilibrium configurations (even if they exist), can be found in this way. Second, the stability of the configuration can only be checked on a time interval as long as the simulation time, which is often short.}. However, we can take the results plotted in Fig.~\ref{fig:MassRadiusPlot} as the clear indication that different couplings in the potential play a crucial role in determining the mass spectrum of moduli stars in the dense regime that in turn affects the GW spectrum produced by the dynamics of moduli stars.\\

\begin{table}[h!]
\centering
\begin{tabular}{ccccc}
\hline
$\delta\tilde\varphi_{\rm core}$ & $\epsilon$ & $\phi(0)$ & $\tilde{M} $ & $\tilde{R}_{90}$ \Tstrut\Bstrut\\
\hline
$10^{-1}$ & $-0.12$ & $-6.5 \times 10^{-2}$ & $23.3$ & $16.7$ \\
$10^{-2}$ & $-7.4 \times 10^{-3}$ & $-3.9 \times 10^{-3}$ & $3.92$ & $47.4$ \\
$10^{-3}$ & $-3.49 \times 10^{-4}$ & $-6.75 \times 10^{-4}$ & $1.17$ & $148.4$ \\
$10^{-4}$ & $-3.5 \times 10^{-5}$ & $-6.75 \times 10^{-5}$ & $0.37$ & $469.2$ \\
$10^{-5}$ & $-3.5 \times 10^{-6}$ & $-6.75 \times 10^{-6}$ & $1.16 \times 10^{-1}$ & $1483.2$ \\
$10^{-6}$ & $-3.5 \times 10^{-7}$ & $-6.75 \times 10^{-7}$ & $3.6 \times 10^{-2}$ & $4688.6$ \\
\hline
\end{tabular}
\caption{Details of the numerical solutions obtained via the shooting method with the LVS potential, using $\langle\tilde{\varphi}\rangle = 14$. Notice that for $\delta\tilde{\varphi}_{\rm core} \lesssim 10^{-3}$ the scaling in eq.~\eqref{eq:Scaling} is manifest: as $\delta\tilde{\varphi}_{\rm core}$ decreases by a factor of $10$, the same happens to $\epsilon$ and $\phi(0)$, while $\tilde{M}$ decreases by a factor of $\sqrt{10}$ and $\tilde{R}_{90}$ increases by a factor of $\sqrt{10}$.}
\label{tab:ParametersShooting}
\end{table}

Assuming that the expression for the mass reported in eq.~\eqref{eq:MassEnhancement} is valid in the real field case, and that effectively the leading interaction terms in the LVS and KKLT potential is the quartic one (i.e.~that the cubic is approximately averaged out also in the dense regime), we get an enhancement of the mass of the star of order $\tilde{g}^{1/2}_{\rm LVS} \simeq 8$ in the LVS case and $\tilde{g}^{1/2}_{\rm KKLT} \simeq 16$ in the KKLT case. In both cases the enhancement factor is much smaller than $M_{\rm P}/m$ due to the smallness of the coupling $g$: the Chandrasekar limit in eq.~\eqref{eq:MassEnhancement} is never achieved.

\subsubsection{Blow-up-like potentials}

\begin{figure}[h!]
\begin{center}
\includegraphics[width=\textwidth]{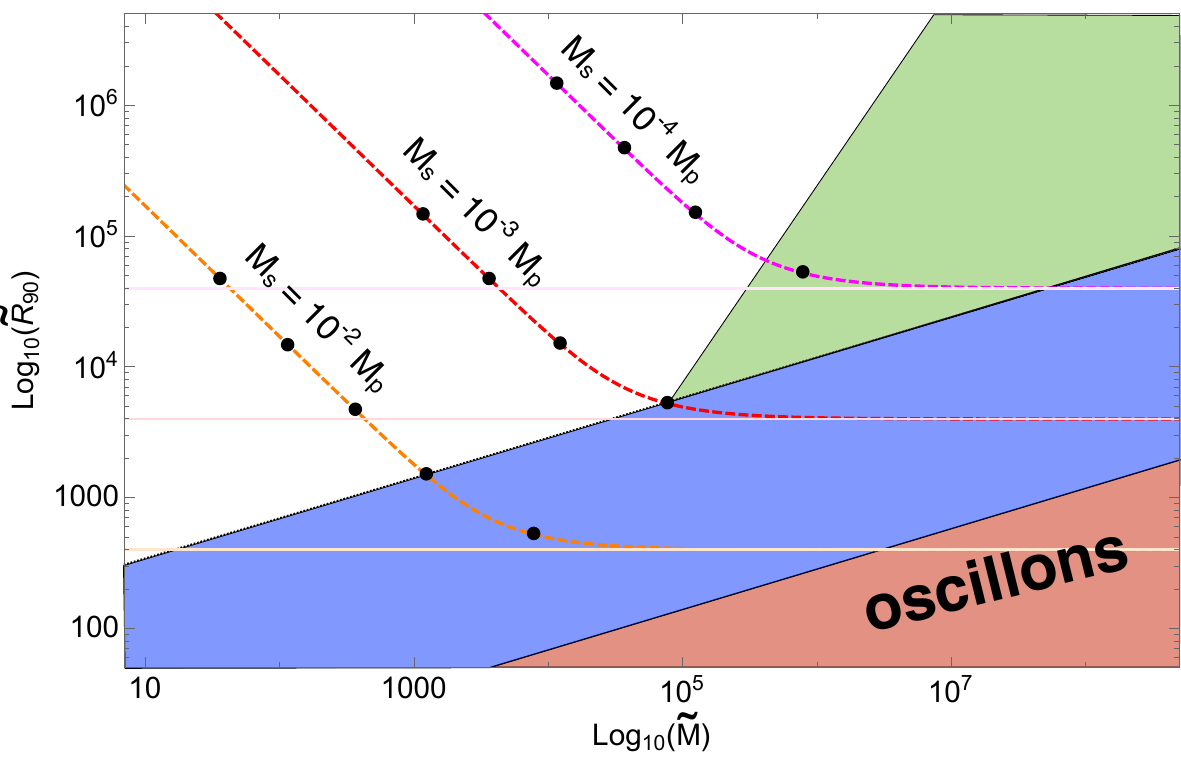}
\caption{We plot the radii of the moduli stars (for blow-up potentials) as a function of the star mass. Different dots correspond to numerical solutions obtained varying the core amplitude $\tilde{\sigma}_{\rm core}$ for different values of the string scale $M_s$. In particular, from left to right the dots represent the solutions for the core amplitudes $\left(10^{-6}, 10^{-5}, 10^{-4}, 10^{-3}, 10^{-2}\right)$ for $M_{\rm s} = 10^{-2} \, M_{\rm P}$, $\left(10^{-6}, 10^{-5}, 10^{-4}, 10^{-3}\right)$ for $M_{\rm s} = 10^{-3} \, M_{\rm P}$ and $\left(10^{-7}, 10^{-6}, 10^{-5}, 10^{-4}\right)$ for $M_{\rm s} = 10^{-4} \, M_{\rm P}$. Dashed lines are found matching the numerical data with the function defined in eq.~\eqref{eq:RStable}. The solid horizontal lines correspond to the asymptotic values defined in eq.~\eqref{eq:AsymptoticRadius}. The blue region corresponds to a background amplitude $\tilde{\sigma}_{\rm core} \gtrsim 10^{-3}$, where the approximations used are not fully reliable, while the green region corresponds to the regime in which interactions become important due to the large mass of the star, as explained in the main text. \label{fig:MassRadiusBlowUp}}
\end{center}
\end{figure}

In this section we study the following phenomenological potential for the canonically normalized modulus $\sigma$
\begin{equation}
V_{\rm bu} (\sigma) = V_0 \left(1 - e^{a\, \sqrt{\V} \,\frac{\sigma}{M_{\rm P}}}\right)^2 \,,
\end{equation}
where $V_0$ is an overall normalization that depends on the details of the compactification, $a$ is typically an $\mathcal{O}(1)$ parameter and the potential has a zero-energy minimum in $\sigma = 0$. This potential mimics that of blow-up moduli in the LVS and $\V$ is the volume of the compactification space. The mass of blow up moduli is $m_{\rm bu} \simeq \mathcal{O}\left(1\right)/\V$ while the scale $\Lambda$ is essentially given by the string scale $M_{\rm s} = M_{\rm P}/\sqrt{\V}$. In terms of the rescaled field
\begin{equation}
\tilde{\sigma} = \frac{\sigma}{M_s} = \sqrt{\V} \,\frac{\sigma}{M_{\rm P}} \,,
\end{equation}
the rescaled scalar potential takes the simple form
\begin{equation}
\tilde{V}_{\rm bu}\left(\tilde{\sigma}\right) = \left(1 - e^{a \tilde{\sigma}}\right)^2 \,,
\end{equation}
where the normalization $V_0$ disappears after the rescalings in eq.~\eqref{eq:Rescalings} are performed and we will take $a=1$ for numerical computations. The Taylor expansion around the minimum of this scalar potential takes again the form
\begin{equation}
V_{\rm bu}(\tilde{\varphi}) \simeq \frac{\tilde{\sigma}}{2} - \frac{\tilde{\lambda}_{\rm bu}}{3!} \tilde\sigma^3 + \frac{\tilde{g}_{\rm bu}}{4!} \tilde\sigma^4 + \dots \,,
\end{equation}
where
\begin{equation}
\tilde{\lambda}_{\rm bu} = 3 \,, \qquad \tilde{g}_{\rm bu} = 7 \,.
\end{equation}
Repeating the analysis outlined in the former section we find the results summarized in Figure~\ref{fig:MassRadiusBlowUp}. These results are essentially equivalent to the dilute regime studied in~\cite{Visinelli:2017ooc} in the case of axion stars (for the QCD axion). The red region correspond to the region where gravity is negligible and the potential supports oscillon formation, as already numerically studied in~\cite{Antusch:2017flz}. The blue region in Fig.~\ref{fig:MassRadiusBlowUp} corresponds to core amplitudes $\tilde\sigma_{\rm core }\gtrsim~10^{-3}$, where we can no longer trust the single harmonic approximation. Finally, the green region corresponds to the case in which interactions become important (even though the background amplitude is small) due to the large mass of the star. In other words the second term under the square root in eq.~\eqref{eq:RStable} becomes dominant over the first one which is suppressed by the large mass of the star:
\begin{equation}
\frac{\left(\Lambda/M_{\rm P}\right)^2}{1/\tilde{M}^2} \gg 1 \quad \text{for} \quad \Lambda/M_{\rm P} \simeq 10^{-4} \quad \text{and} \quad \tilde{M} \gtrsim 10^{5} \,.
\end{equation}
However, we expect that higher order interaction terms will quickly become important in this region, and we hence cannot completely trust the results. Finally, even though we can take the results in the blue and green regions as indications of what happens when interactions become important and the single harmonic approximation breaks down, these regimes need a more careful numerical study that we will present in a forthcoming publication.
\subsubsection{Estimates of masses and sizes}
It is interesting to translate the dimensionless numbers into real masses and radii. In general the masses and radii can be written as
\begin{align}
M/\tilde{M} & = \frac{\Lambda^2}{M_{\rm P}^2} \frac{M_{\rm P}^2}{m} \simeq \frac{\Lambda^2}{M_{\rm P}^2} \, \left(\frac{100 \, \text{TeV}}{m}\right) \times 10^8 \, \text{g} \,, \\
R/\tilde{R} & = 1/m \simeq 2 \times \left(\frac{100 \, \text{TeV}}{m}\right) \times 10^{-19} \, \text{cm} \,,
\end{align}
where $\tilde{M}$ and $\tilde{R}$ are the dimensionless numbers previously determined numerically and can vary by a few orders of magnitude.\\

Being in string compactification we can express all the masses and sizes in terms of the compactification space volume $\V$. We start form the blow-up-like case for which star mass takes the form
\begin{equation}
M_{\rm bu}/\tilde{M} = \left(\frac{M_{\rm s}}{M_{\rm P}}\right)^2 \times \V \times M_{\rm P} = M_{\rm P} \,,
\end{equation}
where $\tilde{M}$ can be read in Fig.~\ref{fig:MassRadiusBlowUp}. The radii on the other hand
\begin{equation}
R_{\rm bu}/\tilde{R} \simeq \left(5 \times \ell_{\rm P}\right) \times \V \,,
\end{equation}
where $\tilde{R} \gtrsim 100$ can be read again in Fig.~\ref{fig:MassRadiusBlowUp}.\\

The mass of the volume modulus in KKLT is
\begin{equation}
 m_{\rm KKLT}\simeq m_{3/2}\log{\left(\frac{M_{\rm P}}{m_{3/2}}\right)} \,,
\end{equation}
where the suppression of the gravitino mass compared to the Planck scale arises from a hierarchically small expectation value of the superpotential
\begin{equation}
 m_{3/2}=M_{\rm P} e^{K/2} |\langle W\rangle| \simeq M_{\rm P} \frac{|W_0|}{\V}\,.
\end{equation}
Assuming that the modulus has already decayed gravitationally before BBN, the mass of the volume modulus is constrained to be larger than $\simeq 100 \,{\rm TeV}.$ Different hierarchies are achieved by different amount of flux tuning. The typical field range corresponds to $\Lambda=M_{\rm P}$ as discussed previously. Hence the estimate for the mass and radius become in this case
\begin{align}
 M_{\rm KKLT}/\tilde{M} & \simeq M_{\rm P} \V |W_0|^{-1}\,,\\
 R_{\rm KKLT}/\tilde{R} & \simeq M_{\rm P}^{-1} \V |W_0|^{-1}\,.
\end{align}
The mass of the volume modulus $\V$ in the LVS takes the form~\cite{Balasubramanian:2005zx, Conlon:2005ki}
\begin{equation}
m_{\V} = \mathcal{O}(1) \frac{|W_0|}{\sqrt{4 \pi} g_s^{1/4}} \frac{M_{\rm P}}{\V^{3/2}} \,,
\end{equation}
and for the following estimates we take $\frac{\mathcal{O}(1)|W_0|}{\sqrt{4 \pi} g_s^{1/4}} = 1$ which is easily achievable in the landscape, we can rewrite $M$ in terms of the volume $\V$ as
\begin{equation}
M_{\V}/\tilde{M} \simeq \, M_{\rm P} \, \V^{3/2}  \,,
\end{equation}
where $\tilde{M} \lesssim 2$. In the same way, the radius of the star is given in term of the volume $\V$ by
\begin{equation}
R_{\V}/\tilde{R} \simeq \left(5 \times \ell_{\rm P}\right) \times \V^{3/2} \,,
\end{equation}
where $\tilde{R} \gtrsim 30$.\\

\subsubsection*{Phenomenology of the LVS volume modulus}

There are two phenomenologically allowed windows for $\V$ in the LVS. The first case arises by assuming that the modulus (and hence the star) has already decayed gravitationally. In order not to spoil BBN we require that the volume modulus decays before its start. Since $\V$ is coupled gravitationally, this condition translates into
\begin{equation}
\tau_{\V} =\Gamma^{-1} \lesssim H_{\rm BBN}^{-1} \,,
\end{equation}
for the lifetime of the volume modulus $\tau_{\V}$, where $\Gamma \simeq m_{\V}^3/M_{\rm P}^2$ and $H_{\rm BBN}^2 \simeq T_{\rm BBN}^{2}/M_{\rm P}$ with $T_{\rm BBN} \simeq 3 \, \text{MeV}$. We also need to require that the volume is large enough to trust the effective field theory. Summarizing these conditions in terms of the volume we can write
\begin{equation}
10^3 \lesssim \V \lesssim 2 \times 10^{9} \,,
\end{equation}
that corresponds to the following windows for the masses and the radii of the stars
\begin{equation}
0.1 \, \text{g} \lesssim M_{\V}/\tilde{M} \lesssim 3.8 \times 10^8 \, \text{g} \,, \qquad 2.6 \times 10^{-28} \, \text{cm} \lesssim R_{\V}/\tilde{R} \lesssim  7.2 \times 10^{-19} \, \text{cm}\,.
\end{equation}
These objects turn out to be very massive microscopic objects.\\

The second window corresponds to values of the compactification volume $\V$ such that the volume modulus has not decayed yet. We require that
\begin{equation}
\tau_{\V} \gtrsim H_0^{-1}\,,
\end{equation}
where $H_0 \simeq 10^{-33} \, \text{eV}$. Such condition translates into $m_{\V} \lesssim 10^{-2} \, \text{GeV}$ that can be rewritten in terms of the volume as $\V \gtrsim 10^{13}$. For $\mathcal{V} \sim 10^{13}$, the string scale is $M_s \sim 10^{12} \, \text{GeV}$, the KK scale is $M_{\rm KK} \sim M_{\rm P}/\mathcal{V}^{2/3} \sim 10^9 \, \text{GeV}$ and the gravitino mass is $m_{3/2} \sim M_{\rm P}/\V \sim 10^5 \, \text{GeV}$, in accordance with LHC findings. The mass and radii of the stars for $\V = 10^{13}$ are
\begin{equation}
\label{eq:MassSizeNYD}
M_{\V}/{\tilde{M}} \simeq 1.3 \times 10^{14} \, \text{g} \,, \qquad R_{\V}/{\tilde{R}} \simeq 2.5 \times 10^{-13} \, \text{cm}\,.
\end{equation}

Since an oscillating massive scalar field redshifts as pressureless dust, if the volume modulus has not decayed yet it constitutes dark matter. For this reason we need to ensure that the presence of this field does not overclose the Universe. Assuming that the volume modulus is displaced from the minimum of the potential after inflation, it start oscillating when $m_{\V} \sim H$, which translates into a temperature of $T_i \sim 10^8 \, \text{GeV} \gg T_{\rm BBN} \sim 1 \, \text{MeV}$. Since its energy density redshifts as matter, it ends up dominating the energy density of the Universe. The moment in which it starts dominating depends on the initial energy density stored in the modulus, which is roughly $\rho_i \sim m_{\V}^2 \varphi_i^2$, where $\varphi_i$ is the value of the initial displaced field. Moreover the energy density stored in the field redshifts as
\begin{equation}
\rho_\varphi(t) = m_{\V}^2 \varphi_i^2 \left(\frac{a_i}{a(t)}\right)^3 \,,
\end{equation}
while the energy density in radiation
\begin{equation}
\rho_{\rm rad}(t) = m_{\V}^2 M_{\rm P}^2 \left(\frac{a_i}{a(t)}\right)^4 \,,
\end{equation}
where we assumed that at $a(t_i) = a_i$ the energy density is dominated by radiation, $\rho \simeq H^2 M_{\rm P}^2$ and $m_{\V} \sim H$ ($t_i$ denotes the time at the beginning of the oscillations). Now consider $t = t_{\rm eq}$, i.e.~the moment in which the energy densities stored in radiation and matter are equal. Then the ratio of $\rho_{\varphi}$ and $\rho_{\rm rad}$ is
\begin{equation}
\label{eq:RatioEDMErad}
\frac{\rho_{\varphi}(t_{\rm eq})}{\rho_{\rm rad}(t_{\rm eq})} \simeq \frac{\varphi_i^2}{M_{\rm P}^2} \frac{a_{\rm eq}}{a_i} \sim 1\,,
\end{equation}
where we made the rough approximation that all the matter energy density is stored in the dark matter candidate field $\varphi$ (in the reality part of it is composed of baryons). From eq.~\eqref{eq:RatioEDMErad} we get for the initial displacement
\begin{equation}
\varphi_i \simeq \left(\frac{a_i}{a_{\rm eq}}\right)^{1/2} M_{\rm P} \simeq \left(\frac{T_{\rm eq}}{T_i}\right)^{1/2} M_{\rm P} \simeq 3 \times 10^{-9} \, M_{\rm P} \,,
\end{equation}
where we used that $T_{\rm eq} \simeq 1 \, \text{eV}$, that during radiation domination $a(t) \sim t^{1/2}$ and we assumed radiation domination from the start of the oscillations to matter-radiation equality. Clearly the initial displacement needs to be very fine-tuned. If there is a mechanism that leads to the growth of quantum fluctuations of the volume modulus, compact objects like the previously studied stars could be formed, see Section~\ref{sec:formationall}, with a core amplitude even larger than $\varphi_i$. In such a case part or even the full abundance of dark matter could be composed of microscopic solitonic objects with mass and size given in eq. \eqref{eq:MassSizeNYD}.

\subsubsection{GW production}

The production of GWs would need a careful non-linear analysis of the dynamics of formation and dynamics of the compact objects described in the previous sections, and we leave it for future work. In particular, a numerical study is needed to get the amplitude of the stochastic GW background generated by the dynamics of stars. However we can make some estimates about the frequency of the produced GWs. If the single harmonic approximation holds and the star profile is exactly spherically symmetric as described in the previous section, a single star cannot produce GWs\footnote{It is however expected that a single star in the dense regime can produce GWs as it happens for oscillons~\cite{Antusch:2016con, Antusch:2017flz}.}. However, a possible source for GW production is given by binaries: after their formation in the early Universe, moduli stars can decouple from the Universe expansion and form binary systems. The energy loss in GWs is compensated by a decrease in the distance beween the compact objects and an increase of the frequency. If the time available between the formation of the binary system and the decay of the corresponding modulus is sufficiently long, the orbiting compact objects merge, producing a burst of GWs. If the time is not sufficient for the stars to merge, they will orbit until they disappear due to the decay of the modulus. Since we do not know the initial distance between the moduli stars at their formation, we can only put bounds on the maximum frequency of the GWs produced by the system. This is given by the frequency associated to the \textit{innermost stable circular orbit} (ISCO) orbit (see for instance~\cite{Giudice:2016zpa})
\begin{equation}
\label{eq:FreqISCO}
f_{\rm ISCO} \simeq \frac{1}{\left(6 \pi\right)^{3/2}} \frac{C^{3/2}}{M_{\rm tot} M_{\rm P}} = \frac{1}{2 \left(6 \pi\right)^{3/2}} \frac{\tilde{C}^{3/2}}{\tilde{M}} \left(\frac{\Lambda}{M_{\rm P}}\right) m \,,
\end{equation}
where we defined the (dimensionless) compactness parameter in eq. \eqref{eq:CompactnessDef} (and below) and we used that $M_{\rm tot} = 2 M$. The factor $\Lambda/M_{\rm P}$ in eq.~\eqref{eq:FreqISCO} suppresses the GW frequency at emission with respect to the value of the mass, if $\Lambda < M_{\rm P}$. However, in order for the compact objects to merge it is necessary that the coalescence time $t_{\rm coal}$ is smaller than the available time between the formation of the binary and the decay of the modulus:
\begin{equation}
\label{eq:MergerCond}
t_{\rm coal} < t_{\rm dec} - t_{\rm form} \,.
\end{equation}
The coalescence time for two stars with equal mass $M$ under the assumption of circular orbit is given by~\cite{Maggiore:1900zz}
\begin{equation}
t_{\rm coal} \simeq  \frac{d_0^4 \, M_{\rm P}^6}{M^3} = \frac{\alpha^4}{m} \, \frac{\tilde{M}}{\tilde{C}^4} \left(\frac{M_{\rm P}}{\Lambda}\right)^6 \,,
\end{equation}
where $d_0$ is the initial distance between the two compact objects, that we wrote in terms of the star radius\footnote{The correct value of the parameter $\alpha$ is expected to be the outcome of numerical studies of the formation of the stars, and is related to the distribution of these compact objects.} $R$ as $d_0 = \alpha R$. This time has to be compared with\footnote{The formation time of the binary is larger than the formation time of the star which is in turn larger than $1/m$. However the the estimate in eq.~\eqref{eq:TdecTform} is valid for most scales that go non-linear and form compact objects, see Section~\ref{sec:formationall}.}
\begin{equation}
\label{eq:TdecTform}
t_{\rm dec} - t_{\rm form} \simeq t_{\rm dec} \approx \frac{M_{\rm P}^2}{m^3} \,,
\end{equation}
where we used that $t_{\rm form} \ll t_{\rm dec} \simeq M_{\rm P}^2/m^3$. In the case of the LVS volume modulus $\Lambda = M_{\rm P}$ and the condition that the merger of the stars take place is
\begin{equation}
\label{eq:AlphaBound}
\alpha < \left(\frac{\V^3}{\tilde{M}}\right)^{1/4} \tilde{C} \,.
\end{equation}
that can be satisfied by the denser objects and for large values of the volume $\V$. In the case of blow-up-like moduli the scale $\Lambda$ is the string scale $M_s \simeq M_{\rm P}/\sqrt{\V}$ and the condition in eq.~\eqref{eq:MergerCond} translates into
\begin{equation}
\frac{\alpha^4 \tilde{M}}{\tilde{C}^4} \left(\frac{M_{\rm P}}{M_{\rm s}}\right)^6 \sim \frac{\alpha^4 \tilde{M}}{\tilde{C}^4} \V^3  < \frac{M_{\rm P}^2}{m^2} \sim \V^2 \,.
\end{equation}
This is clearly never satisfied for $\alpha \gtrsim 1$: the blow-up-like moduli stars do not have time to merge before disappearing due to the decay of the modulus.\\

We find that the dimensionless compactness for the LVS and KKLT volume moduli is included in the range
\begin{equation}
8 \times 10^{-6} \lesssim \tilde{C} \lesssim 1.39 \,,
\end{equation}
for the numerical solutions reported in Fig. \ref{fig:MassRadiusPlot}\footnote{In~\cite{Giudice:2016zpa} conventions this should be divided by $8 \pi$, giving a compactness $\sim 0.055 < 0.16$ which is the maximum compactness for interacting boson stars~\cite{AmaroSeoane:2010qx}. However, recall that solutions with $\delta \tilde\varphi_{\rm core} \gtrsim 10^{-3}$ are not fully reliable for the reasons explained in Sec. \ref{sec:ModuliStars}.}. The maximum compactness in the dilute region $\delta\tilde\varphi_{\rm core} \lesssim 10^{-3}$ is $\tilde{C} \simeq 7.8 \times 10^{-3}$. In terms of the modulus mass $m$, the maximum frequency from mergers $f_{\rm ISCO}$ is contained in the range
\begin{equation}
\label{eq:FrequencyWindow}
1.3 \times 10^{-6} \lesssim f_{\rm ISCO}/m \lesssim 3.6 \times 10^{-4} \,,
\end{equation}
where the upper bound corresponds to $\delta\tilde{\varphi}_{\rm core} = \delta\tilde\chi_{\rm core} = 10^{-1}$. The upper bound corresponding to the largest core amplitude in the dilute region $\delta\tilde\varphi_{\rm core} = \delta\tilde\chi_{\rm core} = 10^{-1}$ would be $f_{\rm ISCO}/m \simeq 4.1 \times 10^{-5}$. We recall that, in Hertz units
\begin{equation}
m \simeq 1.5 \times \left(\frac{m}{100 \, \text{TeV}}\right) \times 10^{29} \, \text{Hz}\,.
\end{equation}
Since the blow-up-like moduli stars do not have time to merge, they emit GWs with frequency twice the orbital period, under the assumption of stationary orbit. Its value would typically be much smaller than the corresponding ISCO frequency but in order to compute it, the distribution of distances between stars is needed, and we leave it for future work.\\

The frequency values obtained from mergers refer to the emission time and have to be redshifted to take into account the expansion of the Universe. Assuming that the emission takes at $t_{\rm e}$, the dilution factor is
\begin{equation}
\mathcal{R} = \frac{a(t_{\rm e})}{a(t_{\rm R})} \frac{a(t_{\rm R})}{a(t_0)} \,,
\end{equation}
where $t_{\rm e}$ is the GW production time, $t_{\rm R}$ is the reheating time (given by the decay of the modulus) and $t_0$ is today. We can estimate
\begin{equation}
\label{eq:Redshift}
\frac{a(t_{\rm R})}{a(t_0)} = \left(\frac{\rho_{\rm rad}(t_0)}{\rho_{\rm rad}(t_{\rm R})}\right)^{1/4} = \left(\frac{\pi^2}{30} g_* (t_{\rm R})\right)^{-1/4} \frac{M_{\rm P}^{1/2}}{m^{3/2}} \, \rho_{\rm rad}^{1/4}(t_0) \,,
\end{equation}
where we used that radiation redshifts as $\rho_{\rm rad} \propto a^{-4}$, $g_* (t_{\rm R})$ is the number of relativistic degrees of freedom at reheating and the decay rate of a gravitationally coupled modulus of mass $m$ is $\Gamma \simeq m^3/M_{\rm P}^2$. Moreover $\rho_{\rm rad} (t_0) \simeq 4.3 \times 10^{-5} \rho_{\rm crit}$. Neglecting the numerical prefactor in eq. \eqref{eq:Redshift} the suppression factor is
\begin{equation}
\label{eq:Suppression1}
\frac{M_{\rm P}^{1/2}}{m^{3/2}} \rho_{\rm rad}^{1/4}(t_0) \approx \left(\frac{100 \, \text{TeV}}{m}\right)^{3/2} \, 10^{-11} \,.
\end{equation}
In order to compute the factor $a(t_{\rm e})/a(t_{\rm R})$ a numerical computation of the emission time $t_{\rm e}$ is needed. To give an estimate, this factor is bound to be
\begin{equation}
\label{eq:Suppression2}
\frac{a(t_{\rm e})}{a\left(t_{\rm R}\right)} > \left(\frac{t_{\rm e}}{t_{\rm R}} \right)^{2/3} \simeq \left(\frac{m}{M_{\rm P}}\right)^{4/3} \simeq  1.5 \times \left(\frac{m}{100 \, \text{TeV}}\right)^{4/3} \times 10^{-18} \,,
\end{equation}
where the right hand side is computed assuming that the stars are formed immediately after the modulus starts oscillating, that GWs are produced immediately after the formation of the stars and that the Universe is always matter dominated from the start of the oscillations to the modulus decay. This is of course not the realistic situation in which after the start of matter domination the stars have to be first formed (see Sec. \ref{sec:formationall}), then they have to decouple from the expansion of the Universe and form binaries and then they can start emitting GWs. However, the combination of eq. \eqref{eq:Suppression1} and eq. \eqref{eq:Suppression2} gives the indication that it is in principle possible to lower the frequency down to the LIGO range and even lower to the LISA range (taking into account the window given in eq. \eqref{eq:FrequencyWindow}). Of course, as GWs redshift as radiation ($\rho_{\rm rad} \propto a^{-4}$), the more the frequency is lowered during matter domination, the more also the GW background amplitude is suppressed and is hard to be observed.\\

Besides the sources mentioned in this section, moduli stars can also produce GWs via other mechanisms that need a careful numerical analysis, such as Bremsstrahlung~\cite{Dolgov:2011cq}. In particular it will be exciting to explore which features the generic decay of the modulus leaves in the GW spectrum. Clearly, a numerical analysis of these phenomena, although highly interesting in the GW astronomy era, is beyond the scope of this article. Such GW signals could shed light on the very first instants of the Universe's history, not accessible within optical astronomy.

\subsection{Q-Balls from Open Strings}
\label{sec:OpenStringsQBalls}

The space of open string moduli is vast, model dependent and much unexplored yet.  But there are concrete cases that can be considered. The typical examples are moduli corresponding to the position of D-branes in type II string compactification but also Wilson lines. In the four-dimensional effective field theory they appear as chiral matter multiplets that do not appear in the superpotential  but they may be charged under  Abelian and/or non-Abelian gauge interactions. They can be  part of the observable sector containing the standard model fields or be part of a hidden sector which is coupled only gravitationally to the standard model. 

If the fields do not have holomorphic superpotential couplings the main source of the scalar potential are D-terms. Generically there are many  supersymmetry preserving D-flat directions that correspond to the open string moduli.

In order to explore the possibility of boson stars from open string moduli, a first attempt is to look for non-topological solitons such as Q-balls. At first, the general string theoretical property that no-global symmetries are present in string theory seems to be an obstacle to have Q-balls. There is however a concrete way to have low-energy Abelian symmetries as remnants of anomalous or non-anomalous gauge U$(1)$s for which the gauge field gets a mass by the St\"uckelberg mechanism in which the gauge field absorbs an axion-like field to get a mass but no Higgs field charged under the U$(1)$ gets a vev (see for instance~\cite{Ibanez:2012zz}). In this case a perturbatively exact global U$(1)$ symmetry remains at low-energies which can be the basis of Q-ball solutions.

Following a procedure analogous to an analysis in the MSSM~\cite{Kusenko:1997ad} case, let us consider a number of canonically normalised scalar fields $\Phi_i$ with positive, negative or zero charges under the global U$(1)$. The source of their potential are supersymmetric D-terms of the original local U$(1)$:
\be
U_{\rm D}=g^2 \left(\xi-\sum_{i} q_i| \Phi_i|^2\right)^2
\ee
where the Fayet-Iliopoulos coefficient $\xi$ depends on the closed string moduli. In particular for branes at singularities it is proportional to the size of the cycle, i.e.~the resolution of the singularity, and may hence be arbitrarily small. In this case there are solutions of the D-term equations that have vanishing $\Phi_i$ vevs: after the breaking of supersymmetry these fields get potentials from the standard soft-supersymmetry breaking terms:
\be
U_{\rm soft}= \sum_{i}m_i^2\, |\Phi_i|^2 + \left(\sum_{ijk} A_{ijk}\, \Phi_i\Phi_j\Phi_k +\sum_{ij}  B_{ij}\, \Phi_i\Phi_j + h.c.\right),
\ee
where the coefficients $m_i, A_{ijk}, B_{ij}$ are functions of the closed string moduli which are assumed to be stabilised at the supersymmetry breaking minimum~\cite{Choi:2005ge, hep-th/0505076, Aparicio:2014wxa, Aparicio:2015psl}. Since supersymmetry is assumed to be broken in the closed string sector the global U$(1)$ symmetry remains unbroken and these terms are such that only U$(1)$ preserving combinations are allowed. The condition for the existence of Q-balls can be stated as the search for a non-vanishing minimum for the quantity:
\be
E^2=\frac{2U}{\sum_{i} q_i|\Phi_i|^2}=\frac{2(U_D+U_{soft})}{\sum_{i} q_i|\Phi_i|^2}
\ee
Notice that for small enough $\xi$ the point $\Phi_i=0$ is a minimum of the scalar potential $U$. But it is straightfoward to see that there is a nonvanishing minimum of $E$ above. To see this explicitly we can follow~\cite{Kusenko:1997ad} and consider the time dependent  $\Phi$ fields: $\Phi_i=\rho_i e^{iq_iwt}$ and use  `spherical' coordinates  with the overall radial coordinate $\rho^2=\sum_{i} q_i\rho_i^2=\sum_{i} q_i|\Phi_i|^2$. It is clear that the potential above is time-independent and quadratic in $\rho$ and then there is generically a minimum for $\rho\neq 0$ which is the condition for the existence of Q-balls. This argument applies to both flat directions from the observable sector (as it was argued for the MSSM in~\cite{Kusenko:1997ad}) but also for the $\Phi_i$ fields in a hidden sector coupled to the standard model fields only through gravitational interactions. The properties of the corresponding boson stars differ substantially: Q-balls from the observable sector have been considered to have important phenomenological implications, especially if they carry lepton or baryon number. Then they can play an important role for baryogenesis and constitute part of dark matter~\cite{Kusenko:1997si, Enqvist:2003gh}. 

Since global symmetries are rare in string models it may be easier to consider solutions for gauged symmetries (charged Q-balls). However there is a bound on the strength of the corresponding gauge coupling compared to gravity. Solutions tend to exist if gravity is stronger than the corresponding gauge interactions (see for instance~\cite{0801.0307, 1202.5809}). This may be in conflict with the weak gravity conjecture~\cite{ArkaniHamed:2006dz} in string theory. In general the open string sector of string compactifications is the most model dependent  and it is difficult to establish model independent conclusions. However, even if non-topological solutions may not exist, the attractive nature of gravity makes it very generic that the corresponding boson star solutions will exist.

\subsection{PQ-balls}
\label{sec:PQBalls}

We consider now the possibility to have Q-ball like solution from the PQ shift-symmetry of closed string axions. This symmetry is usually broken by non-perturbative effects giving rise to non-trivial potentials for the corresponding axion field as we have discussed before. However, in special cases its breaking is hierarchically suppressed compared to the potential for the real part and it may be considered as a good approximate symmetry. This is the case in the LVS for the overall volume where the volume axion receives a potential which is doubly exponentially suppressed (i.e.~terms proportional to $e^{-a\tau}$ for which $\tau$ is itself exponentially large whereas the rest of the Lagrangian is only suppressed by powers of $1/\tau$).

In the general case of an exact PQ shift-symmetry for the axion we consider the two-fields system described by the following action
\begin{equation}
S = \int d^4 x\, \mathcal{L} = \int d^4x\, \left[- f(\tau) \left[\partial_\mu \tau \partial^\mu \tau + \partial_\mu \theta \partial^\mu \theta\right] - V(\tau)\right] \,,
\end{equation}
where the two fields can be identified as the real and imaginary parts of a complex modulus $T = \tau + i \theta$ and $f(\tau) = K_{T \overline{T}}$ is the second derivative of the \Kahler potential $K$\footnote{In the following we will leave the $\tau$-(or $\varphi$-)dependence understood in the functions $f(\tau)$ and $f(\varphi)$.}. In the following we take the standard assumption in a Q-ball analysis with a flat Minkowski metric, i.e.~neglecting gravitational effects, and we further assume that the potential $V(\tau)$ has a runaway to zero at $\tau\rightarrow \infty.$\footnote{The potential may also feature another minimum at finite $\tau$ as in LVS.} This runaway, in terms of the canonically normalised field $\varphi$, is assumed to be exponential which is precisely realised for the overall volume in the LVS. The action is then invariant under a PQ shift-symmetry, i.e.~a constant shift of the axion field
\begin{equation}
\label{eq:ShiftSymmetry}
\theta \rightarrow \theta + \text{const.}
\end{equation}

The equation of motion for the axion field $\theta$ takes the current conservation form
\begin{equation}
\label{eq:ConservedCurrent}
\partial_\mu \left(f \partial^\mu \theta\right) \equiv \partial_\mu J^\mu = 0\,,
\end{equation}
where $J^\mu$ is a conserved current associated to the symmetry in eq.~\eqref{eq:ShiftSymmetry}. The conserved current and charge are then
\begin{equation}
J^\mu = f \partial^\mu \theta \,, \qquad Q = \int d^3x\, J^0 = \int d^3x\, f \dot\theta\,.
\end{equation}
Expanding eq.~\eqref{eq:ConservedCurrent} we get
\begin{equation}
\label{eq:EomTheta}
f \ddot\theta - f \nabla^2 \theta + f_\tau \dot\tau\dot\theta - f_\tau \nabla\tau\,\nabla\theta = 0 \,,
\end{equation}
while the equation of motion for $\tau$ is
\begin{equation}
\label{eq:EomTau}
2 f \ddot\tau + f_\tau \dot\tau^2 - 2 f \nabla^2 \tau - f_\tau \left(\nabla \tau\right)^2 + f_\tau \left(\nabla \theta\right)^2 - f_\tau \dot\theta^2  + \partial_\tau V = 0 \,.
\end{equation}

In the regime in which gravity is negligible we can consider the possibility that the PQ shift-symmetry can play a similar role as the U$(1)$ global symmetry in Coleman's Q-balls. After all, redefining the field $T$ in terms of $\Phi=e^{-T}$, the PQ shift-symmetry $T\to T+i\alpha$ becomes $\Phi\to e^{-i\alpha}\Phi$. as in the Q-balls case. However this field redefinition is not that straightforward as we will see now.

Formally, to extremise the energy keeping $Q$ constant we can consider the quantity:
 \bea
 E_\omega & = & \int d^3x\left[ f(\tau)\left( \dot{\theta}^2 + \dot{\tau}^2 + (\nabla\theta)^2+(\nabla\tau)^2\right) + V(\tau)\right] + 2 \omega\left(Q-\int d^3x \partial_0(f\theta)\right) = \nonumber \\
 & = &  \int d^3x\left[ f(\tau)\left( (\dot{\theta}-\omega)^2 + \dot{\tau}^2 + (\nabla\theta)^2+(\nabla\tau)^2\right) + \hat{V}(\tau) \right] + 2\omega Q \,.
 \eea
Here again $\omega$ starts as a Lagrange multiplier. The effective potential is now:
\be
\label{eq:Vhat}
\hat{V}(\tau)=V(\tau) -\omega^2 f(\tau) \,,
\ee
and the $\theta$-dependent terms are   minimised for:
\be
\dot{\theta}=\omega\,,  \qquad \nabla{\theta}=0\,.
\ee
Assuming a stationary solution in which $\dot{\tau}=0$. We arrive then at a similar situation as with Q-balls. A time dependence in the axion field $\theta$ that allows a time translation to be compensated by a constant PQ shift making all physical quantities time-independent.\\

We may try to extend the comparison noticing that for $f=\alpha/\tau^2$ and the original potential $V(\tau)$ vanishing at $\tau\to\infty$ we have that at this limit the charge and the potential vanish, similar to what happens at $\Phi=0$ in the Q-ball case.
 
Notice that $\dot{\theta}=\omega, \dot{\tau}=\nabla{\theta}=0$ automatically satisfy the equation of motion for $\theta$. The one for $\tau$  simplifies considerably if $\tau$ is represented in terms of the canonically normalised field $\varphi$ for which $\partial_\mu\varphi=\sqrt{2f} \partial_\mu \tau$. In this case the equation appears to be of the standard form:
\be
\nabla^2\varphi - \partial_\varphi \hat{V}=0 \,,
\ee
which in spherical coordinates can be written as:
\be
\varphi{''}+\frac{2}{r}\varphi{'}-\partial_\varphi \hat{V}=0 \,.
\label{eqmotion}
\ee
As usual, this leads to an equation equivalent to the motion of a particle in three dimensions under a potential $-\hat{V}$ with $r$ playing the role of time and the second term can be seen as a friction term. Assuming
\be
f(\tau)=\frac{\alpha}{\tau^2}=\alpha e^{-\sqrt{2/\alpha}\, \varphi} \,,
\ee
the potential (and  $\hat{V}$) vanishes asymptotically at $\varphi\to\infty$ corresponding to  decompactification (if $\tau$ determines the overall volume). This is the analogue of the $\Phi=0$ minimum for the Q-balls. At first sight the two systems look very similar. We can also notice that in the Q-balls case, the polar decomposition  $\Phi=\varphi e^{i\theta}$ is not appropriate at the minimum in which $\varphi=0$ since the kinetic term for $\theta$ is $\varphi^2(\partial\theta)^2$ which is singular at $\varphi=0$. Similar in the PQ case the kinetic term for the axionic $\theta$ field: $(\partial\theta)^2/\tau^2$ vanishes at $\tau\to \infty$. 

However, in the decompactification limit $\tau\to \infty$ an infinite number of degrees of freedom are excited and the effective 4D field theory is not the appropriate description. Even independent of this geometric interpretation the fact that all the derivatives of the potential vanish at this minimum renders this setup very different from the Q-balls case for which the second derivative is  already nonzero (mass).\\

Still, mathematically,  the overshoot/undershoot argument by Coleman can be used to look for a bounce solution of the scalar field equation as long as $\hat{V}$ has a finite minimum at negative $\hat{V}$ for which the analogue of the rolling particle in the inverted potential guarantees that there will be initial conditions such that the particle can start close to that minimum and end at $\infty$. The field profile would be increasing for $r\to \infty$, instead of vanishing and then there is no thin wall approximation. Depending on how fast the field increases with $r$ the charge $Q$ and energy $E$ may or may not be finite. The condition for a finite charge is that $r^2 f\to 0$ when $r\to \infty$. If $Q$ and $E$ were finite we may still claim a localised object interpretation,  otherwise the solution is not localised at least in four dimensions and a full ten-dimensional uplift of the solution would be needed. Furthermore the stability argument for Q-balls based on charge conservation and the fact that the Q-ball  is the configuration of minimal energy for a fixed charge is not clearly extended for PQ-balls since both quantities are not finite and there do not seem to be perturbative states charged under this symmetry.\\

For an exponential runaway behaviour of the potential $\hat{V}\sim  -\gamma e^{-b\varphi}$ (e.g. for large $\varphi$ the potential $\hat{V}$ is dominated by the $\omega^2 f(\tau)$ term with $b^2=2/\alpha$ and $\alpha\omega^2=\gamma$ ), then an asymptotic solution of eq.~\eqref{eqmotion} is:
\be
\varphi=A+B\ln r + \dots \,,
\ee
where $\dots$ denotes $\mathcal{O} (1/r)$ terms. For this $\partial_\varphi \hat{V} \sim  b\gamma e^{-b\varphi}$ and the constants $A,B$ can be determined by 
\be
B=2/b, \qquad A=\frac{1}{b}\,\ln\left(\frac{\gamma b^2}{2}\right),
\ee
and the charge density $f$ is proportional to $r^{-bB}=1/r^2$.  The total charge diverges proportionally to the radius
\be
Q=\omega\int d^3x f(\tau) \propto  \int 4\pi r^2 dr \frac{\omega}{r^2}\to\infty \,.
\ee
Similarly the total energy would be dominated  by the $\omega^2 f$ term and would also diverge. However, both charge and energy density are finite at finite $r$ and decrease asymptotically as $1/r^2$, while their ratio is proportional to $\omega$.\\

Notice that the charge of this solution is an axionic charge and can be written in terms of its dual field in four-dimensions, an antisymmetric tensor $B_{\mu\nu}$. Roughly, $f\partial_\mu\theta=\epsilon_{\mu\nu\rho\sigma}\partial_{\nu} B_{\rho\sigma}$ and so:
\be
Q=\int d^3x f\dot{\theta}\propto \int d^3x\, \epsilon_{ijk}H_{ijk} \,,
\ee
where $i,j,k$ denote spatial indices and $H=dB.$ Spherical symmetry implies that $B$ depends only on $r$. This expression is of the standard RR-flux. In fact recall that for the volume modulus the corresponding axion comes from the RR-field $C_{MNPQ}$ and the $B$ field is essentially $B_{\mu\nu}=C_{\mu\nu mn} J_{mn}$ with $m,n$ internal indices and $J_{mn}$ the canonical two-form for Calabi-Yau spaces. From these expressions it is natural to identify the PQ-ball charge as a flux from the ten-dimensional theory. Notice that the PQ-ball charge is similar to the  charge of axionic black holes~\cite{Bowick:1988xh} for which $H$ is exact and $Q=\int_{S^2} B$.\\

\begin{figure}
\includegraphics[width=1\textwidth]{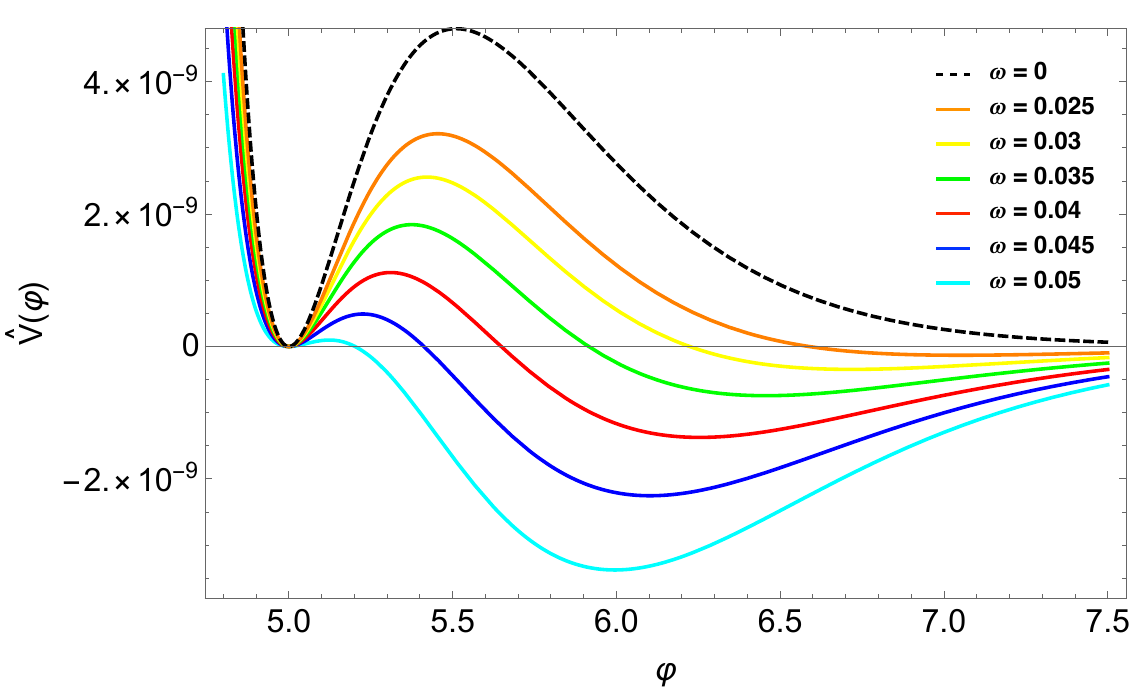}
\caption{Modified potential $\hat{V}$ for different values of the constant $\omega$.}
\label{fig:ModifiedPotential}
\end{figure}

\begin{figure}
\subfigure{\includegraphics[width=7.8cm]{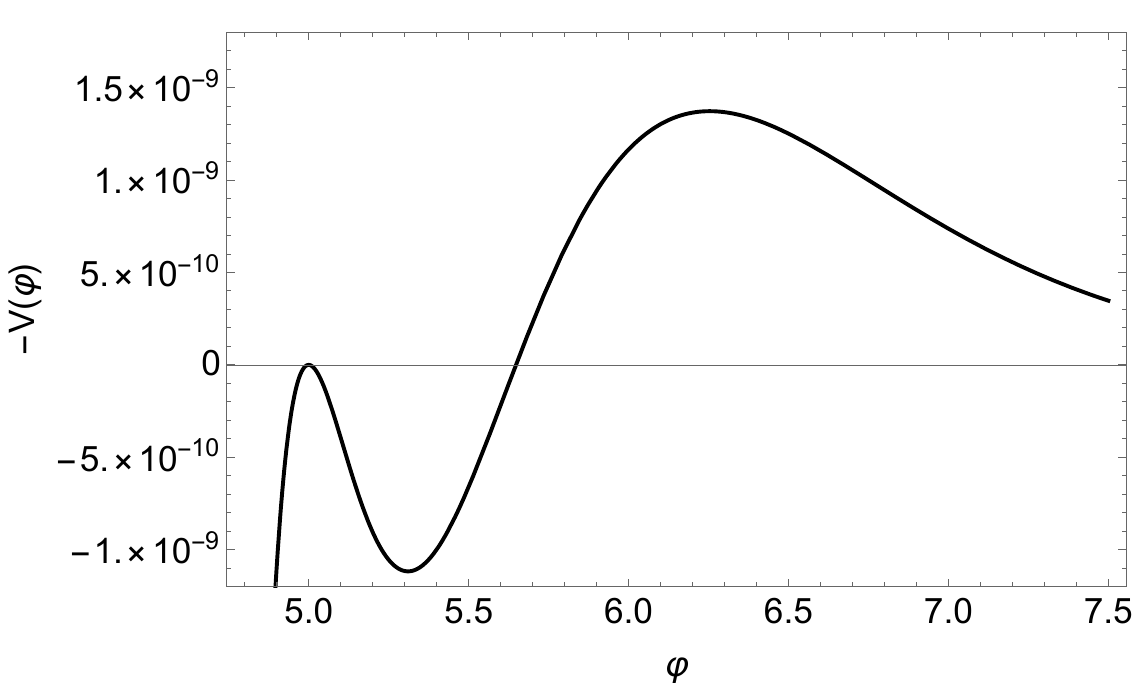}}
\hfill
\subfigure{\includegraphics[width=7.cm]{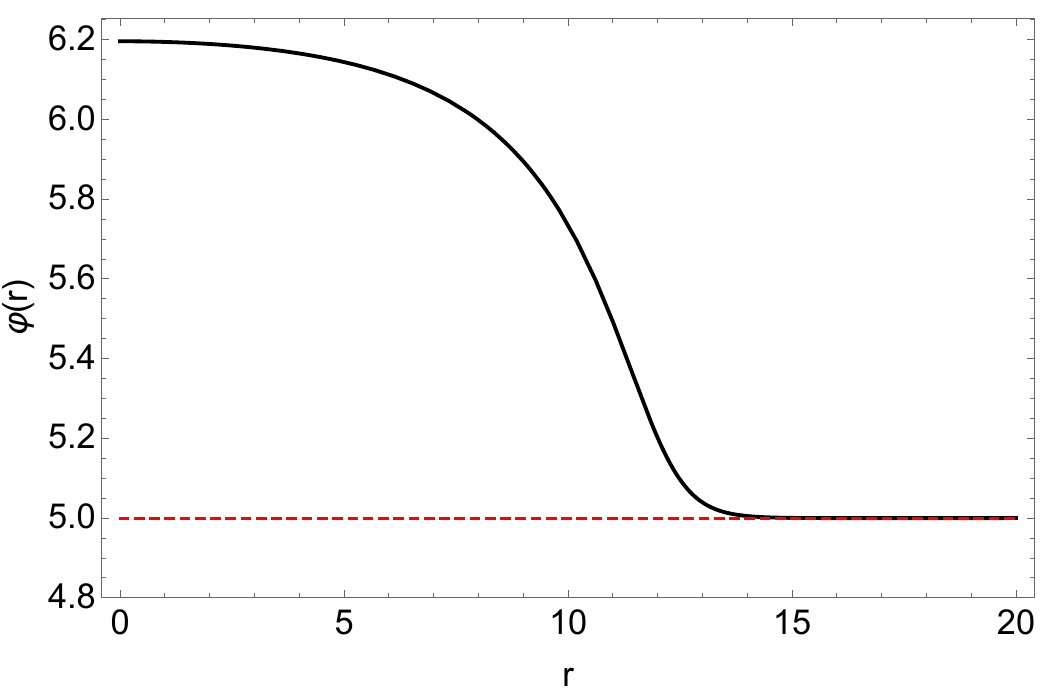}}
\caption{\textit{Left panel}: Inverted potential in the case $\omega=0.025$. The two minima become two maxima and the bounce solution corresponds to the classical motion of a point particle that starts close to the larger maximum and stops exactly in the lower maximum. The effective mass computed around the fake minimum in $\varphi \simeq 5.3$ sets the timescale for such classical motion and hence also the natural spatial scale for the radius of the bounce solution plotted in the right panel. \textit{Right panel:} bounce solution interpolating between the fake AdS vacuum and the Minkowski vacuum of the modified potential for $\omega = 0.025$. The red dashed line represents the position of the Minkowski minimum.}
\label{fig:Bounce}
\end{figure}

Besides the decompactification minimum, if the original scalar  potential ($\omega=0$) also has a second minimum, corresponding to a four-dimensional spacetime, we may also consider the possibility for `transitions' from the $\omega\neq 0 $ minimum of $\hat{V}$ and the finite $\tau$ minimum of $V(\tau)$. We consider for simplicity the following potential
\begin{equation}
\hat{V}(\varphi) = a_1 e^{-5 \varphi} - a_2 e^{-4 \varphi} + a_3 e^{-3 \varphi} - \omega^2 e^{-2 \varphi} \,,
\end{equation}
where the last term come from $\omega^2 f$ in eq. \eqref{eq:Vhat} with $\alpha=1/2$. It is possible to impose that such a potential has a vanishing energy stationary point in $\varphi_0$ requiring $V(\varphi_0) = V'(\varphi_0) = 0$, that translates into two requirements for the coefficients
\begin{equation}
a_2 = \omega^2 e^{2 \varphi_0} + 2 a_1 e^{-\varphi_0}  \,, \qquad a_3 = 2 \omega^2 e^{\varphi_0} + a_1 e^{-2 \varphi_0} \,.
\end{equation}
Requiring that $\varphi_0$ is also a minimum leads to another condition on $a_1$. For all purposes of the subsequent discussion we take $\varphi_0 = 5$ and $a_1 = 10^4$ that ensure that $\varphi_0$ is a minimum. Varying the value of $\omega$ leads to a modification of the potential that feature a second AdS minimum, as shown in Fig. \ref{fig:ModifiedPotential}.
Classical paths can be found connecting these two points which would correspond to symmetry breaking points in the Q-balls case. We solve the bounce equation in eq. \eqref{eqmotion} for the case $\omega = 0.025$ (see left panel of Fig. \ref{fig:Bounce} for the inverted potential) and we get a thick-wall solution as shown in the right panel of Fig. \ref{fig:Bounce}\footnote{The radial distance in the plot of the solution is given in units of the fake mass around the fake minimum (at $\varphi \simeq 5.3$) of the inverted potential in the left panel of Fig. \ref{fig:Bounce}, that sets the natural timescale.}. Notice that from the inverted potential it is possible to start from the new minimum to either the (shifted) compactified vacuum or to the decompactified vacuum at infinity. Clearly, the total charge is infinite in this case also.\\

We have seen that the PQ-balls solutions are mathematically very similar to the original Q-balls, however they have different physical properties. In particular the fact that Q-balls correspond to the minimum energy configurations for a fixed charge $Q$ does not extend to the PQ-balls case since the total charge is infinite.  Once gravity is included we can simply see them as extensions  of the $\omega=0$ case discussed above for the volume modulus to arbitrary values of $\omega$. Gravity, rather than the properties of the symmetric potential provides the attractive force to generate the boson star solutions.

\section{Formation Mechanisms}
\label{sec:formationall}

In Sec.~\ref{sec:StringCOs} we have pointed out that moduli potentials support many different types of compact objects. However, whether they are actually formed during the history of the Universe is a different question. The formation of compact objects typically requires that the following two conditions are satisfied:
\begin{itemize}
\item[I)] There is some initial localized overdensity;
\item[II)] The initial overdensity collapses due to the effect of attractive interactions.
\end{itemize}
Typical examples include the formation of (pseudo-)solitonic objects like Q-balls or oscillons and the formation of structures in the Universe. Following these two examples we can schematically distinguish between two different classes of formation mechanisms, depending on whether gravity plays a crucial role in the realisation of the above conditions. In this section we will mainly discuss condition I).\\

Condition I) can be achieved immediately after inflation, if there is a quick amplification of the quantum fluctuations of the inflaton (or any other scalar field) that is oscillating around the minimum of its potential~\cite{Amin:2014eta}. As we previously reviewed in~\cite{Antusch:2017flz} there are two main mechanisms for the amplification of the quantum fluctuations for an oscillating scalar field, i.e.~parametric resonance and tachyonic oscillations. As the timescale for these amplifications is typically short, gravity can be neglected during the amplification of these fluctuations. A further possibility is that quantum fluctuations are amplified as the field is rapidly spinning in a U$(1)$ symmetric potential, as described in~\cite{Boyle:2001du}\footnote{The growth of fluctuations in spintessence models can take place even in the absence of gravity, as shown in~\cite{Kasuya:2001pr}.}. In Section~\ref{sec:SpinningAxion} we will show that even if there is no U$(1)$ symmetry, this mechanism could still work for a modulus-axion system, provided that the axionic direction is flat and that the field is spinning at constant speed. \\

A possible alternative to get large initial overdensities is through phase transitions in the early Universe. As briefly reviewed at the beginning of Section~\ref{sec:bosonstarsinfieldtheory} the typical example is a scenario in which the PQ U$(1)$ symmetry breaking takes place after inflation: in this case the field resides in a different vacuum in different regions of the Universe, so that the overdensities are typically large to start with. Numerical simulations have shown that they can lead to the formation of axion miniclusters~\cite{Hogan:1988mp, Kolb:1993zz, Kolb:1993hw, Enander:2017ogx}. An alternative mechanism that could be at work even if the PQ U$(1)$ symmetry is broken before the end of inflation employs a first order phase transition in the sector that generates the axion mass~\cite{Hardy:2016mns}. Despite these mechanisms can be very efficient, they are obviously model-dependent.\\

If gravity is negligible the collapse of the overdensities, i.e.~condition II), can take place due to the attractive self-interaction of the scalar field. The requirement that the self-interaction is attractive translates into the condition that the scalar potential has to be shallower than quadratic~\cite{Amin:2013ika}. The formation of (pseudo-)solitonic objects like Q-balls and oscillons has been intensively studied with the help of lattice codes~\cite{Amin:2013ika, Amin:2010jq, Amin:2010dc, Amin:2011hj, Lozanov:2014zfa, Antusch:2015vna, Antusch:2015ziz, Antusch:2016con}. We already numerically studied the formation of oscillons in string models in~\cite{Antusch:2017flz}. We would like to highlight that in this case the formation of compact objects depends crucially on the choice of the initial conditions of the background scalar field. If the field is the inflaton, the initial conditions are fixed by the inflationary evolution. However, if the field is a modulus displaced during inflation, the Hubble friction provided by the thermal bath that dominates the energy density between the decay of the inflaton and the early phase of matter domination would typically damp the background field very quickly, making the mechanism for the amplification of fluctuations much less efficient.\\

The effects of gravity could provide an alternative way to get large overdensities from the initial scalar perturbations (seeded e.g.~during inflation), as in the case of structure formation. When the overdensities are of order unity, the same gravity could provide the attractive interaction that makes them collapse and produce the large structures that we currently observe in the Universe~\cite{Gorbunov:2011zzc}. In Section~\ref{sec:EarlyMatterEra} we point out that string moduli typically satisfy requirement I) for the formation of compact structures even in the cases of potentials (or initial conditions) that do not support tachyonic oscillations, parametric resonance, spinning axion solutions or phase transitions. Similar results have already been obtained in the past in~\cite{Khlopov:1985jw, 0805.1748, Cembranos:2015oya, Hidalgo:2017dfp}. The fulfillment of condition II) requires detailed and model-dependent numerical studies of the non-linear evolution of the overdensities including the effects of gravity. We are performing such analyses in the case of the moduli stars described in Sec.~\ref{sec:ModuliStars} and we will report these results in forthcoming publications. Similar aspects of early structure formation have already been studied in a few papers. In~\cite{Easther:2010mr, Jedamzik:2010dq}  the authors focused on an early matter era caused by the inflaton, in~\cite{Assadullahi:2009nf} the authors focused on the GW production due to an early matter era and in~\cite{Erickcek:2011us} the authors studied the formation of small compact minihalos due to the presence of an early matter era (possibly detectable by LISA). These papers show, in the light of the discovery of GWs, how promising the study of this early matter domination era can be.

\subsection{Spinning Axion}
\label{sec:SpinningAxion}

In this section we extend the previous discussion on the effects of the mixing kinetic terms between the volume modulus and its axionic partner to other cosmological implications apparently independent of boson stars. As we will observe at the end of the section the spinning axion scenario could be an interesting option to trigger the growth of fluctuations, that eventually could clump and form compact objects.\\

We work in Newtonian gauge
\begin{equation}
ds^2 = - \left(1+2 \phi\right)  dt^2 + a^2 \left(1-2\phi\right) \mathbf{dx}^2 \,,
\end{equation}
The action is
\begin{equation}
\label{eq:ActionSpintessence}
\mathcal{S} = \int d^4x \, \sqrt{-g} \mathcal{L} = \int d^4x \, \sqrt{-g} \left[-f g^{\mu \nu} \left(\partial_\mu \tau \partial_\nu \tau + \partial_\mu \theta \partial_\nu \theta\right) - V(\tau)\right] \,.
\end{equation}
Canonically normalizing the radial field $\tau$:
\begin{equation}
\sqrt{2 f} d \tau = d \varphi \,,
\end{equation}
the action reduces to
\begin{equation}
\mathcal{S} = \int d^4x \, \sqrt{-g} \left[-g^{\mu \nu} \left(\partial_\mu \varphi \partial_\nu \varphi + f(\varphi) \partial_\mu \theta \partial_\nu \theta\right) - V\right] \,.
\end{equation}
As discussed before, this generalises the case for a U$(1)$ invariant Lagrangian for a complex field for which the kinetic term is $f(\varphi)=\varphi^2$. For the case of string moduli we will use
\begin{equation}
\label{eq:FModuli}
f=\alpha/\tau^2=\alpha e^{-\sqrt{2/\alpha}\varphi} \,.
\end{equation}
This kinetic mixing can have other implications. In particular the proposal of `spintessence' \cite{Boyle:2001du} relies on this kinetic mixing for the U$(1)$ case. The idea is that (as for Q-balls)  the phase of $\varphi$ is linear in time and its  kinetic term  provides an extra term to the equation for the modulus of $\varphi$. This spinning of the scalar field modifies substantially the equation of state and therefore the cosmological implications of $\varphi$. We may wonder if a similar situation happens for the closed string moduli with an approximate shift-symmetry for the axionic component allowing a linear time-dependence of the axion field which may be denoted 'spinning axion'.
The equations of motion are
\begin{equation}
\label{eq:EOMtau}
- f_\tau g^{\mu \nu} \left(\partial_\mu \tau \partial_\nu \tau + \partial_\mu \theta \partial_\nu \theta\right) + \frac{2}{\sqrt{-g}} \partial_\mu \left(\sqrt{-g} f g^{\mu \nu} \partial_\nu \tau\right) - \partial_\tau V = 0 \,,
\end{equation}
\begin{equation}
\label{eq:EOMTheta}
\partial_\mu \left(\sqrt{-g} f g^{\mu \nu} \partial_\nu \theta\right) = 0 \,.
\end{equation}
The last equation is in the form of a conservation law, with current $J^\mu = \sqrt{-g} f g^{\mu \nu} \partial_\nu \theta$, from which we can define a conserved charge
\begin{equation}
Q = - \int d^3x\, \sqrt{-g} g^{00} f \dot\theta \equiv q \, \text{Vol} \,,
\end{equation}
where $q$ is the charge density defined as
\begin{equation}
\label{eq:ChargeDensity}
q = 2 f a^3 \dot\theta \,,
\end{equation}
and $\text{Vol} = \int d^3x \sqrt{-g}$. Taking just the homogeneous part of the field $\tau$ we get the following equation of motion
\begin{equation}
\ddot\tau + \frac{\partial_\tau f}{2 f} \dot\tau^2 + 3 H \dot\tau + \frac{\partial_\tau V}{2 f} - \frac{q^2}{8 a^6} \frac{\partial_\tau f}{f^3} = 0 \,.
\end{equation}
A spintessence-like solution is obtained if there is a regime in which the first three terms in this equation of motion are negligible, such that
\begin{equation}
\label{eq:SpintessenceConstraint}
\frac{\partial_\tau V}{2 f} = \frac{q^2}{8 a^6} \frac{\partial_\tau f}{f^3} \,.
\end{equation}
Notice that for the $f$ in eq. \eqref{eq:FModuli} (of for any $f \propto \tau^{-y}$ with $y > 0$), $\partial_\tau f < 0$ and in order to satisfy eq.~\eqref{eq:SpintessenceConstraint} the classical field has to lie in the $\partial_\tau V < 0$ region of the potential. 

In terms of the canonically normalized homogeneous field $\varphi$ defined as
\begin{equation}
\frac{\sqrt{\alpha} d\tau}{\tau} = \frac{d \varphi}{\sqrt{2}} \,,\qquad  \tau=e^{\frac{\varphi}{\sqrt{2\alpha}}} \,,
\end{equation}
the equation of motion for the homogeneous field can be written as
\be
\ddot{\varphi}+3H\dot{\varphi}+\partial_{\varphi} V=\frac{q^2\partial_\varphi f}{4 a^6 f^2} \,.
\label{FieldEvolution}
\ee

To illustrate matters let us take the  simple  run-away potential and gauge kinetic function:
\begin{equation}
\label{eq:RunAwayPotential}
V =V_0  \,e^{-\kappa_1\varphi} \,, \qquad f=\alpha\, e^{-\kappa_2\varphi} \,,
\end{equation}
where we have redefined $\kappa_2 = \sqrt{2/\alpha}$ for later convenience. Contrary to what happens to the $|\Phi|^n$ potential  for the complex field $\Phi$ in spintessence, the first two terms in eq.~\eqref{FieldEvolution} are not directly negligible for the exponential potential since for a constant equation of state all terms in this equation scale as $1/t^2$. The explicit solution with constant equation of state and `charge' $q$ is 

\be
\varphi(t)=B\, \ln t - C\,, \qquad a(t)=t^{\frac{\kappa_1+\kappa_2}{3\kappa_1}} 
\ee
with $B,C$ and $q$ related by:

\be
B=\frac{2}{\kappa_1}\,, \qquad   q^2\, =\, \frac{4\kappa_1 V_0}{\kappa_2}\, e^{(\kappa_1+\kappa_2)C}\, -\,  \frac{8}{\kappa_1^2}\,  e^{\kappa_2 C}\,,
\ee
and the equation of state is
\begin{equation}
w(\varphi) = \frac{\kappa_1-\kappa_2}{\kappa_1 +\kappa_ 2} \,.
\end{equation}
This is  similar to the  equation of state  for an oscillating field with  potential $V=|\Phi|^n$~\cite{0805.1748} for which the time average $\langle w\rangle=(n-2)/(n+2)$. But in our case the result holds without the time average.\\

Notice that the term coming from the kinetic mixing is crucial for this solution to exist since there is no way to satisfy the relations above for $q=\kappa_2=0$. Furthermore, the first two terms in eq.~\eqref{FieldEvolution} can be neglected even though they have the same time dependence of the other two as long as $q^2\gg 1$ and $C,\kappa_1,\kappa_2>0$.  Notice also that $\kappa_1=\kappa_2$ leads to matter domination ($w = 0$) whereas $\kappa_1=2\kappa_2$ gives radiation domination ($w = 1/3$). Also the limit $\kappa_1\gg \kappa_2$ gives kinetic domination ($w \rightarrow 1$) and $\kappa_2 \gg \kappa_1$ leads to dark energy domination ($w \rightarrow -1$).\\
 
It has not escaped our notice that this runaway potential, once dressed with the kinetic contribution from $\theta$, leads to an effective potential (in flat spacetime ($a(t)$=constant)):
\be
V_{\rm eff} = V+ \frac{q^2}{4f} \,,
\ee
 which, unlike the runaway $V$ above, has a minimum. Therefore kinetic axion terms can stabilise the real part of the modulus field. In particular  flux compactifications of IIB string theory lead to a no-scale flat potential at tree level but with a runaway potential for negative Euler with $\alpha=3/4$ and $\kappa=9/2$ leading to a minimum for $\tau\propto q^{-4/13}$ which is in the effective field theory regime for $q\ll 1$. Notice, however, that since the equation of state is not standard even though the minimum is at non zero vacuum energy this does not correspond to $w=-1$.\\
 
We have also investigated the stability of such solution following the procedure of~\cite{Boyle:2001du, Kasuya:2001pr}. To make the equations cleaner, we neglect gravity in the study of the evolutions of perturbations, setting $\phi = 0$. It has been shown that the conclusions should not change by including gravity in the discussion~\cite{Boyle:2001du}. We can compute the equations of motion for the perturbations using
\begin{equation}
\varphi = \varphi_0 + \delta \varphi \,, \qquad \theta = \theta_0 + \delta \theta \,.
\end{equation}
Denoting $f_0 = \alpha \, e^{-\kappa_2 \varphi_0}$, we get respectively
\begin{equation}
\label{eq:EOMdeltavarphi}
\delta \ddot\varphi + 3 H \delta \dot\varphi - \kappa_2^2 \, f_0 \dot\theta_0^2 \delta \varphi - \frac{\nabla^2}{a^2} \delta \varphi + 2 \kappa_2 \, f_0 \dot\theta_0 \delta \dot\theta + V'' \delta \varphi = 0 \,,
\end{equation}
\begin{equation}
\label{eq:EOMdeltatheta}
\delta \ddot\theta + 3 H \delta \dot \theta - \frac{\nabla^2}{a^2} \delta \theta - \kappa_2 \, \dot\varphi_0 \delta \dot\theta - \kappa_2 \, \dot\theta_0 \delta \dot\varphi = 0 \,,
\end{equation}
where $V'' = \partial_\varphi \partial_\varphi V$.\\

It is possible to study the stability of the system against perturbations by just using the ansatz
\begin{equation}
\label{eq:FlucAnsatz}
\delta \varphi = \delta \varphi_0 \, e^{\Omega t + i \mathbf{k \cdot x}} \,, \quad \delta \theta = \delta \theta_0\, e^{\Omega t + i \mathbf{k \cdot x}} \,.
\end{equation}
Plugging this ansatz in eq.s~\eqref{eq:EOMdeltavarphi} and~\eqref{eq:EOMdeltatheta}, working in Fourier space and using that $a H \ll k$ and that $\varphi_0$ varies slowly we get a quadratic equation in $\Omega$
\begin{equation}
\label{eq:OmegaEquation}
\Omega^4 + \Omega^2 \left[2 \frac{k^2}{a^2} + V'' + \kappa_2^2 f_0 \,\dot\theta_0^2\right] + \frac{k^2}{a^2} \left(\frac{k^2}{a^2} + V'' - \kappa_2^2 f_0\, \dot\theta_0^2\right) = 0 \,.
\end{equation}
The fluctuations in eq.~\eqref{eq:FlucAnsatz} grow if $\Omega$ is real and positive, which is ensured if the last term in eq.~\eqref{eq:OmegaEquation} is negative, i.e.~for modes that satisfy
\begin{equation}
0 < \frac{k^2}{a^2} < \frac{k_{\rm J}^2}{a^2} \equiv \kappa_2^2 f_0 \, \dot\theta_0^2 - V'' \,,
\end{equation}
where $k_{\rm J}$ is the Jeans mode that can be rewritten as
\begin{equation}
\frac{k_{\rm J}^2}{a^2} = \frac{\kappa_2^2\, q^2}{4 f_0 a^6} - V'' \,.
\end{equation}
As in string models we expect $V''$ to be positive, the existence of such an instability band has to be checked on a case by case basis. For instance, if we assume that the potential is dominated by the run away potential in eq.~\eqref{eq:RunAwayPotential} in the region where the motion of the field $\varphi_0$ is taking place, then the Jeans mode becomes
\begin{equation}
\frac{k_{\rm J}^2}{a^2} = \frac{\kappa_2^4 \, q^2}{8 a^6} e^{\kappa_2 \varphi_0} - V_0 e^{-\kappa_1 \varphi_0} \,,
\end{equation}
that, depending on the parameters of the model can stay positive for some time (despite the $a^{-6}$ suppression in the first term), leading to a significant growth of the fluctuations. Let us stress that we neglected gravity in the stability analysis, but it is expected that its inclusion would not change the conclusion, as it happens in the original spintessence model~\cite{Boyle:2001du}. If the spinning axion field has to provide dark matter (that could be the case if $\kappa_1 = \kappa_2$), one should explicitly check that the Jeans length is such that it allows the formation of large scale structures in agreement with observation. We leave a detailed scan of the potentials for which the spinning axion provides a good dark matter candidate for the future. Moreover, the growth of the fluctuations can lead to the formation of non-topological solitons. Unlike the case of spintessence, Q-balls cannot form, due to the absence of an unbroken U$(1)$ vacuum. However, oscillons can be formed if the potential in the radial direction has a minimum and it is shallower than quadratic around it.

\subsection{Early matter era}
\label{sec:EarlyMatterEra}

\begin{figure}
\includegraphics[width=1\textwidth]{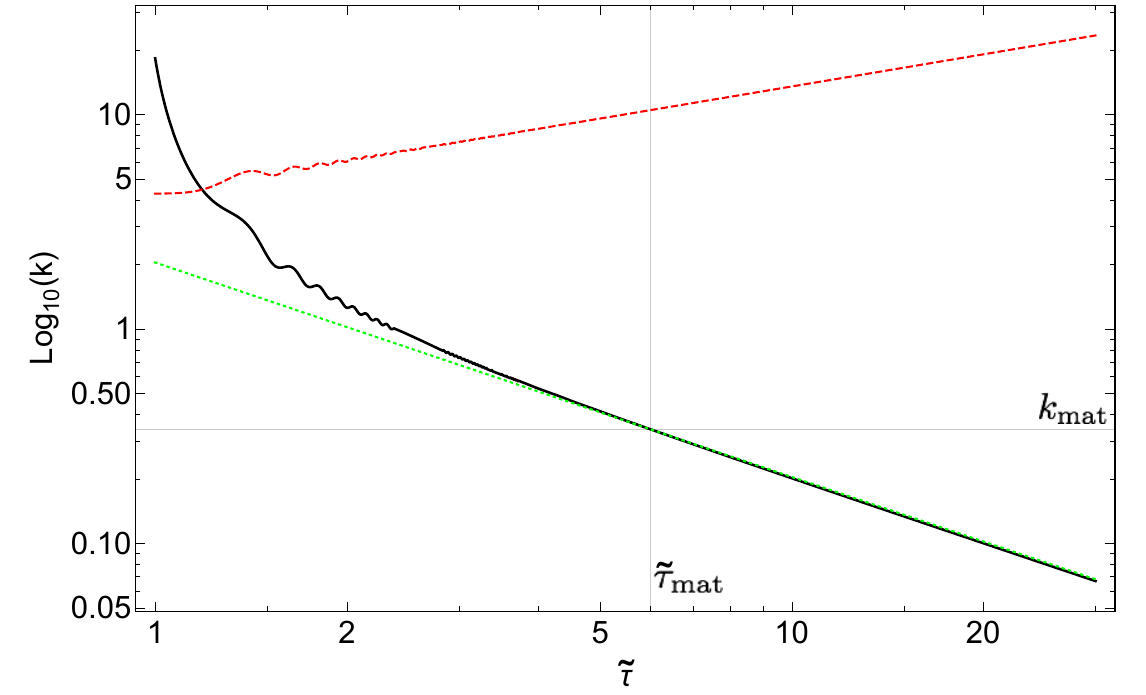}
\caption{Evolution of the comoving horizon $a H$ (black) and of the comoving Jeans mode $k_{\rm J}$ (dashed red). The green dotted line represents the relation between $k$ and $\tilde{\tau}$ in eq.~\eqref{eq:TauHE}. The horizontal and vertical lines correspond to $k_{\rm mat}$ and $\tilde{\tau}_{\rm mat}$ respectively.}
\label{fig:InterestingModes}
\end{figure}

\begin{figure}
\subfigure{\includegraphics[width=7.cm]{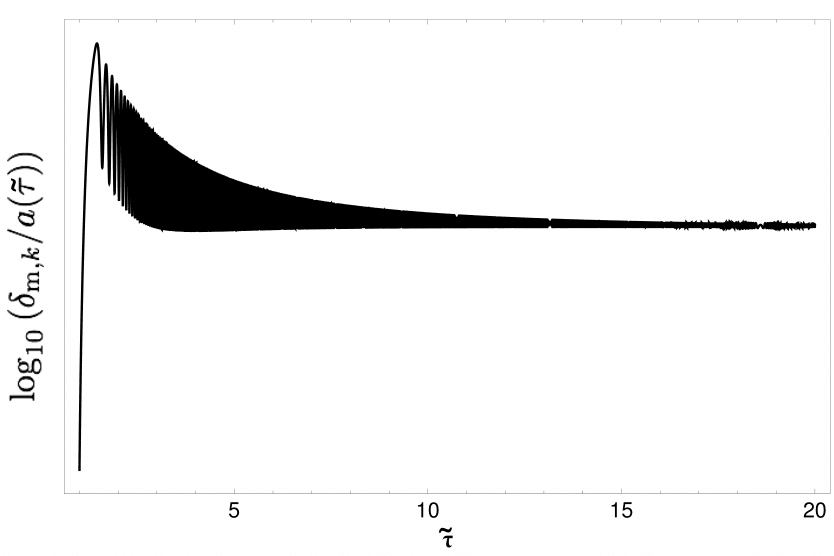}}
\hfill
\subfigure{\includegraphics[width=7.5cm]{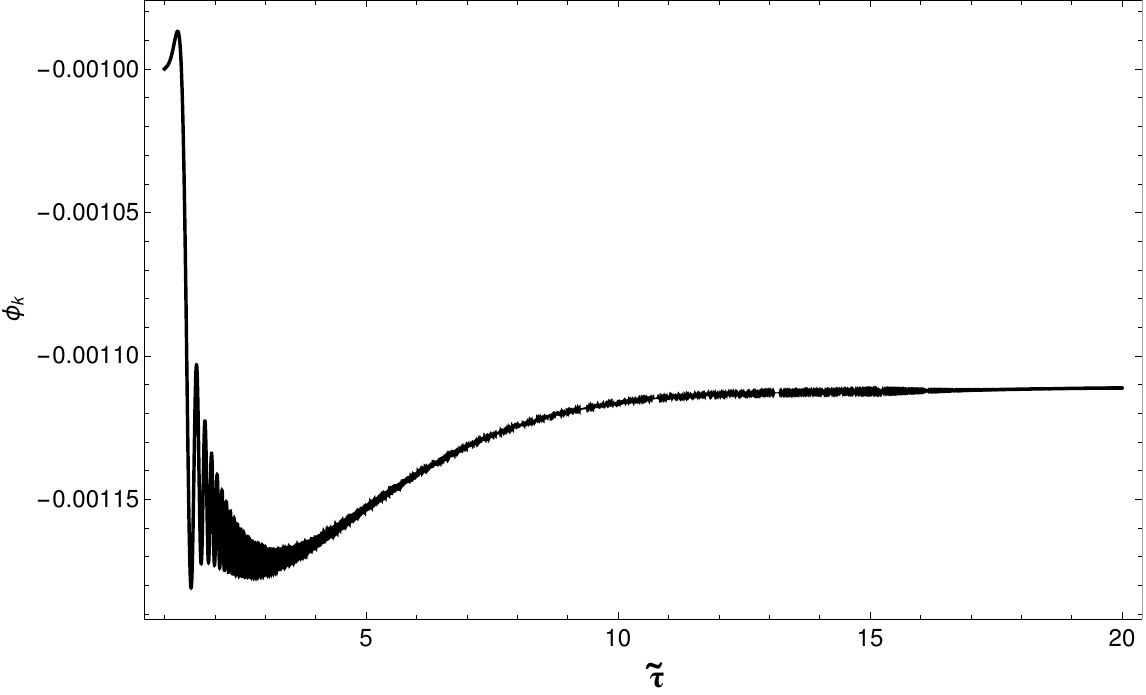}}
\caption{\textit{Left panel}: Overdensity for the mode $k=1$ normalized by the scale factor $\delta_{\text{m,k}}/a(\tilde{\tau})$, in log scale. Since during matter domination the matter overdensity increases as the scale factor, the curve tends to a constant value. We use arbitrary units on the y-axis. \textit{Right panel}: Newtonian potential for the mode $k=1$. The initial value is normalized to $\phi_k = -1$. As expected during matter domination the Newtonian potential tends to a constant value.}
\label{fig:OverD}
\end{figure}

In this section we point out that the requirement I) at the beginning of Section~\ref{sec:formationall} is generically satisfied by string models before the beginning of BBN. The main observation relevant to this section is that during matter domination sub-horizon matter density perturbations modes $\delta_{\rm m,k}$ grow linearly with the scale factor\footnote{In terms of conformal time $\delta_{\rm m,k} \propto \tau^2$. The reader should not confuse conformal time $\tau$ with moduli.}
\begin{equation} 
\delta_{\rm m,k} \equiv \frac{\delta \rho_{\rm m,k}}{\langle \rho \rangle} \propto a(t) \sim t^{2/3} \,, \qquad k \gg a H \,.
\end{equation}
Since an oscillating modulus can be well described as pressureless dust, a growth of the matter perturbations is expected also during the early matter domination prior to BBN\footnote{A modulus driven early matter era generically leads to a rich phenomenology, see~\cite{Kane:2015jia, Acharya:2008bk, Acharya:2009zt, Acharya:2010af, Aparicio:2015sda, Aparicio:2016qqb, Allahverdi:2013noa, Allahverdi:2016yws}.}. Density perturbations can roughly grow as much as
\begin{equation}
\Psi = \frac{\delta_{\rm m,k}(t_{\rm dec})}{\delta_{\rm m,k}(t_{\rm mat})} \approx \left(\frac{t_{\rm dec}}{t_{\rm mat}}\right)^{2/3} \approx \left(\frac{H_{\rm mat}}{H_{\rm dec}}\right)^{2/3} \approx \left(\frac{m}{\Gamma}\right)^{2/3} \approx \left(\frac{M_{\rm P}}{m}\right)^{4/3} \,,
\end{equation}
where the subscripts \textit{dec} and \textit{mat} denote the modulus decay time and the moment at which the early matter domination starts.  We used that during matter domination $H \propto t^{-1}$, that the modulus starts oscillating when $H_{\rm mat} \sim m$ and that the decay rate is $\Gamma \simeq m^3/M_{\rm P}^2$ for a gravitationally coupled modulus. In the case of the volume modulus the enhancement can then be as large as
\begin{equation}
\Psi = \left.\frac{\delta_{\rm m,k}(\tau_{\rm dec})}{\delta_{\rm m,k}(\tau_{\rm mat})}\right|_{\V}  \approx \left(\frac{M_{\rm P}}{M_{\rm P}/\V^{3/2}}\right)^{4/3} = \V^2 \,,
\end{equation}
since the volume modulus mass is $m \simeq M_{\rm P}/{\V^{3/2}}$. For the blow-up moduli the maximum enhancement is $\Psi \simeq \V^{4/3}$, for fibre moduli $\Psi \simeq \V^{20/9}$ while for KKLT is $\Psi \simeq \left(\frac{M_{\rm P} \V}{|W_0|}\right)^{4/3}$. The maximum enhancement comes independently of the value of $\mathcal{V}$ or $W_0$ from requiring that $m \gtrsim 100 \, \text{TeV}$, in order to avoid the cosmological moduli problem. Such bound still gives a huge possible enhancement~\cite{Assadullahi:2009nf}
\begin{equation}
\label{eq:MaxEnhancement}
\Psi_{\rm max} \simeq 10^{20}\,.
\end{equation}
This behaviour can be easily checked numerically by solving the linearized evolution equations for scalar perturbations in Newtonian gauge~\cite{Mukhanov:1990me} derived from general relativity. The energy density of the Universe is initially dominated by a thermal bath\footnote{In the numerics we chose $\rho_{\rm radiation} = 10^4 \, \rho_{\varphi} \gg \rho_{\varphi}$ (where $\rho_{\varphi}$ is the energy density stored in the displaced scalar field) at the initial time.}, while a scalar field is displaced from its minimum\footnote{We consider a quadratic potential: including corrections to the quadratic potential slightly changes only the transient evolution of the perturbations.} and stuck due to Hubble friction. In Figure~\ref{fig:InterestingModes} we show the evolution of the comoving horizon $a H$ and of the comoving Jeans mode\footnote{$\tilde\tau$ and $k$ are in units of the mass of the field $m$.} $k_{\rm J} = a \sqrt{m H}$~\cite{Gorbunov:2011zzc}. All the matter overdensity modes $k$ that enter the horizon (i.e.~$k > a H$) and such that $k < k_{\rm J}$, grow like the one shown in the left panel of Figure~\ref{fig:OverD}, where we show the evolution of $\delta_{\rm m,k}$ for a mode ($k = 1$) that enters the horizon immediately after the beginning of the evolution. After an initial brief transient in which the Universe is going from being radiation dominated to being matter dominated, the overdensity starts growing linearly with the scale factor. In the right panel of Figure~\ref{fig:OverD} we show the corresponding Newtonian potential, that tends to a constant in the matter dominated era, as expected. Finally, radiation perturbations oscillate around the constant value of $\phi$.\\

It is possible to get an analytical solution for modes that are still superhorizon while the Universe is already matter dominated\footnote{In the numerics we adapted the discussion below to take into account that the Universe is initially radiation dominated, hence $w = 1/3$.}. The initial conditions needed to solve the linearized equations can be found using the relation between the comoving curvature perturbation $\mathcal{R}$ and the Newtonian potential $\phi$\footnote{In the approximation of constant equation of state $w$.} for modes that are superhorizon
\begin{equation}
\label{eq:RelationCurvaturePotential}
\mathcal{R}_k \simeq - \frac{5 + 3 w}{3 + 3 w} \phi_k \,.
\end{equation}
We can set the initial conditions directly in the matter dominated phase ($w=0$) so that for super-horizon modes
\begin{equation}
\label{eq:SHModes}
\delta_{\text{m},k} \approx- 2 \phi_k = \frac{6}{5} \mathcal{R}_k \,.
\end{equation}
Assuming a flat spectrum of perturbations from inflation
\begin{equation}\Delta_{\mathcal{R}}^2 (k) = \frac{k^3}{2 \pi^2} \left|\mathcal{R}_k\right|^2\simeq 2.4 \times 10^{-9} \,,
\end{equation}
and from eq. \eqref{eq:SHModes} the initial condition at horizon re-entry of the $k$ mode $\tilde\tau_{\rm HE}(k)$ is
\begin{equation}
\delta_{\text{m},k}^0 \equiv \delta_{\text{m},k} (\tilde\tau_{\rm HE}(k)) \simeq \left(3 \times 10^{-9}\right)^{1/2} \,.
\end{equation}
and the evolution of $\delta_{\text{m},k}$ after horizon re-entry is
\begin{equation}
\delta_{\text{m},k} (\tilde\tau) = \delta_{\text{m},k}^0 \left(\frac{\tilde\tau}{\tilde\tau_{\rm HE}(k)}\right)^2 \,.
\end{equation}
$\tilde\tau_{\rm HE}(k)$ can be computed by noting that
\begin{equation}
\label{eq:HE}
\frac{k}{a(\tilde\tau_{\rm HE}(k))} = H(\tilde\tau_{\rm HE}(k)) \,.
\end{equation}
Denoting by $\tilde\tau_{\rm mat}$ the conformal time at the start of matter domination (i.e.~when the scale factor starts evolving as $a(\tilde\tau) \simeq \tilde\tau^2$), from eq.~\eqref{eq:HE} and using that during matter domination $a(\tilde\tau) \propto \tilde\tau^2$ and that $H = a'/a^2$, we can write
\begin{equation}
\label{eq:TauHE}
\tilde\tau_{\rm HE}(k) = \frac{k_{\rm mat} \tilde\tau_{\rm mat}}{k} \,.
\end{equation}
For the example at hand we numerically find $\tilde\tau_{\rm mat} \simeq 6$ and $\log_{10} \left(k_{\rm mat}\right) = 0.34$, see Fig. \ref{fig:InterestingModes}. The typical size of the density constrast fluctuations on a comoving scale $k^{-1}$ can be written as the square root of the variance of the density perturbations
\begin{equation}
\label{eq:SpaceVariance}
\delta_{k^{-1}}^2 \simeq \langle \delta(\mathbf{x}) \delta(\mathbf{x}) \rangle = \int_0^k d \log q \, \frac{q^3}{2 \pi^2} \left|\delta_{\text{m},q}\right|^2 = \frac{\left(\delta_{\text{m},k}^0\right)^2}{2 \pi^2} \left(\frac{\tilde\tau}{\tilde\tau_{\rm mat}}\right)^4 \int_0^k \frac{dq}{k_{\rm mat}} \left(\frac{q}{k_{\rm mat}}\right)^3\,,
\end{equation}
from which the requirement that fluctuations go non-linear on a comoving scale $k^{-1}$ (with $k \ll k_{\rm mat}$) can be related to a requirement on $\tilde\tau/\tilde\tau_{\rm mat}$. For instance if we take as the upper limit of integration $k = k_{\rm mat}/10$, the integral gives a factor $0.25 \times 10^{-4}$ and the requirement on $\tilde\tau/\tilde\tau_{\rm mat}$ in order for $\delta^2_{k^{-1}}$ to go non-linear is
\begin{equation}
\Psi(\tilde\tau) = \left(\frac{\tilde\tau}{\tilde\tau_{\rm mat}}\right)^2 \gtrsim \left(\frac{8 \pi^2}{3} \times 10^{13}\right)^{1/2} \simeq 1.6 \times 10^7 \ll \Psi_{\rm max} \,.
\end{equation}
This analytical estimate takes into account modes $k \ll k_{\rm mat}$ and it is enough to show that in general it is quite hard to avoid some scales to become non-linear. Of course, modes $k \simeq k_{\rm mat}$ go non-linear much earlier. Assuming a flat spectrum of fluctuations as initial conditions $\Delta_{\mathcal{R}}^2(k) \simeq 2 \times 10^{-9}$, scales around $k_{\rm mat}$ (for which the result of the integral in eq. \eqref{eq:SpaceVariance} is of $\mathcal{O}\left(1\right)$) go non-linear as long as the mass of the gravitationally coupled modulus satisfies
\begin{equation}
\label{eq:EnhConstraint}
10^{-9} \left(\frac{M_{\rm P}}{m}\right)^{8/3} \gtrsim 1 \quad \Rightarrow \quad m \lesssim 4 \times 10^{-4} \, M_{\rm P} \,, \quad \Rightarrow \quad \V \gtrsim 10^3 \,,
\end{equation}
which is a rather weak constraint\footnote{The last implication of eq. \eqref{eq:EnhConstraint} is computed for the LVS volume modulus, for which $m \simeq M_{\rm P}/\V^{3/2}$.}. Once the fluctuations satisfy $\delta_{k^{-1}} \simeq 1$ on a given comoving scale $k^{-1}$, the linear analysis breaks down and a fully non-linear simulation that includes the effects of gravity is needed. The implication is rather strong: the existence of a scalar field like the volume modulus in the LVS always implies that non-linear physics has to be taken into account. A more detailed study is beyond the scope of this paper and left for future work. However, it is reasonable to expect that once fluctuations go non-linear the formation of compact structures (such as moduli stars or primordial black holes) could start~\cite{Jedamzik:2010dq}. The formation of primordial black holes through this mechanism could be particularly interesting for subsequent evolution of the Universe. Since they would be extremely light, they would evaporate very quickly providing for instance the initial conditions needed for instance for \textit{Hawking genesis}~\cite{Lennon:2017tqq}. Moreover, even if they do not form compact objects, such non-linearities could still give rise to a stochastic background of gravitational radiation, as pointed out in~\cite{Assadullahi:2009nf}.

\section{Conclusions}
\label{sec:Conclusions}

We have started a systematic study and  found the first concrete examples of boson stars in string compactifications. Both hidden and visible open string sectors essentially reproduce the previous field theoretical studies regarding Q-balls and boson stars, taking into account that the appearance of global symmetries is very restrictive in string theory. Closed string moduli offered a more stringy realisation of boson stars with particular properties that may be eventually identified and compared with observations. Concrete scenarios of moduli stabilisation such as the LVS or KKLT provide different examples of boson stars with different properties. Even though in the boson stars literature the case of boson stars with masses $M_{\rm P}^3/m^2$ are much studied we find that they are not naturally realised in the closed string sector for which only the original mini-boson stars of mass $M_{\rm P}^2/m$ are obtained. This is due to the fact that the particularly strong self-interactions are not obtained since all terms in the moduli interactions come from the string scale and coefficients are not many orders of magnitude larger than one, as it would be needed to get boson stars with masses $M_{\rm P}^3/m^2$. 

Axion stars are one of the most promising and generic outcomes of string compactifications. Different axionic particles provide different sources of axion stars. Typically axions obtain masses of the same order as their moduli partners and can give rise to light microscopic stars. However the large volume scenario predicts that there is at least one axion field with mass of order $m\sim e^{-\alpha\mathcal{V}^{2/3}}$ and therefore the corresponding stars have exponentially large masses $M\sim e^{\alpha \mathcal{V}^{2/3}}$ in Planck units. If the mass of this light axion is in the right ballpark ($1\text{-}10 \times 10^{-22} \,$ eV) it can be fuzzy dark matter. In this case, an axion star would form at the core of dwarf galaxies possibly addressing the cusp-core problem of cold dark matter. A similar situation occurs for any other string modulus which is stabilised by perturbative effects: its axionic partner would be the source of macroscopic stars.

Even though axion stars have been very much studied in the  literature independent of string theory, in string theory the corresponding scalar modulus field can also be a source for moduli stars. Since they are components of a complex scalar field together with an axion, in string theory both fields should be considered together. If they are both of similar masses they have to be considered together with the corresponding axion. In the small amplitude regime the two fields oscillate independently and their effect is simple to consider. But in other regimes the two-field system needs to be considered. This we will leave for a future study.

\begin{figure}
\begin{center}
\includegraphics[width=1\textwidth]{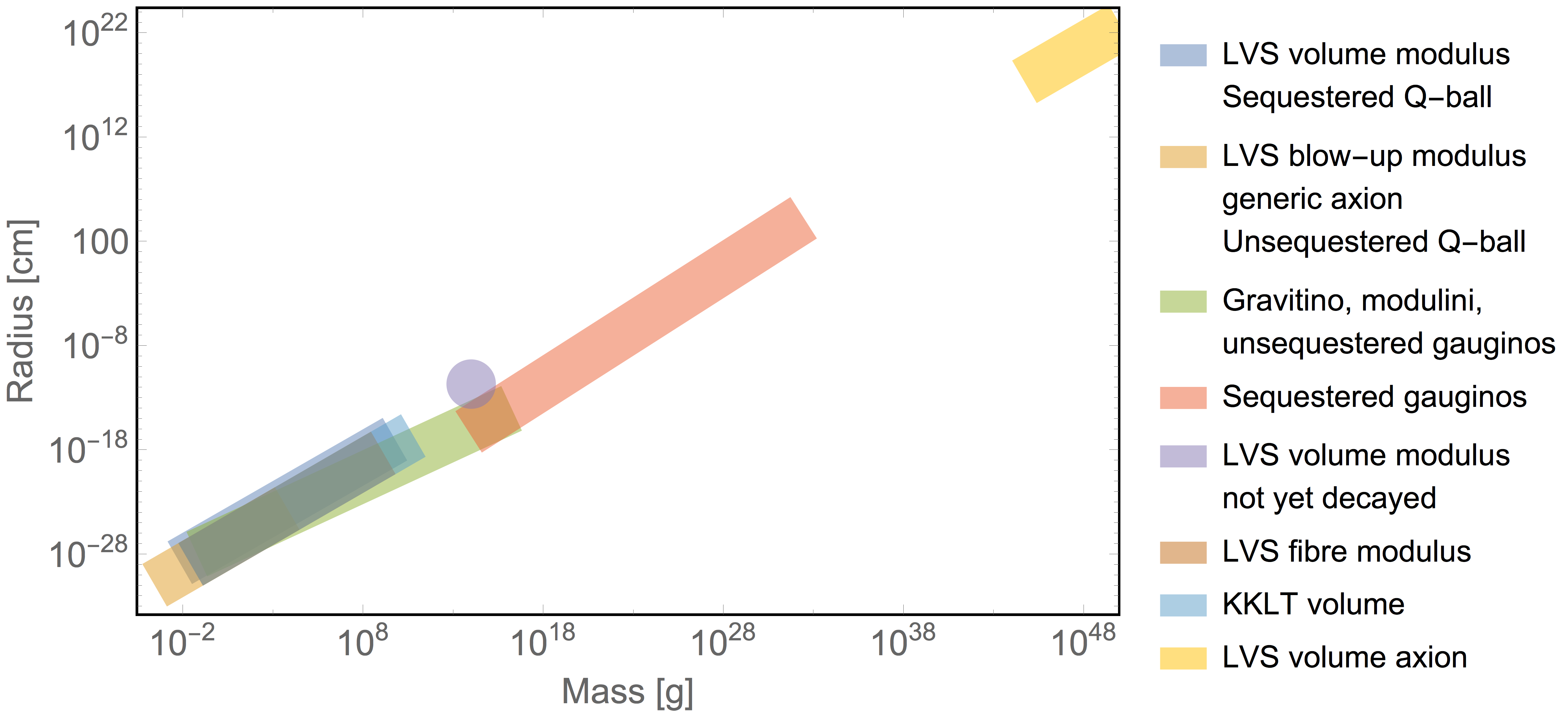}
\end{center}
\caption{Summary of expected mass ranges and radii for typically considered values of the overall volume and flux parameter.\label{fig:massvsradiussummary}}
\end{figure}

The moduli stars we considered vary if the corresponding field is the overall modulus, a blow-up mode or a fibre modulus. In each case the star mass is of order $M_{\rm P} \mathcal{V}^{n}$ with $n=1, 3/2, 5/3$ respectively. In the KKLT case the star mass is $M_{\rm P} |W_0|^{-1}\mathcal{V}.$ We determined the field profile and the mass-radius behaviour for each case in the dilute regime\footnote{The relation of the mass of the object with the inverse mass of the corresponding particle is reminiscent of the fact that in gauge theories elementary particles have masses proportional to couplings and that of the solitonic objects to powers of the inverse couplings. This played a role in uncovering strong-weak dualities in field and string theories. In our case the interaction is gravity and for moduli stars the relation corresponds to inverse powers of the volume as can be seen in Tab.~\ref{tab:summary}.}. A summary of the configurations obtained in this article can be seen in Table~\ref{tab:summary} and a summary of expected mass ranges and radii for typically considered values of the overall volume and flux parameter are shown in Figure~\ref{fig:massvsradiussummary}.
\begin{table}
\begin{center}
{\tabcolsep=5pt
\renewcommand{\arraystretch}{1.5}
\begin{tabular}{p{4cm} || c | c | c | c}
Particle & State mass & Star mass & Star radius & Enhancement\\ \hline \hline
LVS volume modulus & $M_{\rm P}/{{\cal V}^{3/2}}$& $M_{\rm P}{\cal V}^{3/2}$ &  $l_{\rm P} {\cal V}^{3/2}$& ${\cal V}^2$ \\ \hline
LVS blow-up modulus Generic axion & $M_{\rm P}/{{\cal V}}$ &$M_{\rm P}{\cal V}$ & $l_{\rm P} {\cal V}^{5/3}$ & ${\cal V}^{4/3}$\\ \hline
LVS fibre moduli & $M_{\rm P}/{{\cal V}^{5/3}}$  & $M_{\rm P}{\cal V}^{5/3}$&$l_{\rm P}{\cal V}^{5/3}$&${\cal V}^{20/9}$\\ \hline
LVS volume axion & $M_{\rm P} e^{-\alpha {\cal V}^{2/3}}$ & $M_{\rm P}e^{\alpha {\cal V}^{2/3}}$&    $l_{\rm P} e^{\alpha {\cal V}^{2/3}}$& $e^{4/3\alpha {\cal V}^{2/3}}$\\ \hline
KKLT volume modulus & $M_{\rm P}|W_0|/{\cal V}$& $M_{\rm P}|W_0|^{-1}{\cal V}$&$l_{\rm P}|W_0|^{-1}{\cal V}$&$(|W_0|^{-1}{\cal V})^{4/3}$\\ \hline
Gravitino, modulini, unsequestered gauginos & $M_{\rm P} |W_0|/{{\cal V}}$ & $M_{\rm P}{\cal V}^2/|W_0|^2$& $l_{\rm P}{\cal V}^2/|W_0|^2$& ${\cal V}^{4/3}/|W_0|^{4/3}$\\ \hline
Sequestered gauginos & $M_{\rm P}/{{\cal V}^2}$&$M_{\rm P}{\cal V}^4$&$l_{\rm P}{\cal V}^4$&${\cal V}^{8/3}$\\ \hline
Unsequestered Q-balls& $M_{\rm P}/{{\cal V}}$&$M_{\rm P}{\cal V}$&$l_{\rm P}{\cal V}$&${\cal V}^{4/3}$\\ \hline
Sequestered Q-balls & $M_{\rm P}/{{\cal V}^{3/2}}$&$M_{\rm P}{\cal V}^{3/2}$&$l_{\rm P}{\cal V}^{3/2}$&${\cal V}^{2}$
\end{tabular}}
\end{center}
\caption{Summary of the compact configurations with the scalings of their associated particle and star mass scalings, radii, and their respective enhancement factors. \label{tab:summary}}
\end{table}
Similarly studying the formation of these objects we pointed out that the first condition listed at the beginning of Section~\ref{sec:formationall} (i.e.~the existence of a localized overdensity) is always satisfied independently of the initial conditions and of the details of the model. The enhancement of the density perturbations takes place if there is an early matter dominated era due to an oscillating modulus. In the examined cases the enhancement is of order $\mathcal{V}^l$ with $l=4/3, 2, 20/9$ respectively, illustrating concrete physical differences between the different objects.

We have also studied the interesting limiting case in which an axion is so light that the PQ shift-symmetry can be considered exact for the study of the physics of the associated modulus field.
This case resembles the Q-ball case for a complex scalar. We call the corresponding configurations PQ-balls and started to study their properties. Even though they do not provide localised solitonic configurations in 4D they extend the solutions for a single modulus field to non-zero values of the angular velocity of the axionic component. This spinning axion configurations also generalises the spintessence proposal with potential cosmological implications which we hope to further study in the future.

Potential signatures of this study hints naturally at GWs. First, independently of the formation of boson stars, the substantial enhancement of density perturbations during an early matter dominated era (up to factors of order $10^{20}$) can lead to the production of a stochastic spectrum of GWs~\cite{Assadullahi:2009nf}. Furthermore, the possible formation of boson stars and even of primordial black holes could leave a distinctive signal in the GW spectrum. Other potential signatures have been studied in the past~\cite{0801.0307}.

We are only in the early stages of an ambitious programme to explore the formation, evolution and dynamics of the many inhomogeneities distinctive of string cosmology after inflation due to the rich structure of moduli fields. Their imprint in the GW spectrum may eventually provide important information about the structure and composition of the extra dimensions in string theory.

\section*{Acknowledgements}
We appreciate useful discussions with Ricardo Becerril, Diego Blas, Francesco Cefal\'a, Michele Cicoli, Katy Clough, Gary Gibbons, Nemanja Kaloper, Jin U Kang, Alexander Kusenko, Edward Hardy, Luis Lehner, David J. E. Marsh, David M. C. Marsh, Sonia Pab\'an, Ashoke Sen, David Tong, Paul Townsend, Luis A. Ure\~na-L\'opez, Cumrun Vafa, Roberto Valandro, Giovanni Villadoro, Luca Visinelli, David Wands. SK thanks ICTP for hospitality. SK's research is funded by ERC Advanced Grant ``Strings and Gravity'' (Grant No. 320040).

\bibliography{modulistars}

\providecommand{\href}[2]{#2}\begingroup\raggedright\begin{thebibliography}{100}

\bibitem{Abbott:2016blz}
{\scshape Virgo, LIGO Scientific} collaboration, B.~P. Abbott et~al.,
  \emph{{Observation of Gravitational Waves from a Binary Black Hole Merger}},
  \href{https://doi.org/10.1103/PhysRevLett.116.061102}{\emph{Phys. Rev. Lett.}
  {\bfseries 116} (2016) 061102},
  [\href{https://arxiv.org/abs/1602.03837}{{\ttfamily 1602.03837}}].

\bibitem{Abbott:2016nmj}
{\scshape Virgo, LIGO Scientific} collaboration, B.~P. Abbott et~al.,
  \emph{{GW151226: Observation of Gravitational Waves from a 22-Solar-Mass
  Binary Black Hole Coalescence}},
  \href{https://doi.org/10.1103/PhysRevLett.116.241103}{\emph{Phys. Rev. Lett.}
  {\bfseries 116} (2016) 241103},
  [\href{https://arxiv.org/abs/1606.04855}{{\ttfamily 1606.04855}}].

\bibitem{Abbott:2017vtc}
{\scshape VIRGO, LIGO Scientific} collaboration, B.~P. Abbott et~al.,
  \emph{{GW170104: Observation of a 50-Solar-Mass Binary Black Hole Coalescence
  at Redshift 0.2}},
  \href{https://doi.org/10.1103/PhysRevLett.118.221101}{\emph{Phys. Rev. Lett.}
  {\bfseries 118} (2017) 221101},
  [\href{https://arxiv.org/abs/1706.01812}{{\ttfamily 1706.01812}}].

\bibitem{Abbott:2017gyy}
{\scshape Virgo, LIGO Scientific} collaboration, B.~P. Abbott et~al.,
  \emph{{GW170608: Observation of a 19-solar-mass Binary Black Hole
  Coalescence}},
  \href{https://doi.org/10.3847/2041-8213/aa9f0c}{\emph{Astrophys. J.}
  {\bfseries 851} (2017) L35},
  [\href{https://arxiv.org/abs/1711.05578}{{\ttfamily 1711.05578}}].

\bibitem{Abbott:2017oio}
{\scshape Virgo, LIGO Scientific} collaboration, B.~P. Abbott et~al.,
  \emph{{GW170814: A Three-Detector Observation of Gravitational Waves from a
  Binary Black Hole Coalescence}},
  \href{https://doi.org/10.1103/PhysRevLett.119.141101}{\emph{Phys. Rev. Lett.}
  {\bfseries 119} (2017) 141101},
  [\href{https://arxiv.org/abs/1709.09660}{{\ttfamily 1709.09660}}].

\bibitem{TheLIGOScientific:2017qsa}
{\scshape Virgo, LIGO Scientific} collaboration, B.~Abbott et~al.,
  \emph{{GW170817: Observation of Gravitational Waves from a Binary Neutron
  Star Inspiral}},
  \href{https://doi.org/10.1103/PhysRevLett.119.161101}{\emph{Phys. Rev. Lett.}
  {\bfseries 119} (2017) 161101},
  [\href{https://arxiv.org/abs/1710.05832}{{\ttfamily 1710.05832}}].

\bibitem{gaia}
{Gaia Collaboration}, \emph{{The Gaia mission}}, {\emph{aap} },
  [\href{https://arxiv.org/abs/1609.04153}{{\ttfamily 1609.04153}}].

\bibitem{Jetzer:1991jr}
P.~Jetzer, \emph{{Boson stars}},
  \href{https://doi.org/10.1016/0370-1573(92)90123-H}{\emph{Phys. Rept.}
  {\bfseries 220} (1992) 163--227}.

\bibitem{Liddle:1993ha}
A.~R. Liddle and M.~S. Madsen, \emph{{The Structure and formation of boson
  stars}}, \href{https://doi.org/10.1142/S0218271892000057}{\emph{Int. J. Mod.
  Phys.} {\bfseries D1} (1992) 101--144}.

\bibitem{Schunck:2003kk}
F.~E. Schunck and E.~W. Mielke, \emph{{General relativistic boson stars}},
  \href{https://doi.org/10.1088/0264-9381/20/20/201}{\emph{Class. Quant. Grav.}
  {\bfseries 20} (2003) R301--R356},
  [\href{https://arxiv.org/abs/0801.0307}{{\ttfamily 0801.0307}}].

\bibitem{1202.5809}
S.~L. Liebling and C.~Palenzuela, \emph{{Dynamical Boson Stars}},
  \href{https://doi.org/10.12942/lrr-2012-6,
  10.1007/s41114-017-0007-y}{\emph{Living Rev. Rel.} {\bfseries 15} (2012) 6},
  [\href{https://arxiv.org/abs/1202.5809}{{\ttfamily 1202.5809}}].

\bibitem{Narain:2006kx}
  G.~Narain, J.~Schaffner-Bielich and I.~N.~Mishustin,
  ``Compact stars made of fermionic dark matter,''
  Phys.\ Rev.\ D {\bf 74} (2006) 063003
  [astro-ph/0605724].

\bibitem{deCarlos:1993wie}
B.~de~Carlos, J.~A. Casas, F.~Quevedo and E.~Roulet, \emph{{Model independent
  properties and cosmological implications of the dilaton and moduli sectors of
  4-d strings}},
  \href{https://doi.org/10.1016/0370-2693(93)91538-X}{\emph{Phys. Lett.}
  {\bfseries B318} (1993) 447--456},
  [\href{https://arxiv.org/abs/hep-ph/9308325}{{\ttfamily hep-ph/9308325}}].

\bibitem{GomezReino:2006dk}
M.~Gomez-Reino and C.~A.~Scrucca,
``Locally stable non-supersymmetric Minkowski vacua in supergravity,''
JHEP {\bf 0605} (2006) 015
[hep-th/0602246].

\bibitem{Coughlan:1983ci}
G.~D. Coughlan, W.~Fischler, E.~W. Kolb, S.~Raby and G.~G. Ross,
  \emph{{Cosmological Problems for the Polonyi Potential}},
  \href{https://doi.org/10.1016/0370-2693(83)91091-2}{\emph{Phys. Lett.}
  {\bfseries 131B} (1983) 59--64}.

\bibitem{Banks:1993en}
T.~Banks, D.~B. Kaplan and A.~E. Nelson, \emph{{Cosmological implications of
  dynamical supersymmetry breaking}},
  \href{https://doi.org/10.1103/PhysRevD.49.779}{\emph{Phys. Rev.} {\bfseries
  D49} (1994) 779--787},
  [\href{https://arxiv.org/abs/hep-ph/9308292}{{\ttfamily hep-ph/9308292}}].

\bibitem{Vafa:2005ui}
C.~Vafa, \emph{{The String landscape and the swampland}},
  \href{https://arxiv.org/abs/hep-th/0509212}{{\ttfamily hep-th/0509212}}.

\bibitem{Kaup:1968zz}
D.~J. Kaup, \emph{{Klein-Gordon Geon}},
  \href{https://doi.org/10.1103/PhysRev.172.1331}{\emph{Phys. Rev.} {\bfseries
  172} (1968) 1331--1342}.

\bibitem{Ruffini:1969qy}
R.~Ruffini and S.~Bonazzola, \emph{{Systems of selfgravitating particles in
  general relativity and the concept of an equation of state}},
  \href{https://doi.org/10.1103/PhysRev.187.1767}{\emph{Phys. Rev.} {\bfseries
  187} (1969) 1767--1783}.

\bibitem{Colpi:1986ye}
M.~Colpi, S.~L. Shapiro and I.~Wasserman, \emph{{Boson Stars: Gravitational
  Equilibria of Selfinteracting Scalar Fields}},
  \href{https://doi.org/10.1103/PhysRevLett.57.2485}{\emph{Phys. Rev. Lett.}
  {\bfseries 57} (1986) 2485--2488}.

\bibitem{Coleman:1985ki}
S.~R. Coleman, \emph{{Q-Balls}},
  \href{https://doi.org/10.1016/0550-3213(85)90286-X,
  10.1016/0550-3213(86)90520-1}{\emph{Nucl. Phys.} {\bfseries B262} (1985)
  263}.

\bibitem{Halverson:2018xge}
  J.~Halverson and P.~Langacker,
  ``TASI Lectures on Remnants from the String Landscape,''
  arXiv:1801.03503 [hep-th].

\bibitem{Lee:1991ax}
T.~D. Lee and Y.~Pang, \emph{{Nontopological solitons}},
  \href{https://doi.org/10.1016/0370-1573(92)90064-7}{\emph{Phys. Rept.}
  {\bfseries 221} (1992) 251--350}.

\bibitem{Rosen}
G.~Rosen, \emph{Charged particlelike solutions to nonlinear complex scalar
  field theories}, \href{https://doi.org/10.1063/1.1664694}{\emph{Journal of
  Mathematical Physics} {\bfseries 9} (1968) 999--1002},
  [\href{https://arxiv.org/abs/https://doi.org/10.1063/1.1664694}{{\ttfamily
  https://doi.org/10.1063/1.1664694}}].

\bibitem{Lee:1988ag}
K.-M. Lee, J.~A. Stein-Schabes, R.~Watkins and L.~M. Widrow, \emph{{Gauged Q-Balls}}, \href{https://doi.org/10.1103/PhysRevD.39.1665}{\emph{Phys. Rev.}
  {\bfseries D39} (1989) 1665}.

\bibitem{Kusenko:1997vi}
A.~Kusenko, M.~E. Shaposhnikov and P.~G. Tinyakov, \emph{{Sufficient conditions
  for the existence of Q-balls in gauge theories}},
  \href{https://doi.org/10.1134/1.567658}{\emph{Pisma Zh. Eksp. Teor. Fiz.}
  {\bfseries 67} (1998) 229},
  [\href{https://arxiv.org/abs/hep-th/9801041}{{\ttfamily hep-th/9801041}}].

\bibitem{Gleiser:1993pt}
M.~Gleiser, \emph{{Pseudostable bubbles}},
  \href{https://doi.org/10.1103/PhysRevD.49.2978}{\emph{Phys. Rev.} {\bfseries
  D49} (1994) 2978--2981},
  [\href{https://arxiv.org/abs/hep-ph/9308279}{{\ttfamily hep-ph/9308279}}].

\bibitem{Antusch:2017flz}
S.~Antusch, F.~Cefala, S.~Krippendorf, F.~Muia, S.~Orani and F.~Quevedo,
  \emph{{Oscillons from String Moduli}},
  \href{https://doi.org/10.1007/JHEP01(2018)083}{\emph{JHEP} {\bfseries 01}
  (2018) 083}, [\href{https://arxiv.org/abs/1708.08922}{{\ttfamily
  1708.08922}}].

\bibitem{Cohen:1986ct}
A.~G. Cohen, S.~R. Coleman, H.~Georgi and A.~Manohar, \emph{{The Evaporation of
  $Q$ Balls}}, \href{https://doi.org/10.1016/0550-3213(86)90004-0}{\emph{Nucl.
  Phys.} {\bfseries B272} (1986) 301--321}.

\bibitem{Mukaida:2016hwd}
K.~Mukaida, M.~Takimoto and M.~Yamada, \emph{{On Longevity of
  I-ball/Oscillon}}, \href{https://doi.org/10.1007/JHEP03(2017)122}{\emph{JHEP}
  {\bfseries 03} (2017) 122},
  [\href{https://arxiv.org/abs/1612.07750}{{\ttfamily 1612.07750}}].

\bibitem{Segur:1987mg}
H.~Segur and M.~D. Kruskal, \emph{{Nonexistence of Small Amplitude Breather
  Solutions in $\phi^4$ Theory}},
  \href{https://doi.org/10.1103/PhysRevLett.58.747}{\emph{Phys. Rev. Lett.}
  {\bfseries 58} (1987) 747--750}.

\bibitem{Fodor:2009kf}
G.~Fodor, P.~Forgacs, Z.~Horvath and M.~Mezei, \emph{{Radiation of scalar
  oscillons in 2 and 3 dimensions}},
  \href{https://doi.org/10.1016/j.physletb.2009.03.054}{\emph{Phys. Lett.}
  {\bfseries B674} (2009) 319--324},
  [\href{https://arxiv.org/abs/0903.0953}{{\ttfamily 0903.0953}}].

\bibitem{Hertzberg:2010yz}
M.~P. Hertzberg, \emph{{Quantum Radiation of Oscillons}},
  \href{https://doi.org/10.1103/PhysRevD.82.045022}{\emph{Phys. Rev.}
  {\bfseries D82} (2010) 045022},
  [\href{https://arxiv.org/abs/1003.3459}{{\ttfamily 1003.3459}}].

\bibitem{Zhou:2013tsa}
S.-Y. Zhou, E.~J. Copeland, R.~Easther, H.~Finkel, Z.-G. Mou and P.~M. Saffin,
  \emph{{Gravitational Waves from Oscillon Preheating}},
  \href{https://doi.org/10.1007/JHEP10(2013)026}{\emph{JHEP} {\bfseries 10}
  (2013) 026}, [\href{https://arxiv.org/abs/1304.6094}{{\ttfamily 1304.6094}}].

\bibitem{Antusch:2016con}
S.~Antusch, F.~Cefala and S.~Orani, \emph{{Gravitational waves from oscillons
  after inflation}},
  \href{https://doi.org/10.1103/PhysRevLett.118.011303}{\emph{Phys. Rev. Lett.}
  {\bfseries 118} (2017) 011303},
  [\href{https://arxiv.org/abs/1607.01314}{{\ttfamily 1607.01314}}].

\bibitem{Amin:2018xfe}
M.~A. Amin, J.~Braden, E.~J. Copeland, J.~T. Giblin, C.~Solorio, Z.~J. Weiner
  et~al., \emph{{Gravitational waves from asymmetric oscillon dynamics?}},
  \href{https://arxiv.org/abs/1803.08047}{{\ttfamily 1803.08047}}.

\bibitem{Riotto:1999yt}
A.~Riotto and M.~Trodden, \emph{{Recent progress in baryogenesis}},
  \href{https://doi.org/10.1146/annurev.nucl.49.1.35}{\emph{Ann. Rev. Nucl.
  Part. Sci.} {\bfseries 49} (1999) 35--75},
  [\href{https://arxiv.org/abs/hep-ph/9901362}{{\ttfamily hep-ph/9901362}}].

\bibitem{Dine:2003ax}
M.~Dine and A.~Kusenko, \emph{{The Origin of the matter - antimatter
  asymmetry}}, \href{https://doi.org/10.1103/RevModPhys.76.1}{\emph{Rev. Mod.
  Phys.} {\bfseries 76} (2003) 1},
  [\href{https://arxiv.org/abs/hep-ph/0303065}{{\ttfamily hep-ph/0303065}}].

\bibitem{Buchmuller:2005eh}
W.~Buchmuller, R.~D. Peccei and T.~Yanagida, \emph{{Leptogenesis as the origin
  of matter}},
  \href{https://doi.org/10.1146/annurev.nucl.55.090704.151558}{\emph{Ann. Rev.
  Nucl. Part. Sci.} {\bfseries 55} (2005) 311--355},
  [\href{https://arxiv.org/abs/hep-ph/0502169}{{\ttfamily hep-ph/0502169}}].

\bibitem{Lozanov:2014zfa}
K.~D. Lozanov and M.~A. Amin, \emph{{End of inflation, oscillons, and
  matter-antimatter asymmetry}},
  \href{https://doi.org/10.1103/PhysRevD.90.083528}{\emph{Phys. Rev.}
  {\bfseries D90} (2014) 083528},
  [\href{https://arxiv.org/abs/1408.1811}{{\ttfamily 1408.1811}}].

\bibitem{Kusenko:1997si}
A.~Kusenko and M.~E. Shaposhnikov, \emph{{Supersymmetric Q-balls as dark
  matter}}, \href{https://doi.org/10.1016/S0370-2693(97)01375-0}{\emph{Phys.
  Lett.} {\bfseries B418} (1998) 46--54},
  [\href{https://arxiv.org/abs/hep-ph/9709492}{{\ttfamily hep-ph/9709492}}].

\bibitem{Burgess:2000yq}
C.~P. Burgess, M.~Pospelov and T.~ter Veldhuis, \emph{{The Minimal model of
  nonbaryonic dark matter: A Singlet scalar}},
  \href{https://doi.org/10.1016/S0550-3213(01)00513-2}{\emph{Nucl. Phys.}
  {\bfseries B619} (2001) 709--728},
  [\href{https://arxiv.org/abs/hep-ph/0011335}{{\ttfamily hep-ph/0011335}}].

\bibitem{Seidel:1991zh}
E.~Seidel and W.~M. Suen, \emph{{Oscillating soliton stars}},
  \href{https://doi.org/10.1103/PhysRevLett.66.1659}{\emph{Phys. Rev. Lett.}
  {\bfseries 66} (1991) 1659--1662}.

\bibitem{Kolb:1993zz}
E.~W. Kolb and I.~I. Tkachev, \emph{{Axion miniclusters and Bose stars}},
  \href{https://doi.org/10.1103/PhysRevLett.71.3051}{\emph{Phys. Rev. Lett.}
  {\bfseries 71} (1993) 3051--3054},
  [\href{https://arxiv.org/abs/hep-ph/9303313}{{\ttfamily hep-ph/9303313}}].

\bibitem{Kolb:1993hw}
E.~W. Kolb and I.~I. Tkachev, \emph{{Nonlinear axion dynamics and formation of
  cosmological pseudosolitons}},
  \href{https://doi.org/10.1103/PhysRevD.49.5040}{\emph{Phys. Rev.} {\bfseries
  D49} (1994) 5040--5051},
  [\href{https://arxiv.org/abs/astro-ph/9311037}{{\ttfamily
  astro-ph/9311037}}].

\bibitem{Hogan:1988mp}
C.~J. Hogan and M.~J. Rees, \emph{{AXION MINICLUSTERS}},
  \href{https://doi.org/10.1016/0370-2693(88)91655-3}{\emph{Phys. Lett.}
  {\bfseries B205} (1988) 228--230}.

\bibitem{Enander:2017ogx}
  J.~Enander, A.~Pargner and T.~Schwetz,
  ``Axion minicluster power spectrum and mass function,''
  JCAP {\bf 1712} (2017) no.12,  038
  [arXiv:1708.04466 [astro-ph.CO]].

\bibitem{UrenaLopez:2001tw}
L.~A. Urena-Lopez, \emph{{Oscillatons revisited}},
  \href{https://doi.org/10.1088/0264-9381/19/10/307}{\emph{Class. Quant. Grav.}
  {\bfseries 19} (2002) 2617--2632},
  [\href{https://arxiv.org/abs/gr-qc/0104093}{{\ttfamily gr-qc/0104093}}].

\bibitem{UrenaLopez:2002gx}
L.~A. Urena-Lopez, T.~Matos and R.~Becerril, \emph{{Inside oscillatons}},
  \href{https://doi.org/10.1088/0264-9381/19/23/320}{\emph{Class. Quant. Grav.}
  {\bfseries 19} (2002) 6259--6277}.

\bibitem{UrenaLopez:2012zz}
L.~A. Urena-Lopez, S.~Valdez-Alvarado and R.~Becerril, \emph{{Evolution and
  stability $\phi^4$ oscillatons}},
  \href{https://doi.org/10.1088/0264-9381/29/6/065021}{\emph{Class. Quant.
  Grav.} {\bfseries 29} (2012) 065021}.

\bibitem{Alcubierre:2003sx}
M.~Alcubierre, R.~Becerril, S.~F. Guzman, T.~Matos, D.~Nunez and L.~A.
  Urena-Lopez, \emph{{Numerical studies of $\Phi^2$ oscillatons}},
  \href{https://doi.org/10.1088/0264-9381/20/13/332}{\emph{Class. Quant. Grav.}
  {\bfseries 20} (2003) 2883--2904},
  [\href{https://arxiv.org/abs/gr-qc/0301105}{{\ttfamily gr-qc/0301105}}].

\bibitem{Guzman:2004wj}
F.~S. Guzman and L.~A. Urena-Lopez, \emph{{Evolution of the Schrodinger-Newton
  system for a selfgravitating scalar field}},
  \href{https://doi.org/10.1103/PhysRevD.69.124033}{\emph{Phys. Rev.}
  {\bfseries D69} (2004) 124033},
  [\href{https://arxiv.org/abs/gr-qc/0404014}{{\ttfamily gr-qc/0404014}}].

\bibitem{Giudice:2016zpa}
G.~F. Giudice, M.~McCullough and A.~Urbano, \emph{{Hunting for Dark Particles
  with Gravitational Waves}},
  \href{https://doi.org/10.1088/1475-7516/2016/10/001}{\emph{JCAP} {\bfseries
  1610} (2016) 001}, [\href{https://arxiv.org/abs/1605.01209}{{\ttfamily
  1605.01209}}].

\bibitem{Visinelli:2017ooc}
L.~Visinelli, S.~Baum, J.~Redondo, K.~Freese and F.~Wilczek, \emph{{Dilute and
  dense axion stars}},
  \href{https://doi.org/10.1016/j.physletb.2017.12.010}{\emph{Phys. Lett.}
  {\bfseries B777} (2018) 64--72},
  [\href{https://arxiv.org/abs/1710.08910}{{\ttfamily 1710.08910}}].

\bibitem{Guth:2014hsa}
  A.~H.~Guth, M.~P.~Hertzberg and C.~Prescod-Weinstein,
  ``Do Dark Matter Axions Form a Condensate with Long-Range Correlation?,''
  Phys.\ Rev.\ D {\bf 92} (2015) no.10,  103513
  [arXiv:1412.5930 [astro-ph.CO]].

\bibitem{ValdezAlvarado:2011dd}
S.~Valdez-Alvarado, L.~A. Urena-Lopez and R.~Becerril, \emph{{$\Phi^{4}$
  Oscillatons}},  \href{https://arxiv.org/abs/1107.3135}{{\ttfamily
  1107.3135}}.

\bibitem{Chavanis:2016dab}
P.-H. Chavanis, \emph{{Collapse of a self-gravitating Bose-Einstein condensate
  with attractive self-interaction}},
  \href{https://doi.org/10.1103/PhysRevD.94.083007}{\emph{Phys. Rev.}
  {\bfseries D94} (2016) 083007},
  [\href{https://arxiv.org/abs/1604.05904}{{\ttfamily 1604.05904}}].

\bibitem{Helfer:2016ljl}
T.~Helfer, D.~J.~E. Marsh, K.~Clough, M.~Fairbairn, E.~A. Lim and R.~Becerril,
  \emph{{Black hole formation from axion stars}},
  \href{https://doi.org/10.1088/1475-7516/2017/03/055}{\emph{JCAP} {\bfseries
  1703} (2017) 055}, [\href{https://arxiv.org/abs/1609.04724}{{\ttfamily
  1609.04724}}].

\bibitem{Levkov:2016rkk}
D.~G. Levkov, A.~G. Panin and I.~I. Tkachev, \emph{{Relativistic axions from
  collapsing Bose stars}},
  \href{https://doi.org/10.1103/PhysRevLett.118.011301}{\emph{Phys. Rev. Lett.}
  {\bfseries 118} (2017) 011301},
  [\href{https://arxiv.org/abs/1609.03611}{{\ttfamily 1609.03611}}].

\bibitem{Schiappacasse:2017ham}
E.~D. Schiappacasse and M.~P. Hertzberg, \emph{{Analysis of Dark Matter Axion
  Clumps with Spherical Symmetry}},
  \href{https://doi.org/10.1088/1475-7516/2018/03/E01,
  10.1088/1475-7516/2018/01/037}{\emph{JCAP} {\bfseries 1801} (2018) 037},
  [\href{https://arxiv.org/abs/1710.04729}{{\ttfamily 1710.04729}}].

\bibitem{Braaten:2015eeu}
E.~Braaten, A.~Mohapatra and H.~Zhang, \emph{{Dense Axion Stars}},
  \href{https://doi.org/10.1103/PhysRevLett.117.121801}{\emph{Phys. Rev. Lett.}
  {\bfseries 117} (2016) 121801},
  [\href{https://arxiv.org/abs/1512.00108}{{\ttfamily 1512.00108}}].

\bibitem{Hu:2000ke}
W.~Hu, R.~Barkana and A.~Gruzinov, \emph{{Cold and fuzzy dark matter}},
  \href{https://doi.org/10.1103/PhysRevLett.85.1158}{\emph{Phys. Rev. Lett.}
  {\bfseries 85} (2000) 1158--1161},
  [\href{https://arxiv.org/abs/astro-ph/0003365}{{\ttfamily
  astro-ph/0003365}}].

\bibitem{Hui:2016ltb}
L.~Hui, J.~P. Ostriker, S.~Tremaine and E.~Witten, \emph{{Ultralight scalars as
  cosmological dark matter}},
  \href{https://doi.org/10.1103/PhysRevD.95.043541}{\emph{Phys. Rev.}
  {\bfseries D95} (2017) 043541},
  [\href{https://arxiv.org/abs/1610.08297}{{\ttfamily 1610.08297}}].

\bibitem{Viel:2013apy}
M.~Viel, G.~D. Becker, J.~S. Bolton and M.~G. Haehnelt, \emph{{Warm dark matter
  as a solution to the small scale crisis: New constraints from high redshift
  Lyman-$\alpha$ forest data}},
  \href{https://doi.org/10.1103/PhysRevD.88.043502}{\emph{Phys. Rev.}
  {\bfseries D88} (2013) 043502},
  [\href{https://arxiv.org/abs/1306.2314}{{\ttfamily 1306.2314}}].

\bibitem{Schive:2014dra}
H.-Y. Schive, T.~Chiueh and T.~Broadhurst, \emph{{Cosmic Structure as the
  Quantum Interference of a Coherent Dark Wave}},
  \href{https://doi.org/10.1038/nphys2996}{\emph{Nature Phys.} {\bfseries 10}
  (2014) 496--499}, [\href{https://arxiv.org/abs/1406.6586}{{\ttfamily
  1406.6586}}].

\bibitem{Veltmaat:2018dfz}
J.~Veltmaat, J.~C. Niemeyer and B.~Schwabe, \emph{{Formation and structure of
  ultralight bosonic dark matter halos}},
  \href{https://arxiv.org/abs/1804.09647}{{\ttfamily 1804.09647}}.

\bibitem{Levkov:2018kau}
D.~G. Levkov, A.~G. Panin and I.~I. Tkachev, \emph{{Bose Condensation by
  Gravitational Interactions}},
  \href{https://arxiv.org/abs/1804.05857}{{\ttfamily 1804.05857}}.

\bibitem{Marsh:2015wka}
D.~J.~E. Marsh and A.-R. Pop, \emph{{Axion dark matter, solitons and the
  cusp–core problem}},
  \href{https://doi.org/10.1093/mnras/stv1050}{\emph{Mon. Not. Roy. Astron.
  Soc.} {\bfseries 451} (2015) 2479--2492},
  [\href{https://arxiv.org/abs/1502.03456}{{\ttfamily 1502.03456}}].

\bibitem{Kusenko:1997ad}
A.~Kusenko, \emph{{Small Q-balls}},
  \href{https://doi.org/10.1016/S0370-2693(97)00582-0}{\emph{Phys. Lett.}
  {\bfseries B404} (1997) 285},
  [\href{https://arxiv.org/abs/hep-th/9704073}{{\ttfamily hep-th/9704073}}].

\bibitem{Conlon:2005ki}
J.~P. Conlon, F.~Quevedo and K.~Suruliz, \emph{{Large-volume flux
  compactifications: Moduli spectrum and D3/D7 soft supersymmetry breaking}},
  \href{https://doi.org/10.1088/1126-6708/2005/08/007}{\emph{JHEP} {\bfseries
  08} (2005) 007}, [\href{https://arxiv.org/abs/hep-th/0505076}{{\ttfamily
  hep-th/0505076}}].

\bibitem{Balasubramanian:2005zx}
V.~Balasubramanian, P.~Berglund, J.~P. Conlon and F.~Quevedo,
  \emph{{Systematics of moduli stabilisation in Calabi-Yau flux
  compactifications}},
  \href{https://doi.org/10.1088/1126-6708/2005/03/007}{\emph{JHEP} {\bfseries
  03} (2005) 007}, [\href{https://arxiv.org/abs/hep-th/0502058}{{\ttfamily
  hep-th/0502058}}].

\bibitem{Blumenhagen:2009gk}
  R.~Blumenhagen, J.~P.~Conlon, S.~Krippendorf, S.~Moster and F.~Quevedo,
  ``SUSY Breaking in Local String/F-Theory Models,''
  JHEP {\bf 0909} (2009) 007
  [arXiv:0906.3297 [hep-th]].

\bibitem{Cicoli:2008va}
M.~Cicoli, J.~P. Conlon and F.~Quevedo, \emph{{General Analysis of LARGE Volume
  Scenarios with String Loop Moduli Stabilisation}},
  \href{https://doi.org/10.1088/1126-6708/2008/10/105}{\emph{JHEP} {\bfseries
  10} (2008) 105}, [\href{https://arxiv.org/abs/0805.1029}{{\ttfamily
  0805.1029}}].

\bibitem{Peccei:1977hh}
R.~D. Peccei and H.~R. Quinn, \emph{{CP Conservation in the Presence of
  Instantons}}, \href{https://doi.org/10.1103/PhysRevLett.38.1440}{\emph{Phys.
  Rev. Lett.} {\bfseries 38} (1977) 1440--1443}.

\bibitem{Wilczek:1977pj}
F.~Wilczek, \emph{{Problem of Strong p and t Invariance in the Presence of
  Instantons}}, \href{https://doi.org/10.1103/PhysRevLett.40.279}{\emph{Phys.
  Rev. Lett.} {\bfseries 40} (1978) 279--282}.

\bibitem{Weinberg:1977ma}
S.~Weinberg, \emph{{A New Light Boson?}},
  \href{https://doi.org/10.1103/PhysRevLett.40.223}{\emph{Phys. Rev. Lett.}
  {\bfseries 40} (1978) 223--226}.

\bibitem{Conlon:2006tq}
J.~P. Conlon, \emph{{The QCD axion and moduli stabilisation}},
  \href{https://doi.org/10.1088/1126-6708/2006/05/078}{\emph{JHEP} {\bfseries
  05} (2006) 078}, [\href{https://arxiv.org/abs/hep-th/0602233}{{\ttfamily
  hep-th/0602233}}].

\bibitem{Svrcek:2006yi}
P.~Svrcek and E.~Witten, \emph{{Axions In String Theory}},
  \href{https://doi.org/10.1088/1126-6708/2006/06/051}{\emph{JHEP} {\bfseries
  06} (2006) 051}, [\href{https://arxiv.org/abs/hep-th/0605206}{{\ttfamily
  hep-th/0605206}}].

\bibitem{Arvanitaki:2009fg}
A.~Arvanitaki, S.~Dimopoulos, S.~Dubovsky, N.~Kaloper and J.~March-Russell,
  \emph{{String Axiverse}},
  \href{https://doi.org/10.1103/PhysRevD.81.123530}{\emph{Phys. Rev.}
  {\bfseries D81} (2010) 123530},
  [\href{https://arxiv.org/abs/0905.4720}{{\ttfamily 0905.4720}}].

\bibitem{Cicoli:2012sz}
M.~Cicoli, M.~Goodsell and A.~Ringwald, \emph{{The type IIB string axiverse and
  its low-energy phenomenology}},
  \href{https://doi.org/10.1007/JHEP10(2012)146}{\emph{JHEP} {\bfseries 10}
  (2012) 146}, [\href{https://arxiv.org/abs/1206.0819}{{\ttfamily 1206.0819}}].

\bibitem{Cicoli:2012aq}
  M.~Cicoli, J.~P.~Conlon and F.~Quevedo,
  ``Dark radiation in LARGE volume models,''
  Phys.\ Rev.\ D {\bf 87} (2013) no.4,  043520
  [arXiv:1208.3562 [hep-ph]].

\bibitem{Higaki:2012ar}
  T.~Higaki and F.~Takahashi,
  ``Dark Radiation and Dark Matter in Large Volume Compactifications,''
  JHEP {\bf 1211} (2012) 125
  [arXiv:1208.3563 [hep-ph]].

\bibitem{Cicoli:2015bpq}
  M.~Cicoli and F.~Muia,
  ``General Analysis of Dark Radiation in Sequestered String Models,''
  JHEP {\bf 1512} (2015) 152
  [arXiv:1511.05447 [hep-th]].

\bibitem{Hardy:2016mns}
E.~Hardy, \emph{{Miniclusters in the Axiverse}},
  \href{https://doi.org/10.1007/JHEP02(2017)046}{\emph{JHEP} {\bfseries 02}
  (2017) 046}, [\href{https://arxiv.org/abs/1609.00208}{{\ttfamily
  1609.00208}}].

\bibitem{Feng:2008mu}
J.~L. Feng, H.~Tu and H.-B. Yu, \emph{{Thermal Relics in Hidden Sectors}},
  \href{https://doi.org/10.1088/1475-7516/2008/10/043}{\emph{JCAP} {\bfseries
  0810} (2008) 043}, [\href{https://arxiv.org/abs/0808.2318}{{\ttfamily
  0808.2318}}].

\bibitem{Burgess:2016owb}
  C.~P.~Burgess, M.~Cicoli, S.~de Alwis and F.~Quevedo,
  ``Robust Inflation from Fibrous Strings,''
  JCAP {\bf 1605} (2016) no.05,  032
  [arXiv:1603.06789 [hep-th]].

\bibitem{Cicoli:2016olq}
M.~Cicoli, K.~Dutta, A.~Maharana and F.~Quevedo, \emph{{Moduli Vacuum
  Misalignment and Precise Predictions in String Inflation}},
  \href{https://doi.org/10.1088/1475-7516/2016/08/006}{\emph{JCAP} {\bfseries
  1608} (2016) 006}, [\href{https://arxiv.org/abs/1604.08512}{{\ttfamily
  1604.08512}}].

\bibitem{Kachru:2003aw}
S.~Kachru, R.~Kallosh, A.~D. Linde and S.~P. Trivedi, \emph{{De Sitter vacua in
  string theory}},
  \href{https://doi.org/10.1103/PhysRevD.68.046005}{\emph{Phys. Rev.}
  {\bfseries D68} (2003) 046005},
  [\href{https://arxiv.org/abs/hep-th/0301240}{{\ttfamily hep-th/0301240}}].

\bibitem{Clough:2015sqa}
  K.~Clough, P.~Figueras, H.~Finkel, M.~Kunesch, E.~A.~Lim and S.~Tunyasuvunakool,
  ``GRChombo : Numerical Relativity with Adaptive Mesh Refinement,''
  Class.\ Quant.\ Grav.\  {\bf 32} (2015) no.24,  245011
   [Class.\ Quant.\ Grav.\  {\bf 32} (2015) 24]
  [arXiv:1503.03436 [gr-qc]].

\bibitem{Maggiore:1900zz}
M.~Maggiore, \emph{{Gravitational Waves. Vol. 1: Theory and Experiments}}.
\newblock Oxford Master Series in Physics. Oxford University Press, 2007.

\bibitem{AmaroSeoane:2010qx}
P.~Amaro-Seoane, J.~Barranco, A.~Bernal and L.~Rezzolla, \emph{{Constraining
  scalar fields with stellar kinematics and collisional dark matter}},
  \href{https://doi.org/10.1088/1475-7516/2010/11/002}{\emph{JCAP} {\bfseries
  1011} (2010) 002}, [\href{https://arxiv.org/abs/1009.0019}{{\ttfamily
  1009.0019}}].

\bibitem{Dolgov:2011cq}
  A.~D.~Dolgov and D.~Ejlli,
  ``Relic gravitational waves from light primordial black holes,''
  Phys.\ Rev.\ D {\bf 84} (2011) 024028
  [arXiv:1105.2303 [astro-ph.CO]].

\bibitem{Ibanez:2012zz}
  L.~E.~Ibanez and A.~M.~Uranga,
  ``String theory and particle physics: An introduction to string phenomenology,''

\bibitem{Choi:2005ge}
K.~Choi, A.~Falkowski, H.~P. Nilles and M.~Olechowski, \emph{{Soft
  supersymmetry breaking in KKLT flux compactification}},
  \href{https://doi.org/10.1016/j.nuclphysb.2005.04.032}{\emph{Nucl. Phys.}
  {\bfseries B718} (2005) 113--133},
  [\href{https://arxiv.org/abs/hep-th/0503216}{{\ttfamily hep-th/0503216}}].

\bibitem{hep-th/0505076}
J.~P. Conlon, F.~Quevedo and K.~Suruliz, \emph{{Large-volume flux
  compactifications: Moduli spectrum and D3/D7 soft supersymmetry breaking}},
  \href{https://doi.org/10.1088/1126-6708/2005/08/007}{\emph{JHEP} {\bfseries
  08} (2005) 007}, [\href{https://arxiv.org/abs/hep-th/0505076}{{\ttfamily
  hep-th/0505076}}].

\bibitem{Aparicio:2014wxa}
  L.~Aparicio, M.~Cicoli, S.~Krippendorf, A.~Maharana, F.~Muia and F.~Quevedo,
  ``Sequestered de Sitter String Scenarios: Soft-terms,''
  JHEP {\bf 1411} (2014) 071
  [arXiv:1409.1931 [hep-th]].

\bibitem{Aparicio:2015psl}
L.~Aparicio, F.~Quevedo and R.~Valandro,
``Moduli Stabilisation with Nilpotent Goldstino: Vacuum Structure and SUSY Breaking,''
JHEP {\bf 1603} (2016) 036
[arXiv:1511.08105 [hep-th]].

\bibitem{Enqvist:2003gh}
K.~Enqvist and A.~Mazumdar, \emph{{Cosmological consequences of MSSM flat
  directions}},
  \href{https://doi.org/10.1016/S0370-1573(03)00119-4}{\emph{Phys. Rept.}
  {\bfseries 380} (2003) 99--234},
  [\href{https://arxiv.org/abs/hep-ph/0209244}{{\ttfamily hep-ph/0209244}}].

\bibitem{0801.0307}
F.~E. Schunck and E.~W. Mielke, \emph{{General relativistic boson stars}},
  \href{https://doi.org/10.1088/0264-9381/20/20/201}{\emph{Class. Quant. Grav.}
  {\bfseries 20} (2003) R301--R356},
  [\href{https://arxiv.org/abs/0801.0307}{{\ttfamily 0801.0307}}].

\bibitem{ArkaniHamed:2006dz}
N.~Arkani-Hamed, L.~Motl, A.~Nicolis and C.~Vafa, \emph{{The String landscape,
  black holes and gravity as the weakest force}},
  \href{https://doi.org/10.1088/1126-6708/2007/06/060}{\emph{JHEP} {\bfseries
  06} (2007) 060}, [\href{https://arxiv.org/abs/hep-th/0601001}{{\ttfamily
  hep-th/0601001}}].

\bibitem{Bowick:1988xh}
M.~J. Bowick, S.~B. Giddings, J.~A. Harvey, G.~T. Horowitz and A.~Strominger,
  \emph{{Axionic Black Holes and a Bohm-Aharonov Effect for Strings}},
  \href{https://doi.org/10.1103/PhysRevLett.61.2823}{\emph{Phys. Rev. Lett.}
  {\bfseries 61} (1988) 2823}.

\bibitem{Amin:2014eta}
  M.~A.~Amin, M.~P.~Hertzberg, D.~I.~Kaiser and J.~Karouby,
  ``Nonperturbative Dynamics Of Reheating After Inflation: A Review,''
  Int.\ J.\ Mod.\ Phys.\ D {\bf 24} (2014) 1530003
  [arXiv:1410.3808 [hep-ph]].

\bibitem{Boyle:2001du}
L.~A. Boyle, R.~R. Caldwell and M.~Kamionkowski, \emph{{Spintessence! New
  models for dark matter and dark energy}},
  \href{https://doi.org/10.1016/S0370-2693(02)02590-X}{\emph{Phys. Lett.}
  {\bfseries B545} (2002) 17--22},
  [\href{https://arxiv.org/abs/astro-ph/0105318}{{\ttfamily
  astro-ph/0105318}}].

\bibitem{Kasuya:2001pr}
S.~Kasuya, \emph{{Difficulty of a spinning complex scalar field to be dark
  energy}}, \href{https://doi.org/10.1016/S0370-2693(01)00867-X}{\emph{Phys.
  Lett.} {\bfseries B515} (2001) 121--124},
  [\href{https://arxiv.org/abs/astro-ph/0105408}{{\ttfamily
  astro-ph/0105408}}].

\bibitem{Amin:2013ika}
M.~A. Amin, \emph{{K-oscillons: Oscillons with noncanonical kinetic terms}},
  \href{https://doi.org/10.1103/PhysRevD.87.123505}{\emph{Phys. Rev.}
  {\bfseries D87} (2013) 123505},
  [\href{https://arxiv.org/abs/1303.1102}{{\ttfamily 1303.1102}}].

\bibitem{Amin:2010jq}
M.~A. Amin and D.~Shirokoff, \emph{{Flat-top oscillons in an expanding
  universe}}, \href{https://doi.org/10.1103/PhysRevD.81.085045}{\emph{Phys.
  Rev.} {\bfseries D81} (2010) 085045},
  [\href{https://arxiv.org/abs/1002.3380}{{\ttfamily 1002.3380}}].

\bibitem{Amin:2010dc}
M.~A. Amin, R.~Easther and H.~Finkel, \emph{{Inflaton Fragmentation and
  Oscillon Formation in Three Dimensions}},
  \href{https://doi.org/10.1088/1475-7516/2010/12/001}{\emph{JCAP} {\bfseries
  1012} (2010) 001}, [\href{https://arxiv.org/abs/1009.2505}{{\ttfamily
  1009.2505}}].

\bibitem{Amin:2011hj}
M.~A. Amin, R.~Easther, H.~Finkel, R.~Flauger and M.~P. Hertzberg,
  \emph{{Oscillons After Inflation}},
  \href{https://doi.org/10.1103/PhysRevLett.108.241302}{\emph{Phys. Rev. Lett.}
  {\bfseries 108} (2012) 241302},
  [\href{https://arxiv.org/abs/1106.3335}{{\ttfamily 1106.3335}}].

\bibitem{Antusch:2015vna}
S.~Antusch, F.~Cefala, D.~Nolde and S.~Orani, \emph{{Parametric resonance after
  hilltop inflation caused by an inhomogeneous inflaton field}},
  \href{https://doi.org/10.1088/1475-7516/2016/02/044}{\emph{JCAP} {\bfseries
  1602} (2016) 044}, [\href{https://arxiv.org/abs/1510.04856}{{\ttfamily
  1510.04856}}].

\bibitem{Antusch:2015ziz}
S.~Antusch and S.~Orani, \emph{{Impact of other scalar fields on oscillons
  after hilltop inflation}},
  \href{https://doi.org/10.1088/1475-7516/2016/03/026}{\emph{JCAP} {\bfseries
  1603} (2016) 026}, [\href{https://arxiv.org/abs/1511.02336}{{\ttfamily
  1511.02336}}].

\bibitem{Gorbunov:2011zzc}
  D.~S.~Gorbunov and V.~A.~Rubakov,
  ``Introduction to the theory of the early universe: Cosmological perturbations and inflationary theory,''
  Hackensack, USA: World Scientific (2011) 489 p

\bibitem{Khlopov:1985jw}
  M.~Khlopov, B.~A.~Malomed and I.~B.~Zeldovich,
  ``Gravitational instability of scalar fields and formation of primordial black holes,''
  Mon.\ Not.\ Roy.\ Astron.\ Soc.\  {\bf 215} (1985) 575.

\bibitem{0805.1748}
M.~C. Johnson and M.~Kamionkowski, \emph{{Dynamical and Gravitational
  Instability of Oscillating-Field Dark Energy and Dark Matter}},
  \href{https://doi.org/10.1103/PhysRevD.78.063010}{\emph{Phys. Rev.}
  {\bfseries D78} (2008) 063010},
  [\href{https://arxiv.org/abs/0805.1748}{{\ttfamily 0805.1748}}].

\bibitem{Cembranos:2015oya}
  J.~A.~R.~Cembranos, A.~L.~Maroto and S.~J.~Núñez Jareño,
  ``Cosmological perturbations in coherent oscillating scalar field models,''
  JHEP {\bf 1603} (2016) 013
  [arXiv:1509.08819 [astro-ph.CO]].

\bibitem{Hidalgo:2017dfp}
  J.~C.~Hidalgo, J.~De Santiago, G.~German, N.~Barbosa-Cendejas and W.~Ruiz-Luna,
  ``Collapse threshold for a cosmological Klein Gordon field,''
  Phys.\ Rev.\ D {\bf 96} (2017) no.6,  063504
  [arXiv:1705.02308 [astro-ph.CO]].



\bibitem{Easther:2010mr}
R.~Easther, R.~Flauger and J.~B. Gilmore, \emph{{Delayed Reheating and the
  Breakdown of Coherent Oscillations}},
  \href{https://doi.org/10.1088/1475-7516/2011/04/027}{\emph{JCAP} {\bfseries
  1104} (2011) 027}, [\href{https://arxiv.org/abs/1003.3011}{{\ttfamily
  1003.3011}}].

\bibitem{Jedamzik:2010dq}
K.~Jedamzik, M.~Lemoine and J.~Martin, \emph{{Collapse of Small-Scale Density
  Perturbations during Preheating in Single Field Inflation}},
  \href{https://doi.org/10.1088/1475-7516/2010/09/034}{\emph{JCAP} {\bfseries
  1009} (2010) 034}, [\href{https://arxiv.org/abs/1002.3039}{{\ttfamily
  1002.3039}}].

\bibitem{Assadullahi:2009nf}
H.~Assadullahi and D.~Wands, \emph{{Gravitational waves from an early matter
  era}}, \href{https://doi.org/10.1103/PhysRevD.79.083511}{\emph{Phys. Rev.}
  {\bfseries D79} (2009) 083511},
  [\href{https://arxiv.org/abs/0901.0989}{{\ttfamily 0901.0989}}].

\bibitem{Erickcek:2011us}
  A.~L.~Erickcek and K.~Sigurdson,
  ``Reheating Effects in the Matter Power Spectrum and Implications for Substructure,''
  Phys.\ Rev.\ D {\bf 84} (2011) 083503
  [arXiv:1106.0536 [astro-ph.CO]].

\bibitem{Kane:2015jia}
  G.~Kane, K.~Sinha and S.~Watson,
  ``Cosmological Moduli and the Post-Inflationary Universe: A Critical Review,''
  Int.\ J.\ Mod.\ Phys.\ D {\bf 24} (2015) no.08,  1530022
  [arXiv:1502.07746 [hep-th]].

\bibitem{Acharya:2008bk}
  B.~S.~Acharya, P.~Kumar, K.~Bobkov, G.~Kane, J.~Shao and S.~Watson,
  ``Non-thermal Dark Matter and the Moduli Problem in String Frameworks,''
  JHEP {\bf 0806} (2008) 064
  [arXiv:0804.0863 [hep-ph]].

\bibitem{Acharya:2009zt}
  B.~S.~Acharya, G.~Kane, S.~Watson and P.~Kumar,
  ``A Non-thermal WIMP Miracle,''
  Phys.\ Rev.\ D {\bf 80} (2009) 083529
  [arXiv:0908.2430 [astro-ph.CO]].

\bibitem{Acharya:2010af}
  B.~S.~Acharya, G.~Kane and E.~Kuflik,
  ``Bounds on scalar masses in theories of moduli stabilization,''
  Int.\ J.\ Mod.\ Phys.\ A {\bf 29} (2014) 1450073
  [arXiv:1006.3272 [hep-ph]].

\bibitem{Aparicio:2015sda}
  L.~Aparicio, M.~Cicoli, B.~Dutta, S.~Krippendorf, A.~Maharana, F.~Muia and F.~Quevedo,
  ``Non-thermal CMSSM with a 125 GeV Higgs,''
  JHEP {\bf 1505} (2015) 098
  [arXiv:1502.05672 [hep-ph]].

\bibitem{Aparicio:2016qqb}
  L.~Aparicio, M.~Cicoli, B.~Dutta, F.~Muia and F.~Quevedo,
  ``Light Higgsino Dark Matter from Non-thermal Cosmology,''
  JHEP {\bf 1611} (2016) 038
  [arXiv:1607.00004 [hep-ph]].

\bibitem{Allahverdi:2013noa}
  R.~Allahverdi, M.~Cicoli, B.~Dutta and K.~Sinha,
  ``Nonthermal dark matter in string compactifications,''
  Phys.\ Rev.\ D {\bf 88} (2013) no.9,  095015
  [arXiv:1307.5086 [hep-ph]].

\bibitem{Allahverdi:2016yws}
  R.~Allahverdi, M.~Cicoli and F.~Muia,
  ``Affleck-Dine Baryogenesis in Type IIB String Models,''
  JHEP {\bf 1606} (2016) 153
  [arXiv:1604.03120 [hep-th]].

\bibitem{Mukhanov:1990me}
V.~F. Mukhanov, H.~A. Feldman and R.~H. Brandenberger, \emph{{Theory of
  cosmological perturbations. Part 1. Classical perturbations. Part 2. Quantum
  theory of perturbations. Part 3. Extensions}},
  \href{https://doi.org/10.1016/0370-1573(92)90044-Z}{\emph{Phys. Rept.}
  {\bfseries 215} (1992) 203--333}.

\bibitem{Lennon:2017tqq}
O.~Lennon, J.~March-Russell, R.~Petrossian-Byrne and H.~Tillim, \emph{{Black
  Hole Genesis of Dark Matter}},
  \href{https://doi.org/10.1088/1475-7516/2018/04/009}{\emph{JCAP} {\bfseries
  1804} (2018) 009}, [\href{https://arxiv.org/abs/1712.07664}{{\ttfamily
  1712.07664}}].

\end{thebibliography}\endgroup
\bibliographystyle{JHEP}

\end{document}